\documentclass{report}

\usepackage{psfig}

\newcommand{\Mo}{$\mbox{M}_\odot$}

\newcommand{\rxp}{RX~J1824.5$-$2452P}
\newcommand{\rxe}{RX~J1824.5$-$2452E}
\newcommand{\ApJ}{ApJ}

\newcommand{\MNRAS}{MNRAS}

\newcommand{\NS}{neutron star~}

\newcommand{\gapr}{\raisebox{-.6ex}{\mbox{$\stackrel{>}{\mbox{\scriptsize$\sim$}}\:$}}}
\newcommand{\lapr}{\raisebox{-.6ex}{\mbox{$\stackrel{<}{\mbox{\scriptsize$\sim$}}\:$}}}

\begin{document}

 %******************************************************************************
 %                                                                             %  
 %                               Coverpage                                     %
 %                                                                             %
 %******************************************************************************
 %
 \noindent
 \centerline{{\Huge THE CENTURY OF SPACE SCIENCE}}\\[1.5cm]
 \centerline{{\Huge Pulsars and Isolated Neutron Stars}}\\[1cm]
 \centerline{by}\\[1cm]
 \centerline{\Large Werner Becker$^1$ \& George Pavlov$^2$}\\[0.3cm]
 \noindent
 \begin{center}
 ${}^1$Max-Planck Institut f\"ur extraterr. Physik, Giessenbachstrasse 1,\\
 D-85740 Garching, Germany\\[0.5cm]
 ${}^2$Pennsylvania State University, 525 Davey Lab, University Park,\\ PA 16802,
 USA
 \end{center}
 \vspace{11cm}
 \centerline{\parbox{11cm}{
 Editors: Johan Bleeker, Johannes Geiss and Martin Huber. To be published by
 Kluwer Academic Publishers in "The Century of Space Science"}}
 \vspace{3ex}
 \centerline{Original: January 2001}
 \vspace{1ex}
 \centerline{Revision: December 2001}

 \thispagestyle{empty}

\newpage

\tableofcontents
\thispagestyle{empty}
\newpage

\setcounter{page}{1}
\setcounter{chapter}{7}

%******************************************************************************
%                                                                             %
%  SECTION:  Introduction: Historical Overview                                %
%                                                                             %
%******************************************************************************
%
\chapter{The Milky Way $-$ Pulsars and Isolated Neutron Stars}
\section{Introduction: Historical Overview}

  The idea of {\em neutron stars} can be traced back to early 1930's, when
  Subrahmanyan Chandrasekhar, whilst investigating the physics of stellar
  evolution, discovered that there is no way for a collapsed stellar core
  with a mass more than 1.4 times the solar mass, \Mo, to hold itself up
  against gravity once its nuclear fuel is exhausted (Chandrasekhar 1931).
  This implies that a star left with $M>1.4\;$\Mo\ (the {\em Chandrasekhar 
  limit}) would keep collapsing and eventually disappear from view.

  After the discovery of the neutron by James Chadwick in 1932, Lev Landau
  was the first who speculated on the possible existence of a  {\em star 
  composed entirely of neutrons} (Landau 1932; Rosenfeld 1974). Using the 
  newly-established Fermi-Dirac statistics and basic quantum mechanics, he 
  was able to estimate that such a star, consisting of $\sim 10^{57}$ neutrons, 
  would form a giant nucleus with a radius of the order of $R\sim (\hbar/m_n c) 
  (\hbar c/G m_n^2)^{1/2} \sim 3\times 10^5\;\mbox{cm}$, in which $\hbar$,
  $c$, $G$ and $m_n$ are the Planck constant, the speed of light, the gravitation
  constant and the mass of the neutron. In view of the peculiar stellar parameters, 
  Landau called these objects ``unheimliche Sterne'' (weird stars), expecting 
  that they would never be observed because of their small size and expected low 
  optical luminosity.

  Walter Baade and Fritz Zwicky were the first who proposed the idea that neutron 
  stars could be formed in {\em supernovae} (Baade \& Zwicky 1934). First models 
  for the structure of neutron stars were worked out in 1939 by Robert Oppenheimer 
  and George Volkoff, who calculated an upper limit for the neutron star mass. 
  Using general relativistic equilibrium equations and assuming that the star 
  is entirely described by an ideal (i.e.~non-interacting) Fermi gas of neutrons, 
  they found that any star more massive than 3\,\Mo\ (Oppenheimer-Volkoff limit) 
  will suffer runaway gravitational collapse to form a black hole (Oppenheimer \& 
  Volkoff 1939). Unfortunately, their pioneering work did not predict anything
  astronomers could actually observe, and the idea of neutron stars was not taken 
  seriously by the astronomical community. Neutron stars therefore had remained
  in the realm of imagination for nearly a quarter of century, until in the 60's 
  a series of epochal discoveries were made in high-energy and radio astronomy.

  X-rays and gamma-rays can only be observed from above the earth's 
  atmosphere\footnote{X-rays are absorbed at altitudes 20$-$100 km.}, which 
  requires detectors to operate from high flying balloons, rockets or satellites. 
  One of the first X-ray detectors brought to space was launched by Herbert Friedman 
  and his team at the Naval Research Laboratory in order to investigate the 
  influence of solar activity on the propagation of radio signals in the earth's 
  atmosphere (cf.~H.~Friedman, this book). Using simple proportional counters put 
  on old V-2 (captured in Germany after the World War II) and Aerobee rockets, 
  they were the first who detected X-rays from the very hot gas in the solar 
  corona. However, the intensity of this radiation was found to be a factor 
  $10^{6}$ lower than that measured at optical wavelengths. In the late 50's, 
  it was therefore widely believed that all other stars, much more distant 
  than the Sun, should be so faint in X-rays that further observations at 
  that energy range would be hopeless.
  On the other hand, results from high-energy cosmic ray experiments suggested 
  that there exist celestial objects (e.g.~supernova remnants) which produce 
  high-energy cosmic rays in processes which, in turn, may also produce X-rays
  and gamma-rays (Morrison et al.~1954, Morrison 1958). These predictions were
  confirmed in 1962, when the team led by Bruno Rossi and Riccardo Giacconi
  accidentally detected X-rays from Sco X-1.
  With the aim to search for fluorescent X-ray photons from the
  Moon\footnote{The Moon was selected as a target because it was expected that
  a state-of-the-art detector available at that time would not be sensitive
  enough to detect X-rays from extra-solar sources. ``We felt [...] that it
  would be very desirable to consider some intermediate target which could
  yield concrete results while providing a focus for the development of more
  advanced instrumentation which ultimately would allow us to detect cosmic
  X-ray sources'' (Giacconi 1974).}, they launched an Aerobee rocket on 12
  June 1962 from White Sands (New Mexico) with three Geiger counters as
  payload, each having a $\sim 100^\circ$ field of view and an effective
  collecting area of about $10\,\mbox{cm}^2$ (Giacconi 1974). The experiment 
  detected X-rays not from the Moon but from a source located in the constellation
  Scorpio, dubbed as Sco X-1, which is now known as the brightest
  extra-solar X-ray source in the sky. Evidence for a weaker source in the Cygnus
  region and the first evidence for the existence of a diffuse isotropic
  X-ray background was also reported from that experiment (Giacconi et
  al.~1962). Subsequent flights launched to confirm these first results
  detected Tau X-1, a source in the constellation Taurus which coincided
  with the Crab supernova remnant (Bowyer et al.~1964).
  Among the various processes proposed for the generation of the detected
  X-rays was {\em thermal radiation from the surface of a hot neutron star}
  (Chiu \& Salpeter 1964), and searching for this radiation has become a
  strong motivation for further development of X-ray astronomy.  However,
  the X-ray emission from the  Crab supernova remnant was found to be of
  a finite angular size ($\sim 1$ arcmin) whereas a neutron star was expected
  to appear as a point source.
  Thus, the early X-ray observations were not sensitive enough to prove
  the existence of neutron stars. This was done a few years later by 
  radio astronomers.

  In 1967, Jocelyn Bell, a graduate student under the supervision of
  Anthony Hewish at the Cambridge University of England, came across a
  series of pulsating radio signals while using a radio telescope
  specially constructed to look for rapid variations in the radio emission
  of quasars. These radio pulses, 1.32 seconds apart, with remarkable
  clock-like regularity, were emitted from an unknown source in the sky
  at right ascension 19$^{\rm h}$ 20$^{\rm m}$ and declination $+23^\circ$. Further
  observations refined the pulsating period to 1.33730113 seconds. The
  extreme precision of the period suggested at first that
  these signals might be generated by extraterrestrial intelligence. They
  were subsequently dubbed as LGM1, an acronym for ``Little Green Man 1''
  (Bell 1977). However, as a few more similar sources had been detected,
  it became clear that a new kind of celestial objects was discovered.
  The link between these pulsating radio sources, which were
  called {\em pulsars}, and fast spinning neutron stars was provided by Franco
  Pacini (1967, 1968) and Thomas Gold (1968, 1969). Pacini, then a young
  postdoc at the Cornell University, had published a paper a few months
  before the discovery by Bell and Hewish in which he proposed that the
  {\em rapid rotation  of a highly magnetized  neutron star} could be the
  source of energy in the Crab Nebula. This prediction was based on the
  pioneering work of Hoyle, Narlikar and Wheeler (1964), who had proposed
  that a magnetic field of $10^{10}$ Gauss might exist on a neutron star
  at the center of the Crab Nebula. The most fundamental ideas on the
  nature of the pulsating radio sources were published by Gold (1968; 1969)
  in two seminal {\em Nature} papers. In these papers Gold introduced
  the concept of the {\em rotation-powered pulsar} which radiates at
  the expense of its rotational energy (pulsar spins down as rotational
  energy is radiated away) and recognized that the  rotational energy
  is lost via electromagnetic radiation of the rotating magnetic dipole
  and emission of relativistic particles. The particles are accelerated 
  in the pulsar magnetosphere along the curved magnetic field lines and 
  emit the observed intense curvature and synchrotron 
  radiation\footnote{When a charged relativistic particle moves along a 
  curved magnetic field line, it is accelerated tranversely and radiates.
  This {\em curvature radiation} is closely related to {\em synchrotron
  radiation} caused by gyration of particles around the magnetic field lines.}.

  Since those early days of pulsar astronomy more than 1000 radio pulsars
  have been discovered (see, e.g., the catalog by Taylor, Manchester \& Lyne 
  1993 which lists about half of them). The discovery of the first radio pulsar 
  was very soon followed by the discovery of two most famous pulsars, the 
  fast 33 ms pulsar in the Crab Nebula (Staelin \& Reifenstein 1968) and the 
  89 ms pulsar in the Vela supernova remnant (Large et al.~1968). The fact 
  that these pulsars are located within supernova remnants provided striking
  confirmation that neutron stars are born in core collapse supernovae from 
  massive main sequence stars. These exciting radio discoveries triggered 
  subsequent pulsar searches at nearly all wavelengths. 

  Cocke, Disney \& Taylor (1969) discovered optical pulses from the Crab pulsar, 
  whereas its X-ray pulsations in the $1.5-10$ keV range were discovered by 
  Friedman's group at the Naval Research Laboratory (Fritz et al.~1969) and by 
  the team of the Massachusetts Institute of Technology (Bradt et al.~1969) three 
  months later. Using a plastic scintillator platform, Hillier et al.~(1970) flew
  a balloon-born experiment over southern England and detected its pulsed gamma-rays 
  at a $\sim 3.5\sigma$ level at energies greater than 0.6 MeV. These early 
  multi-wavelength observations showed that the pulses are all phase-aligned, 
  with a pulse profile which was very nearly the  same at all wavelengths, 
  suggesting a common emission site for the radiation. Moreover, the power 
  observed at the high photon energies exceeded that in the optical band by 
  more than two orders of magnitude, justifying the need for more sensitive 
  satellite-based X-ray and gamma-ray observatories to perform more detailed 
  investigations of the emission mechanism of pulsars and to survey the sky 
  for other X-ray and gamma-ray sources.

  The first earth-orbiting mission dedicated entirely to celestial X-ray
  astronomy, {\sl SAS-1} (Small Astronomy Satellite 1), was launched by NASA 
  in December 1970 from a launch site in Kenya. The observatory, later named 
  {\sl Uhuru}\footnote{Uhuru means `freedom' in Swahili.}, was sensitive in 
  the range $2-20$ keV and equipped with two sets of proportional counters 
  having a collecting area of 840~cm$^2$ (Giacconi et al.~1971). It was designed 
  to operate in survey mode, allowing for the first time to scan the whole sky 
  with a sensitivity of $1.5 \times 10^{-11}\;\mbox{ergs s}^{-1}\mbox{cm}^{-2}$. In 
  somewhat more than two years of very successful operation,  339 new X-ray sources
  were detected (Forman et al.~1978), belonging to the group of accreting binaries,
  supernova remnants, Seyfert galaxies and clusters of galaxies. By far the largest 
  sample of objects was found to belong to the group of accretion-powered 
  pulsars --- neutron stars in binary systems accreting matter from a companion star. 
  As the matter spirals in onto the neutron star surface or heats up in an accretion 
  disc, strong X-ray radiation is emitted (van den Heuvel et al., this book).

  The next major step in high-energy astronomy was the launch of {\sl SAS-2} in 
  November 1972, the first satellite dedicated exclusively to gamma-ray astronomy 
  (Fichtel et al.~1975). The detector, a spark chamber, was sensitive in the energy 
  range $35-1000$ MeV. Although the mission lasted only seven months and ended by 
  a failure of the low-voltage power supply, its measurements confirmed  the 
  existence of the gamma-ray pulses from the Crab (Kniffen et al.~1974) and 
  discovered the gamma-ray pulses from the Vela pulsar (Thompson et al.~1975), which 
  was found to be the strongest gamma-ray source in the sky. The Vela lightcurve
  was characterized by two relatively sharp peaks, separated by 0.4 in phase
  (as observed for the Crab) but not phase-aligned with the radio and optical
  pulses.

  In addition, a few unidentified gamma-ray sources were detected, among them 
  {\em Geminga}\footnote{The source was dubbed with the name Geminga, a pun 
  in Milanese dialect in which {\em gh'\`e minga} means {\it it is not there} 
  or {\it it does not exist}, by Giovanni Bignami --- see Bignami \& Caraveo (1996) 
  for a comprehensive description of the Geminga story, from the first discovery to 
  the final identification. It is amusing to note, the name {\em Geminga} inspired 
  Eric Cohez to choose the title of his science fiction book {\em Geminga: la 
  civilization perdue}.}, a faint source in the Gemini region from which $\sim 100$ 
  $\gamma$-ray photons had been recorded, but which had to await its final 
  identification about 20 years later.
  Gamma-ray astronomy, from its beginning, was often hampered by the relatively 
  small number of detected photons and large position error boxes, typically 
  $\sim 0.5^\circ - 1^\circ$. This position uncertainty strongly complicated 
  follow-up observations for optical and X-ray counterparts. Scientific publications  
  describing data analysis techniques optimized for  `sparse data', particularly 
  the timing analysis aimed at pulsation search,  were therefore always ranked 
  high on the gamma-ray market.

  The first complete and detailed gamma-ray map of the Galaxy was provided
  by the ESA mission {\sl COS-B}, launched in August 1975. Developed under 
  the responsibility of a group of European research laboratories known as 
  the Caravane Collaboration\footnote{formed of members from MPE-Garching, 
  CEN-Saclay, SRON-Leiden (today Utrecht), IFCAI-Palermo, CNR-Milano and
  SSD-ESTEC.}, the satellite carried two scientific payloads, a digital 
  spark chamber, sensitive in the range $0.03-5$ GeV, and a $2-12$ keV 
  collimated proportional counter which was used as a pulsar synchronizer. 
  Because of a not very accurate on-board clock calibration, the latter 
  was to ensure the synchronization of the X-ray and gamma-ray pulses 
  from isolated pulsars, like the Crab and Vela pulsars,  and accreting 
  pulsars in X-ray binaries. It was further used to determine pulsar 
  ephemeris from the temporal analysis of X-ray data, independently 
  from the availability of exact radio ephemeris. The high sensitivity 
  of the gamma-ray detector allowed Kanbach et al.~(1980) to conduct 
  the first detailed temporal and spectral study of the Vela pulsar
  in the range $0.05-3$ GeV. The pulsar's spectrum was found to be  
  represented by a power-law ${\rm d}N/{\rm d}E\propto E^{-\alpha}$ (with a 
  photon index of $\alpha=1.89\pm 0.06$ for the phase-averaged spectrum), but 
  appreciable differences of the photon index were detected for different 
  pulsar phases (e.g., the inter-pulse emission, first detected in the
  {\sl COS-B} data, was found to have the hardest spectrum). The {\sl COS-B}  
  observations of the Crab pulsar provided much improved photon statistics 
  which resulted in a more accurate pulse profile (Wills et al.~1982) and  
  detailed spectral studies (Clear et al.~1987). 

\begin{figure}[h]
  \begin{picture}(120,130)(-15,0)
  \put(0,0){\psfig{file=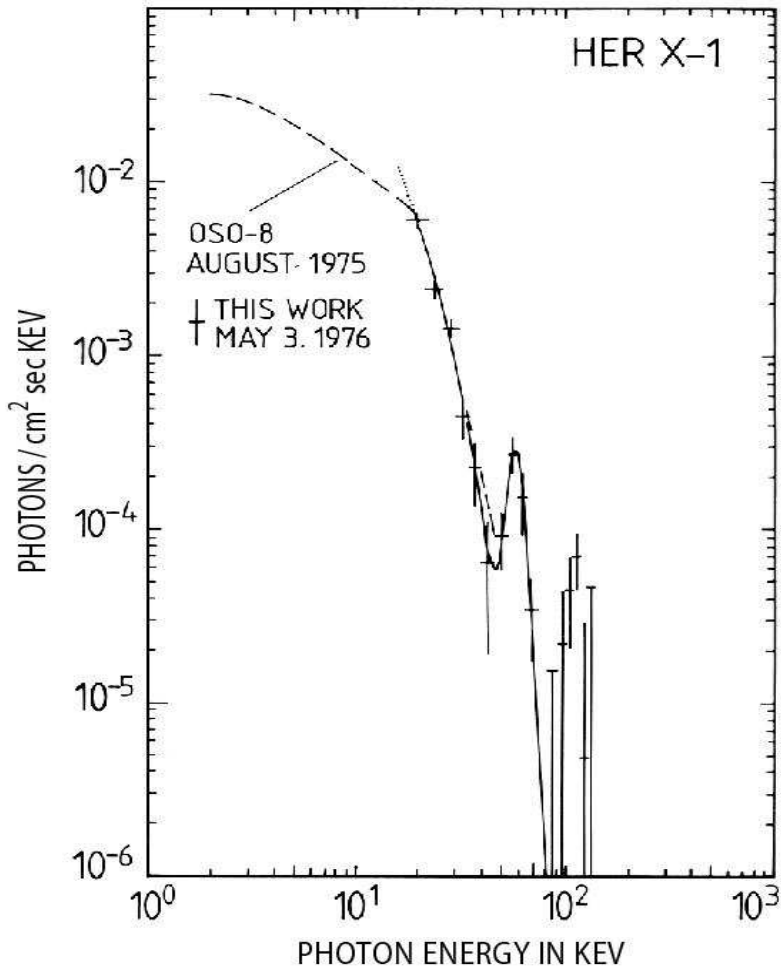,height=13cm,width=11cm,clip=}}
  \end{picture}
  \caption[]{\small Unfolded X-ray spectrum from Hercules X-1 (from Tr\"umper et
   al.~1978), showing the first measurement of a cyclotron line in a pulsed 
   spectrum of an accreting neutron star.}\label{her_X1} 
  \end{figure}

  Many radio pulsars had been observed by mid-seventies, and two of them, the 
  Crab and Vela pulsars, had been detected at high photon energies. Although  
  the interpretation of both isolated and accreting pulsars as neutron stars 
  with enormous magnetic fields, $\sim 10^{12}$ G, had been generally accepted, 
  no direct evidence on the existence of such huge fields had been obtained. 
  This evidence came from a remarkable spectral observation of Hercules X-1, 
  an accreting binary pulsar discovered with {\sl Uhuru} by Tananbaum et 
  al.~(1972). On May 3, 1976, a team of the Max-Planck Institut f\"ur 
  extraterrestrische Physik in Garching and the Astronomische Institut of 
  the University of T\"ubingen, led by Joachim Tr\"umper, launched from 
  Palestine (Texas) a balloon experiment, equipped with a collimated NaI 
  scintillation counter and a NaI-CsI-phoswich detector, sensitive in the 
  range $15-160$ keV. They easily detected the 1.24 s pulsations up to 80 
  keV (Kendziorra et al.~1977). However, when Bruno Sacco and Wolfgang 
  Pietsch attempted to fit the observed spectrum with usual continuum 
  spectral models, they found that a one-component continuum model cannot 
  represent the data --- all fits gave unacceptably large residuals at 
  $\sim 40-60$ keV. Further data analysis confirmed that the spectral 
  feature was not an artifact (e.g., due to incomplete shielding of the 
  in-flight calibration source ${}^{241}$Am, which emitted a spectral line 
  at E=59.5 keV). It was Joachim Tr\"umper who first recognized that the
  excess emission at 58 keV (or an absorption feature at 42 keV, depending 
  on interpretation -- cf.~Fig.\ref{her_X1}) could be associated with the 
  resonant electron cyclotron emission or absorption in the hot polar plasma 
  of the rotating neutron star. The corresponding magnetic field strength would 
  then be $6\times 10^{12}$ or $4\times 10^{12}$ G (Tr\"umper et al.~1978).
  This observation provided the first direct measurement of a neutron
  star magnetic field and confirmed the basic theoretical predictions
  that neutron stars are highly magnetized, fast spinning compact objects.

  Beginning in 1977, NASA launched a series of large scientific payloads called 
  {\sl High Energy Astrophysical Observatories}\footnote{The
  dramatic history of the HEAO project and the experiments on board of
  HEAO satellites are lively described by Wallace Tucker (1984).}:
  HEAO 1 (Aug 1977$-$Jan 1979), HEAO 2 (Nov 1978$-$Apr 1981), and 
  HEAO 3 (Sep 1979$-$May 1981). Particularly important results on 
  isolated neutron stars, among many other X-ray sources, were obtained
  with {\sl HEAO 2}, widely known as the {\sl Einstein} X-ray observatory 
  (Giacconi et al.~1979), which carried the first imaging X-ray telescope
  on a satellite. Among four focal plane detectors of {\sl Einstein}, two 
  proved to be particularly useful for detecting and studying isolated 
  neutron stars. The High Resolution Imager (HRI), a micro-channel plate 
  detector, sensitive in the $0.15-4$ keV energy band, with about 5 arcsec 
  angular resolution, was designed to use the imaging capability of the 
  X-ray telescope. However, it had no energy resolution and its field of
  view was small, $\simeq 25$ arcmin. The Imaging Proportional Counter 
  (IPC), the workhorse of the observatory, could detect weaker sources 
  than the HRI and had a wider field of view, $\simeq 1^\circ$, but its 
  imaging resolution was about 1 arcmin. It was capable of studying spectra 
  with modest energy resolution in the range $0.2-4$ keV.

  {\sl Einstein} investigated the soft X-ray radiation from the previously
  known Crab and Vela pulsars and resolved the compact nebula around
  the Crab pulsar (Harnden \& Seward 1984). It discovered pulsed X-ray 
  emission from two other very young pulsars, PSR B0540$-$69 in the Large
  Magellanic Cloud (Seward, Harnden, \& Helfand 1984) and PSR B1509$-$58 
  (Seward \& Harnden 1982), with periods 50 ms and 150 ms, respectively. 
  Interestingly, these pulsars were the first ones discovered in the X-ray 
  band and only subsequently at radio frequencies. No pulsations from the 
  Vela pulsar were found in the soft X-ray band.

  {\sl Einstein} also detected three middle-aged radio pulsars, PSR B0656+14
  (C\'ordova et al.~1989), B1055$-$52 (Cheng \& Helfand 1983), and B1951+32
  (Wang \& Seward 1984). Also, X-ray counterparts of two nearby old radio 
  pulsars, PSR B0950+08 and  B1929+10, were identified, based on positional 
  coincidence (Seward \& Wang 1988). In addition, many supernova remnants 
  were mapped --- 47 in our Galaxy (Seward 1990) and 10 in the Magellanic 
  Clouds (Long \& Helfand 1979), and several neutron star candidates were 
  detected as faint, soft point sources close to the centers of the supernova 
  remnants RCW 103 (Tuohy \& Garmire 1980), PKS 1209$-$51/52 (Helfand \& 
  Becker 1984), Puppis A (Petre et al.~1982) and Kes 73 (Kriss et al.~1985).

  Some additional information on isolated neutron stars was obtained by {\sl EXOSAT} 
  ({\sl European X-ray Observatory Satellite} --- see Taylor et al.~1981), which was
  equipped with a low-energy detector with imaging capability and grating (0.04$-$2 keV)
  and a medium-energy proportional counter (1.5$-$50 keV). In particular, it measured
  the soft X-ray spectra of the middle-aged pulsar PSR B1055$-$52 (Brinkmann \&
  \"Ogelman 1987) and of a few neutron star candidates in supernova remnants 
  (e.g., PKS 1209$-$51/52 -- Kellett et al.~1987).
  
  In spite of the major advance in the field of high-energy astronomy provided
  by the space observatories (particularly, by {\sl Einstein}) in the 70's$-$80's,
  the results on isolated neutron stars made it clear that more sensitive
  instruments and multi-wavelength observations were required to understand the 
  spatial, temporal and spectral emission properties of these objects.  For 
  instance, {\sl Einstein} was able to detect X-ray pulses only from the young 
  and powerful Crab-like pulsars, whereas only flux estimates could be obtained 
  for the other detected neutron stars. Only two pulsars, Crab and Vela, were 
  detected in the gamma-ray and optical ranges.

  The situation improved drastically in the last decade of the century, which 
  can be seen as the ``decade of space science''. The first X-ray satellite in 
  a series of several launched to explore the Universe from space was the 
  German/US/UK mission {\sl ROSAT} ({\sl R\"ontgen Satellit} -- see Tr\"umper 
  1983), sensitive in the $0.1-2.4$ keV band. Equipped with an imaging X-ray 
  telescope and three detectors, Position Sensitive Proportional Counter (PSPC),
  High Resolution Imager (HRI) and EUV Wide-Field Camera, the observatory 
  performed very successful observations of all kinds of astronomical objects 
  in more than eight years of its life (June 1990 $-$ Feb.~1999). During the 
  first 6 months of the mission, the {\sl ROSAT All-Sky Survey}, with the 
  limiting sensitivity of $\sim 3\times 10^{-12}\mbox{erg s}^{-1}\,\mbox{cm}^{-2}$, 
  provided valuable information on fluxes for all the known radio pulsars.
  This, in particular, made it possible to constrain the neutron star cooling 
  scenarios on a large sample of these objects (Becker, Tr\"umper \& \"Ogelman
  1993; Becker 1995).

  The complement to ROSAT, covering the harder X-ray band $1-10$ keV, was the 
  Japanese/US mission {\sl ASCA} ({\sl Advanced Satellite for Cosmology and 
  Astrophysics} -- see Tanaka et al.~1994), launched in 1993. It was the first 
  X-ray observatory equipped with a charge-coupled-device (CCD) imager -- the 
  Solid-state Imaging Spectrometer (SIS), with a much better spectral resolution 
  than the  ROSAT PSPC. The Gas Imaging Spectrometer (GIS), which was operated 
  in parallel, provided timing information in addition. Launched in 1992, the 
  {\sl EUVE} ({\sl Extreme Ultraviolet Explorer} --  see Bowyer 1990), 
  sensitive in the  range $70-760$ \AA, has been able to observe several 
  neutron stars at very soft X-rays, $0.07-0.2$ keV.
  The contributions to the neutron star research, provided by the instruments
  aboard the Italian/Dutch X-ray mission {\sl BeppoSAX} (Butler \& Scarsi 1990), 
  sensitive in the range of $0.1-200$ keV, and the USA's {\sl RXTE} ({\sl Rossi 
  X-ray Timing Explorer} -- see Bradt, Swank \& Rothschild 1990), both launched 
  in the mid-90's, were particularly useful for studying X-ray binaries, including 
  accretion-powered pulsars (see van den Heuvel et al., this book).
 
  The new advance in the study of gamma-ray emission from neutron stars was
  provided by nine years ($1991-2000$) of operation of {\sl CGRO} ({\sl 
  Compton Gamma-Ray Observatory} -- see Kniffen 1990), which has explored the 
  gamma-ray sky in the broad range from 50 keV to 30 GeV with four instruments. 
  Particularly useful for observations of isolated neutron stars was the Energetic 
  Gamma-Ray Experiment Telescope (EGRET), which detected five new gamma-ray pulsars 
  (Thompson et al.~1999), in addition to the previously observed Crab and Vela 
  pulsars. In particular, the gamma-ray source Geminga, known since 1972,
  was identified as a pulsar (Bertsch et al.~1992) after the discovery of 
  coherent pulsations in X-rays with ROSAT (Halpern \& Holt 1992).

  Finally, the outstanding capabilities of the {\sl Hubble Space Telescope} 
  ({\sl HST}), launched in 1990, enabled astronomers to directly observe neutron 
  stars, despite their extremely small size, in the IR/optical/UV range (see 
  Fig.\ref{0656_spec}), which appeared completely impossible a few decades ago. 
  Of particular interest was the discovery of the (presumably thermal)
  optical-UV radiation from old neutron stars (Pavlov, Stringfellow \& C\'ordova 
  1996a; Walter \& Matthews 1997).
 
  Our current understanding of the high-energy emission of neutron stars, summarized 
  in Section \ref{current_pic}, is largely based on the results obtained with these 
  space observatories. Although some of them have completed their service and rest 
  on the ocean bottom, new and more powerful X-ray missions have taken their place
  just before the onset of the new century --- {\sl Chandra}, with the outstanding 
  imaging capability of its telescope and {\sl XMM-Newton} with its unprecedently high 
  spectral sensitivity and collecting power. It is therefore safe to say that in the 
  very near future a wealth of new X-ray data on various astronomical objects, 
  including isolated neutron stars, will become available and will have a major 
  impact on our current understanding of these objects.

 %******************************************************************************
 %                                                                             %  
 % SECTION:   Physics and Astrophysics of Isolated Neutron Stars               %
 %                                                                             %
 %******************************************************************************

 \section{Physics and Astrophysics of Isolated Neutron Stars} \label{astrophysics}

   Neutron stars represent unique astrophysical laboratories which allow us to 
   explore the properties of matter under the most extreme conditions observable 
   in nature\footnote{Although black holes are even more compact than neutron 
   stars, they can only be observed through the interaction with their 
   surroundings.}. Studying neutron stars is therefore an interdisciplinary 
   field, where astronomers and astrophysicist work together with a broad 
   community of physicists. Particle, nuclear and solid-state physicists
   are strongly interested in the internal structure of neutron stars 
   which is determined by the behavior of matter at densities above the 
   nuclear density $\rho_{\rm nuc} = 2.8\times 10^{14} \mbox{g cm}^{-3}$.
   Plasma physicists are modeling the pulsar emission mechanisms using 
   electrodynamics and general relativity. It is beyond the scope of this 
   section to describe in detail the current status of the theory of 
   neutron star structure or the magnetospheric emission models. 
   We rather refer the reader to the literature (Michel 1991;
   Beskin, Gurevich \& Istomin 1993; Glendenning 1996; Weber 1999) and 
   provide only the basic theoretical background relevant to section 
   \ref{current_pic} which summarizes the observed high-energy emission 
   properties of rotation-powered pulsars and radio-quiet neutron stars.

 %******************************************************************************
 %                                                                             %
 % SubSection:  Rotation-powered Pulsars: The Magnetic Braking Model           %
 %                                                                             %
 %******************************************************************************

  \subsection{Rotation-powered Pulsars: The Magnetic Braking Model} \label{braking_model}

   Following the ideas of Pacini (1967, 1968) and Gold (1968, 1969), the more 
   than 1000 radio pulsars detected so far can be interpreted as rapidly spinning, 
   strongly magnetized neutron stars radiating at the expense of their rotational 
   energy.
   This very useful concept allows one to obtain a wealth of information on 
   basic neutron star/pulsar parameters just from measuring the pulsar's 
   period and period derivative. Using the Crab pulsar as an example will 
   make this more clear. A neutron star with a canonical radius of $R=10$ km 
   and a mass of $M=1.4\,$ \Mo\/ has a moment of inertia $I\approx (2/5)\, M R^2 
   \approx 10^{45}\;\mbox{g cm}^2$. The Crab pulsar spins with a period of 
   $P=33.403$ ms. The rotational energy of such a star is $E_{\rm rot}=2\,
   \pi^2\;I\;P^{-2} \approx 2\times 10^{49}\;\mbox{erg}$. This is comparable 
   with the energy released in thermonuclear burning by a usual star over 
   many million years. Very soon after the discovery of the first radio 
   pulsars it was noticed that their spin periods increase with time. For 
   the Crab pulsar, the period derivative is $\dot{P}= 4.2\times 
   10^{-13}$ s~s$^{-1}$, implying a decrease in the star's rotation 
   energy of ${\rm d}E_{\rm rot}/{\rm d}t \equiv \dot{E}_{\rm rot}=- 4\pi^2 I 
   \dot{P}P^{-3} \approx 4.5 \times 10^{38}\,\mbox{erg s}^{-1}$. Ostriker 
   \& Gunn (1969) suggested that the pulsar slow-down is due to the braking 
   torque exerted on the neutron star by its magneto-dipole radiation, that 
   yields $\dot{E}_{\rm brak}= -(32\pi^4/3c^3)\, B_\perp^2\, R^6\, P^{-4}$ for
   the energy loss of a rotating magnetic dipole, where $B_\perp$ is the 
   component of the  equatorial magnetic field perpendicular to the 
   rotation axis. Equating $\dot{E}_{\rm brak}$ with $\dot{E}_{\rm rot}$,
   we find $B_\perp = 3.2 \times 10^{19} (P\,\dot{P})^{1/2}$ Gauss. 
   For the Crab pulsar, this yields $B_\perp = 3.8\times 10^{12}$ G. From 
   $\dot{E}_{\rm rot}= \dot{E}_{\rm brak}$ one further finds that $\dot{P}
   \propto P^{-1}$, for a given $B_\perp$. This relation can be generalized 
   as $\dot{P} = k\,P^{2-n}$, where $k$ is a constant, and $n$ is the 
   so-called braking index ($n=3$ for the magneto-dipole braking).  
   Assuming that the initial rotation period $P_0$ at the time $t_0$ of 
   the neutron star formation was much smaller than today, at $t=t_0+\tau$, 
   we obtain $\tau= P/[(n-1)\dot{P}]$, or $\tau = P/(2\dot{P})$ for $n=3$. 
   This quantity is called the characteristic spin-down age. It is a measure 
   for the time span required to lose the rotational energy 
   $E_{\rm rot}(t_0)-E_{\rm rot}(t)$ via magneto-dipole radiation. For the 
   Crab pulsar one finds $\tau= 1258$ yrs. As the neutron star in the Crab 
   supernova remnant is the only pulsar for which its historical age is known 
   (the Crab supernova was observed by Chinese astronomers in 1054 AD), we
   see that the spin-down age exceeds the true age by about 25\%. Although 
   the spin-down age is just an  estimate for the true age of the pulsar, it 
   is the only one available for pulsars other than the Crab, and it is
   commonly used in evolutionary studies (e.g., neutron star cooling). 

\begin{figure}[h]
   \centerline{\psfig{file=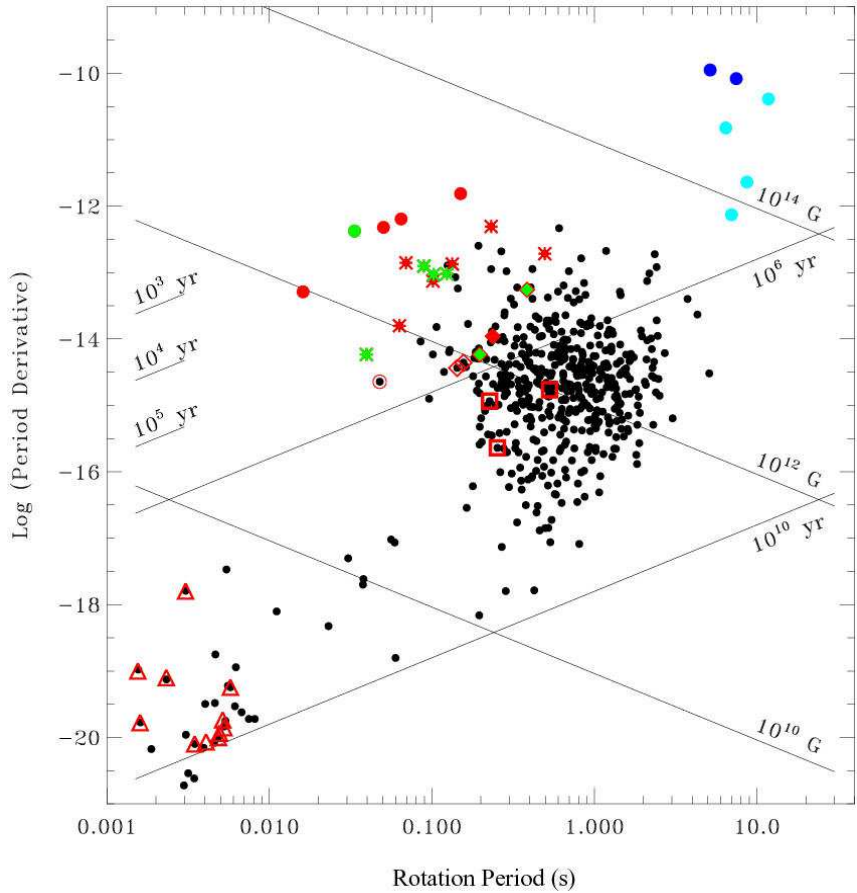,height=12cm}}
   \caption[]{\small The $P - \dot{P}$ diagram --- distribution of 
   rotation-powered pulsars (small black dots) over their spin parameters.
   The straight lines correspond to constant ages $\tau=P/(2\dot{P})$ and magnetic 
   field strengths $B_\perp=3.2\times 10^{19}(P\dot P)^{1/2}$. Separate from the 
   majority of ordinary-field pulsars are the millisecond pulsars in the lower left 
   corner and the putative magnetars --- soft gamma-ray repeaters (dark blue) and 
   anomalous X-ray pulsars (light blue) in the upper right. Although magnetars 
   and anomalous X-ray pulsars are not rotation-powered, they are included in this 
   plot to visualize their estimated superstrong magnetic fields. X-ray detected 
   pulsars are indicated by colored symbols.  Green symbols indicate gamma-ray 
   pulsars. \label{p_pdot}}
   \end{figure}

   A plot of observed periods versus period derivatives is shown in Figure
   \ref{p_pdot}, using the pulsars from the Princeton Pulsar Catalog (Taylor
   et al.~1993). Such a $P$-$\dot{P}$ diagram is extremely useful for 
   classification purposes. The colored symbols represent those 35 pulsars 
   which, by the end of 2000, have been detected at X-ray energies. Among them 
   are the 7 gamma-ray pulsars indicated by green color. The objects in the 
   upper right corner represent the soft-gamma-ray repeaters (SGRs) and 
   anomalous X-ray pulsars (AXPs) which have been suggested to be magnetars 
   (see \ref{magnetars}).\\[-4ex]

   Although the magnetic braking model is generally accepted, the {\em  observed} 
   spin-modulated emission, which gave pulsars their name, is found to account only 
   for a small fraction of $\dot{E}$. The efficiencies, $\eta = L/\dot{E}$, observed 
   in the radio and optical bands are typically in the range $\sim 10^{-7}-10^{-5}$,  
   whereas they are about $10^{-4}-10^{-3}$ and $\sim 10^{-2}-10^{-1}$ at X-ray and  
   gamma-ray  energies, respectively. It has therefore been a long-standing question 
   how rotation-powered pulsars lose the bulk of their rotational energy.
 
   The fact that the energy loss of rotation-powered pulsars cannot be fully
   accounted for by the magneto-dipole radiation is known from the investigation 
   of the pulsar braking index, $n=2-P{\ddot P}  {\dot P}^{-2}$.  Pure dipole
   radiation  would imply  a braking index  $n=3$,   whereas the values
   observed  so  far   are $n=2.515 \pm   0.005$  for  the  Crab (Lyne
   et al.~1988), $n=2.8 \pm 0.2$  for PSR B1509$-$58 (Kaspi et al.~1994),
   $n=2.28\pm 0.02$ for PSR B0540$-$69 (Boyd et al.~1995), and $n=1.4 \pm
   0.2$  for the Vela  pulsar  (Lyne et  al.~1996).
   The deviation from $n=3$ is usually taken as evidence that a significant
   fraction of the pulsar's rotational energy is carried off  by a pulsar 
   wind, i.e., a mixture of charged particles and electromagnetic fields, 
   which, if the conditions are  appropriate,  forms a {\em pulsar-wind nebula} 
   observable at optical, radio and  X-ray energies.
   Such  pulsar-wind nebulae  (often   called plerions  or  synchrotron
   nebulae)  are known so far {\em only} for few young and powerful
   (high $\dot{E}$) pulsars and for some center-filled supernova remnants,  
   in which a  young neutron star  is expected, but only emission from its 
   plerion is  detected. The mechanisms of pulsar wind generation and its 
   interaction with the ambient medium are poorly understood.

   Thus, the popular model of magnetic braking provides plausible estimates
   for the neutron star magnetic field $B_\perp$, its rotational energy loss
   $\dot{E}$, and characteristic age $\tau$, but it does not provide any detailed
   information about the physical processes which operate in the pulsar
   magnetosphere and which are responsible for the broad-band spectrum, from the
   radio to the X-ray and gamma-ray bands (see Fig.\ref{nu_f_nu}). As a 
   consequence, there exist a number of magnetospheric emission models, 
   but no generally accepted theory.

 %******************************************************************************
 %                                                                             %
 % SubSection:  High-energy emission models                                    %
 %                                                                             %
 %******************************************************************************

 \subsection{High-energy Emission Models} \label{emission_models}

  Although rotation-powered pulsars are most widely known for their radio 
  emission, the mechanism of the radio emission is poorly understood. However, 
  it is certainly different from those responsible for the high-energy
  (infrared through gamma-ray) radiation observed with space observatories.
  It is well known that the radio emission of pulsars is a coherent
  process, and the coherent curvature radiation has been proposed as the
  most promising mechanism (see Michel 1991, and references therein). On 
  the other hand, the optical, X-ray and gamma-ray emission observed in
  pulsars must be incoherent. Therefore, the fluxes in these energy bands 
  are directly proportional to the densities of the  radiating high-energy 
  electrons in the acceleration regions, no matter which radiation process
  (synchrotron radiation, curvature radiation or inverse Compton scattering)
  is at work at a given energy. High-energy observations thus provide the
  key for the understanding of the pulsar emission mechanisms.
  So far, the high-energy radiation detected from rotation-driven pulsars 
  has been attributed to various thermal and non-thermal emission processes 
  including the following:

 \begin{itemize}

  \item  Non-thermal emission from charged relativistic particles 
         accelerated in the pulsar magnetosphere. As the energy 
         distribution of these particles follows a power-law, 
         the emission is also characterized by power-law-like 
         spectra in broad energy bands. The emitted radiation 
         can be observed from optical to the gamma-ray band.

  \item  Extended emission from pulsar-driven synchrotron nebulae. 
         Depending on the local conditions (density of the ambient 
         interstellar medium), these nebulae can be observed from 
         radio through hard X-ray energies.

  \item  X-ray and gamma-ray emission from interaction of relativistic 
         pulsar winds with a close companion star or with the wind of 
         a companion star, in binary systems (see Arons \& Tavani 1993).

  \item  Photospheric emission from the hot surface of a cooling neutron
         star. In this case a modified black-body spectrum and smooth, 
         low-amplitude intensity variations with the rotational period 
         are expected, observable from the optical through the soft X-ray 
         range (cf.~Greenstein \& Hartke 1983; Romani 1987; Pavlov et al.~1995).

  \item  Thermal soft X-ray emission from the neutron star's polar caps 
         which are heated by the bombardment of relativistic particles 
         streaming back to the surface from the pulsar magnetosphere 
         (Kundt \& Schaaf 1993; Pavlov et al.~1994). 

\end{itemize}

   In the following subsections we will briefly present the basic ideas 
   on the magnetospheric emission models as well as material relevant to 
   neutron star cooling and thermal emission from the neutron star surface.

 %******************************************************************************
 %                                                                             %
 % SubSubSection:  Magnetospheric emission models                              %
 %                                                                             %
 %******************************************************************************

 \subsubsection{Magnetospheric Emission Models} \label{magnetospheric_emission_models}

\begin{figure}[h]
  \begin{picture}(150,65)(0,0)
  \put(3,0){\psfig{file=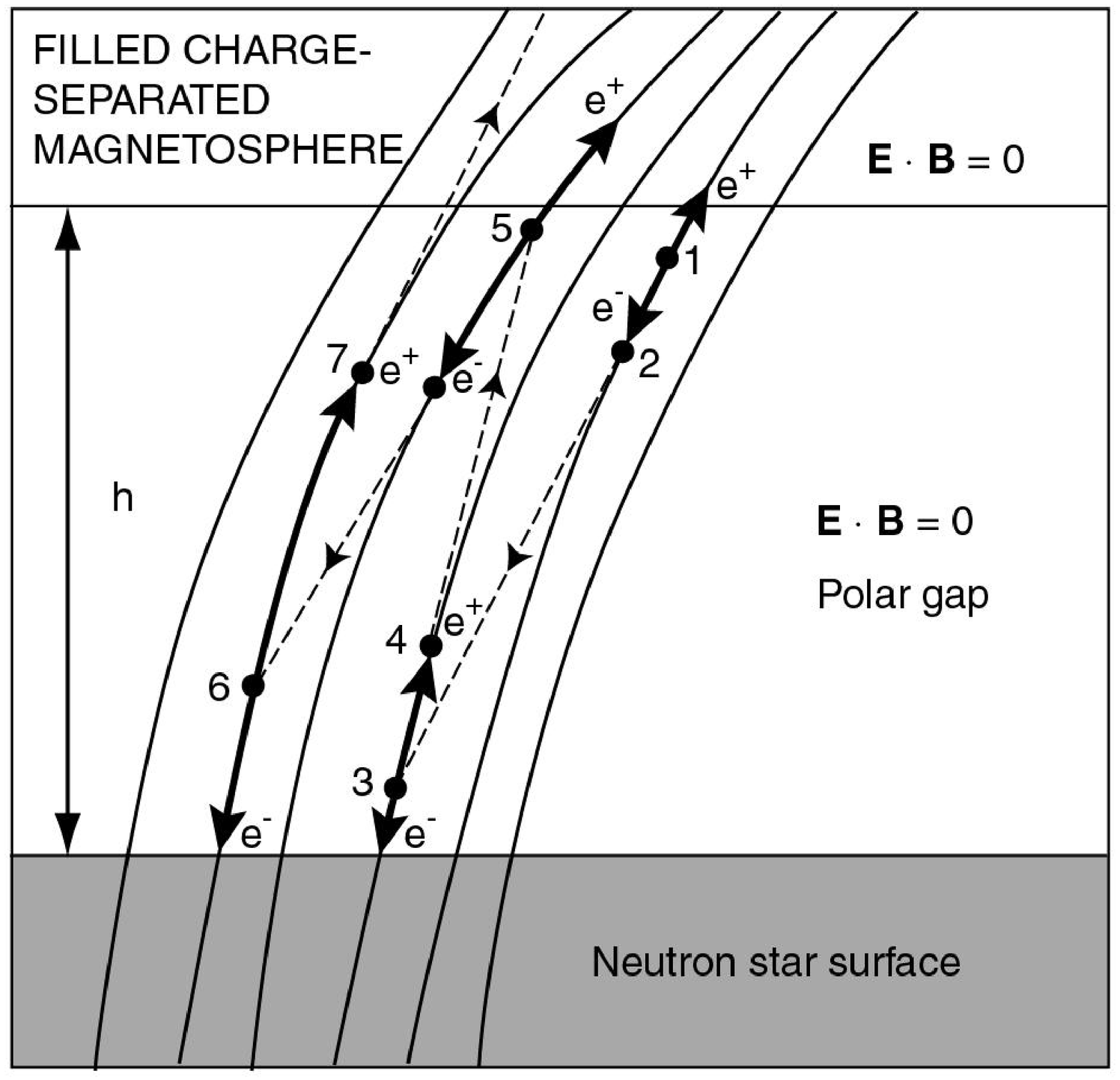,width=6.3cm,clip=}}
  \put(77,0){\psfig{file=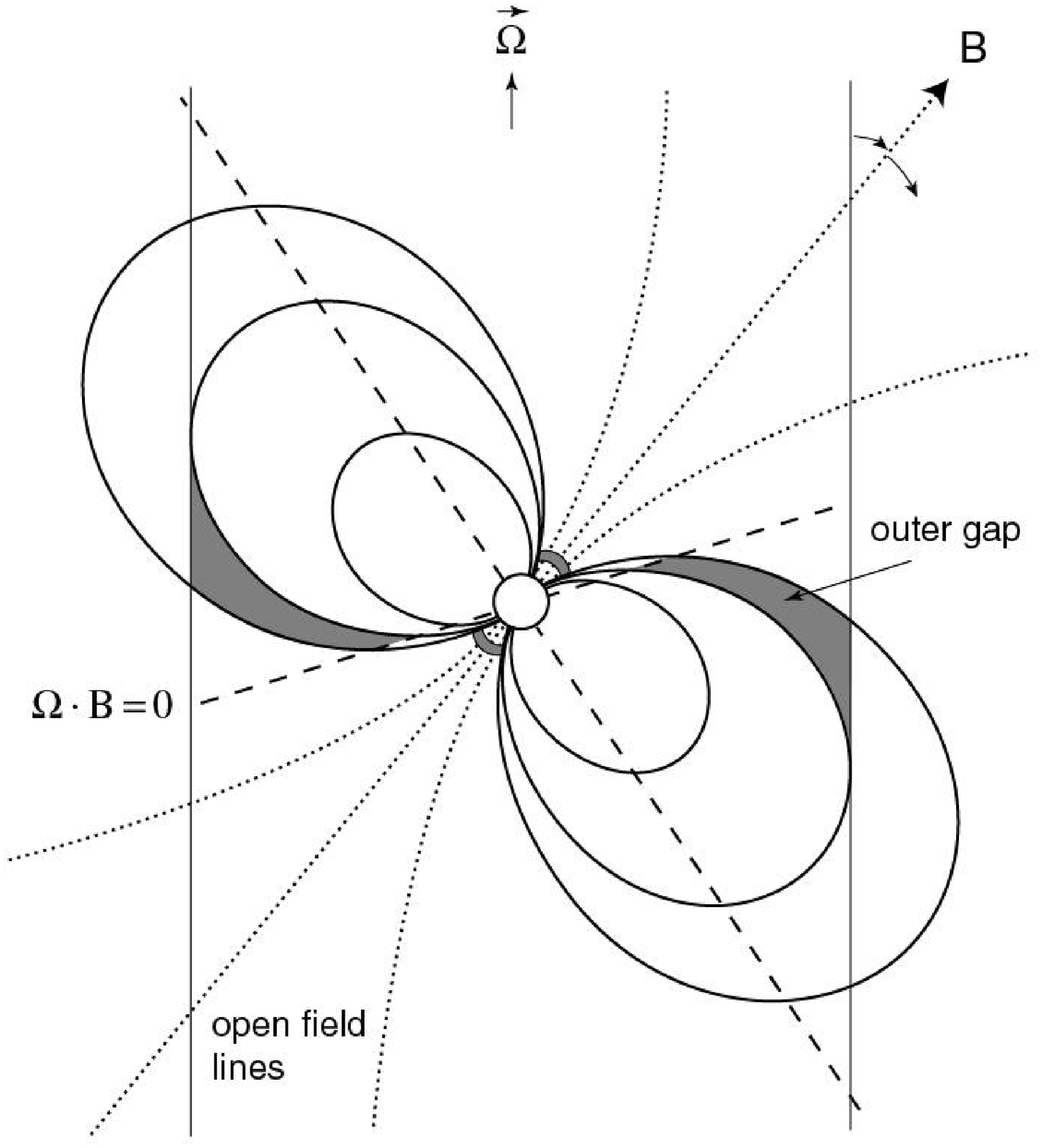,width=6.8cm,height=6.6cm,clip=}}
  \end{picture}
   \caption[]{\small Geometry of the acceleration zones as they are defined in the
   polar cap model (left), according to Ruderman \& Sutherland (1975), and 
   outer gap model (right), according to Cheng, Ho \& Ruderman (1986a,b). 
   The polar cap model predicts ``pencil'' beams emitted by particles accelerated
   along the curved magnetic field lines. According to the outer gap model, the 
   pulsar radiation is emitted in ``fan'' beams.  Being broader, the latters can 
   easier explain two (and more) pulse components observed in several gamma-ray 
   pulsars. \label{acceleration_zones}}
   \end{figure}

  So far, there is no consensus as to where the pulsar high-energy radiation 
  comes from (see for example Michel 1991; Beskin et al.~1993 and discussion 
  therein). There exist two main types of models --- the {\em polar cap models}, 
  which place the emission zone in the immediate vicinity of the neutron star's
  polar caps, and the {\em outer gap models}, in which this zone is assumed to 
  be close to the pulsar's light cylinder\footnote{The light cylinder is a virtual 
  cylinder whose radius, $R_L=cP/(2\pi)$, is defined by the condition that the 
  azimuthal velocity of the co-rotating magnetic field lines is equal to the 
  speed of light.} to prevent materializing of the photons by the one-photon 
  pair creation in the strong magnetic field, according to $\gamma + B 
  \rightarrow e^+ + e^-$ (see Fig.\ref{acceleration_zones}). The gamma-ray emission 
  in the polar cap models (Arons \& Scharlemann 1979; Daugherty \& Harding 1996; 
  Sturner \& Dermer 1994) forms a hollow cone centered on the magnetic pole, 
  producing either double-peaked or single-peaked pulse profiles, depending on the 
  observer's line of sight. The outer gap model was originally proposed to explain 
  the bright gamma-ray emission from the Crab and Vela pulsars (Cheng, Ho \& 
  Ruderman 1986a,b) as the efficiency to get high-energy photons out of the
  high B-field regions close to the surface is rather small. Placing the gamma-ray 
  emission zone at the light cylinder, where the magnetic field strength is reduced 
  to $B_L=B\,(R/R_L)^3$,  provides higher gamma-ray emissivities which are in somewhat 
  better agreement with the observations. 
  In both types of models, the high-energy radiation is emitted by relativistic 
  particles accelerated in the very strong electric field, ${\cal E} \sim (R/cP) B$, 
  generated by the magnetic field co-rotating with the neutron star. These particles
  are generated in cascade (avalanche) processes in charge-free gaps, located either 
  above the magnetic poles or at the light cylinder. The main photon emission 
  mechanisms are synchrotron/curvature radiation and inverse Compton scattering of 
  soft thermal X-ray photons emitted from the hot neutron star surface.

   In recent years the polar-cap and outher-gap models have been further 
   developed (e.g., Sturner, Dermer \& Michel 1995; Harding \& Muslimov 1998; 
   Zhang \& Harding 2000; Romani \& Yadigaroglu 1995; Romani 1996), incorporating 
   the new results on gamma-ray emission from pulsars obtained with the Compton
   Gamma-Ray Observatory. At the present stage, the observational data can be 
   interpreted with any of the two models, albeit under quite different assumptions 
   on pulsar parameters (e.g., on the direction of the magnetic and rotational axes). 
   The critical observations to distinguish between the two models include measuring 
   the relative phases between the peaks of the pulse profiles at different energies. 
   We expect that multi-wavelength timing of a large sample of pulsars with the aid 
   of the Chandra, XMM-Newton, Integral and the Hubble Space Observatory will resolve 
   this problem in a few years.

 %******************************************************************************
 %                                                                             %
 % SubSection:  Thermal evolution of neutron stars                             %
 %                                                                             %
 %******************************************************************************

 \subsubsection{Thermal Evolution of Neutron Stars} \label{photospheric}

  Neutron stars are formed at very high temperatures, $\sim 10^{11}$ K, in the 
  imploding cores of supernova explosions. Much of the initial thermal energy 
  is radiated away from the interior of the star by various processes of 
  neutrino emission (mainly, Urca processes and neutrino bremsstrahlung), 
  leaving a one-day-old neutron star with an internal temperature of about 
  $10^9-10^{10}$ K. After $\sim 100$ yr (typical time of thermal relaxation),
  the star's interior (densities $\rho>10^{10}\;\mbox{g cm}^{-3}$) becomes 
  nearly isothermal, and the energy balance of the cooling neutron star is 
  determined by the following equation (e.g., Glen \& Sutherland 1980): 

   \[ 
   C(T_i)\,\frac{{\rm d}\,T_i}{{\rm d}\,t} = - L_\nu(T_i) - L_\gamma(T_s) 
   + \sum_k H_k~,
   \]

 \noindent
  where $T_i$ and $T_s$ are the internal and surface temperatures,
  $C(T_i)$ is the heat capacity of the neutron star. Neutron star cooling 
  thus means a decrease of thermal energy, which is mainly stored in the 
  stellar core, due to energy loss by neutrinos from the interior 
  ($ L_\nu=\int Q_\nu\,{\rm d}V$, $Q_\nu$ is the neutrino emissivity) 
  plus energy loss by thermal photons from the surface 
  ($L_\gamma=4\pi R^2 \sigma T_s^4$). The relationship between $T_s$ and 
  $T_i$ is determined by the thermal insulation of the outer envelope 
  ($\rho<10^{10}\;\mbox{g cm}^{-3}$), where the temperature gradient is formed. 
  The results of model calculations, assuming that the outer envelope is 
  composed of iron, can be fitted with a simple relation (Gudmundsson, 
  Pethick \& Epstein 1983)

  \[
   T_s = 3.1\, (g/10^{14}~{\rm cm}~{\rm s}^{-2})^{1/4}\,\, 
   (T_i/10^9~{\rm K})^{0.549} ~\mbox{MK},   
  \]

   \noindent
   where $g$ is the gravitational acceleration at the neutron star surface,
   $1~{\rm MK} = 1\times 10^6$ K.  The cooling rate might be reduced by heating
   mechanisms $H_k$, like frictional heating of superfluid neutrons in the inner 
   neutron star crust or some exothermal nuclear reactions.

\begin{figure}[h]
  \begin{picture}(90,75)(0,0)
  \put(25,0){\psfig{file=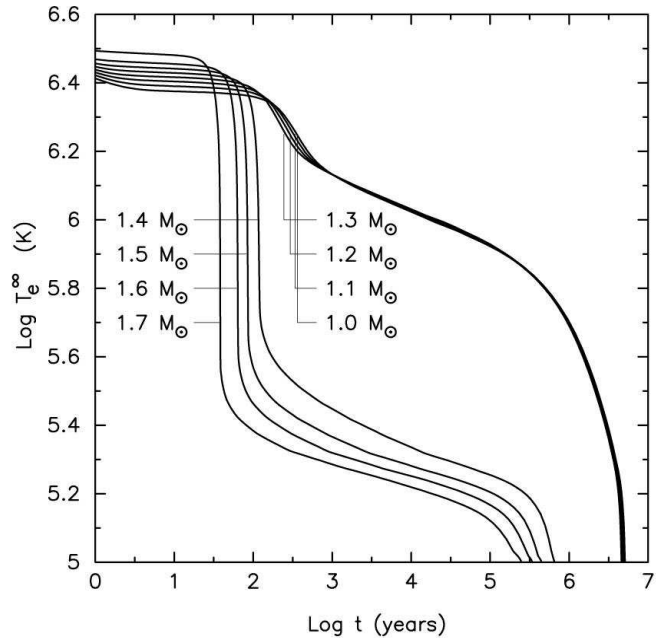,width=9cm}}
  \end{picture}
  \caption[]{\small Fast cooling vs.~standard cooling for neutron stars 
   with different masses. $T_e^\infty$ is the effective surface temperature
   as observed at infinity (i.e.~the gravitational redshift is 
   taken into account), $t$ is the age. Stars of higher masses have a very 
   high core density such that the direct Urca reactions (e.g., direct beta 
   decay) are allowed. This causes a higher neutrino emissivity
   and hence a faster energy loss (more efficient cooling) by neutrino emission. 
   The sharp temperature drop at an age of $50-100$ yrs represents the temperature
   inversion point. Here, the interior of the star, from which the neutrinos
   have escaped without interaction, is cooler than outer neutron star layers 
   which causes the outer regions to heat up the inner parts of the star.
   (From Page \& Applegate 1992). \label{cooling_model}}
  \end{figure}

   Neutrino emission from the neutron star interior is the dominant cooling 
   process for at least the first $10^5$ years. After $\sim 10^6$ years, 
   photon emission from the neutron star surface takes over as the main cooling  
   mechanism. Thermal evolution of a neutron star after the age of $\sim 10-100$ yr,
   when the neutron star has cooled down to $T_s=1.5-3$ MK, can follow two different 
   scenarios, depending on the still poorly known properties of super-dense matter
   (see Fig.\ref{cooling_model}).  According to the so-called {\em standard cooling
   scenario}, the temperature decreases gradually, down to $\sim 0.3-1$ MK, by the  
   end of the neutrino cooling era and then falls down exponentially, becoming lower 
   than $\sim 0.1$ MK in $\sim 10^7$ yr. In this 
   scenario, the main neutrino generation processes are the modified Urca 
   reactions, $n+N \rightarrow p+N+e+\bar{\nu}_e$ and $p+N+e \rightarrow n+N+\nu_e$, 
   where $N$ is a nucleon (neutron or proton) needed to conserve momentum of 
   reacting particles. In the {\em accelerated cooling scenarios}, associated 
   with higher central densities (up to $10^{15}\;\mbox{g cm}^{-3}$) and/or 
   exotic interior composition (e.g., pion condensation, quark-gluon plasma), 
   a sharp drop of temperature, down to $0.3-0.5$ MK, occurs at an age 
   of $\sim 10-100$ yr, followed by a more gradual decrease, down to the same 
   $\sim 0.1$ MK at $\sim 10^7$ yr. The faster cooling is caused by the direct 
   Urca reactions, $n \rightarrow p+e+\bar{\nu}_e$ and $p+e \rightarrow n+\nu_e$, 
   allowed at very high densities (Lattimer et al.~1991). An example of standard 
   and accelerated cooling curves is shown in Figure \ref{cooling_model}.
   The neutron star models used in these calculations are based on a ``moderate''
   equation of state which opens the direct Urca process for $M>1.35 M_\odot$; the 
   stars with lower $M$ undergo the standard cooling. Recent studies have shown that
   both the standard and accelerated cooling can be substantially affected by
   nucleon superfluidity in the stellar interiors (see Tsuruta 1998 and Yakovlev,
   Levenfish \& Shibanov 1999 for comprehensive reviews). In particular,
   there can exist many cooling curves intermediate between those of the standard
   and accelerated scenarios, depending on properties of nucleon superfluidity,
   which are also poorly known.

   Thus, the thermal evolution of neutron stars between $\sim 10$ and $\sim 10^6$ yr 
   is very sensitive to the composition and structure of their interiors, in 
   particular, to the equation of state at super-nuclear densities. Therefore,
   measuring surface temperatures of neutron stars is an important tool to study 
   the super-dense matter. Since typical temperatures of such neutron stars 
   correspond to the extreme UV -- soft X-ray range, the thermal radiation from 
   cooling neutron stars can be observed with X-ray detectors sufficiently 
   sensitive at $E\lapr 1$ keV.

  %******************************************************************************
  %                                                                             %
  % SubSection:  Photospheric emission from cooling neutron stars               %
  %                                                                             %
  %******************************************************************************

  \subsubsection{Photospheric Emission from Cooling Neutron Stars} \label{photospheric}

  Thermal radiation has been observed from about a dozen isolated neutron stars. 
  Much more detailed  data on thermal radiation from these and other neutron stars 
  are expected from the X-ray observatories Chandra and XMM-Newton.
  To interpret these observations, detailed and accurate models for spectra and 
  light curves of thermal radiation from neutron stars are needed. Properties of 
  the neutron star thermal radiation are determined, as in usual stars, by a thin, 
  partially ionized atmosphere with temperature growing inward. As a result, 
  the neutron star thermal radiation may be substantially different from  
  blackbody radiation (Pavlov \& Shibanov 1978). Modeling of neutron star 
  atmospheres requires a special approach because neutron stars possess 
  very strong magnetic fields, $B\sim 10^{11}-10^{13}$~G. In such fields the 
  electron cyclotron energy, $E_{ce} = 11.6\;(B/10^{12}~{\rm G})$ keV, 
  strongly exceeds the thermal energy, $kT\sim 0.01-1$ keV. As a result, the 
  atmospheres are essentially anisotropic, so that the absorption and emission 
  of photons depend on the direction of the photon wavevector, and the radiation 
  propagates there as two normal (polarization) modes with nearly orthogonal
  polarizations and quite different opacities (Gnedin \& Pavlov 1974; Bulik 
  \& Pavlov 1996). The energy dependences of these opacities are substantially  
  different from each other and  from the opacity at $B=0$. Since the ratio 
  $\beta$ of the cyclotron energy to the Coulomb energy\footnote{e.g., 
  $\beta=E_{ce}/(Z^2{\rm Ry}) = 850\, Z^{-2} (B/10^{12}~{\rm G})$ for one-electron 
  ions; $Ze$ is the ion charge, ${\rm Ry} = me^4/(2\hbar^2) = 13.6$ eV is the 
  ionization potential of the hydrogen atom.} is very large, the structure 
  of atoms and ions is distorted by the strong magnetic field, which changes 
  the energies and strengths of spectral features and ionization equilibrium 
  of the atmospheric plasma. As a result, the spectrum, angular 
  distribution and polarization of thermal radiation depend on the magnetic 
  field.

  Another important effect is that the nonuniform magnetic field leads to a 
  nonuniform surface temperature distribution because of anisotropic heat 
  conduction (Greenstein \& Hartke 1983; Shibanov \& Yakovlev 1996), which 
  enhances  pulsations of thermal radiation due to the \NS rotation.
  The high density of the atmospheric matter ($\sim 1-100\;\mbox{g cm}^{-3}$ 
  at unit optical depth), caused by the  immense gravitational 
  acceleration, $g\sim 10^{14}-10^{15}\;\mbox{cm s}^{-2}$, poses additional 
  complications. In particular, the non-ideality (pressure) effects lead to
  pressure ionization and smooth out the spectral dependences of the opacities. 
  The huge surface gravity also leads to chemical stratification of \NS atmospheres,
  so that upper layers, which determine the properties of the emitted radiation, 
  are comprised of the lightest element present. This means, in particular, 
  that if a \NS has accreted some amount of hydrogen (e.g., from the circumstellar
  medium or from the envelope ejected during the supernova explosion), its 
  radiative properties are determined by the hydrogen atmosphere.

  A convenient approach to modeling of \NS atmospheres was described by Pavlov et 
  al.~(1995). It includes, as for usual stars, solving of a set of equations for 
  hydrostatic equilibrium, energy balance, ionization equilibrium, and radiative 
  transfer, complemented by calculations of spectral opacities for partially ionized, 
  nonideal plasma. For atmospheres with strong magnetic fields, two coupled equations 
  of radiative transfer for the intensities of two polarization modes have to be solved. 
  The input parameters for the modeling are the chemical composition, effective temperature
  $T_{\rm eff}$ (or total radiative flux $\propto T_{\rm eff}^4$), magnetic field $B$ 
  (including the field orientation at the radiating \NS surface), and gravitational 
  acceleration $g$ (or the \NS mass $M$ and radius $R$).

  %******************************************************************************
  %                                                                             %
  % SubsubsubSection:  Low-field Neutron Star Atmospheres                       %
  %                                                                             %
  %******************************************************************************

  \subsubsection*{Low-field Neutron Star Atmospheres}

  It is commonly accepted that very old neutron stars, like the $10^{8}-10^{10}$
  years old millisecond pulsars, have  ``low'' surface magnetic fields, $B\sim 
  10^8-10^9$~G, which do not affect the X-ray opacities of the atmospheric
  plasma at temperatures of interest (at $E_{ce}\ll E$, $E_{ce}\ll kT$, and 
  $\beta\ll 1$). First models of the low-field neutron star atmospheres were 
  calculated by Romani (1987). Further works (Rajagopal and Romani 1996; 
  Zavlin, Pavlov, \& Shibanov 1996) used improved opacities (Iglesias and Rogers  
  1996) for pure hydrogen, helium and iron compositions. These works have shown
  that the spectra of radiation emerging from a light-element (H or He) 
  atmosphere are much harder (less steep) than the blackbody spectra
  at $E\gapr kT_{\rm eff}$ (see Fig.~\ref{H_vs_Fe_low_field}). 
  The reason for such behavior is that the hydrogen and helium opacities decrease 
  with increasing $E$, so that the radiation of higher energies is formed in 
  deeper and hotter layers. As a result, fitting observed spectra with the 
  standard blackbody model yields spectral (blackbody) temperatures exceeding
  the true effective temperatures by a factor of $1.5-3$, which makes a great 
  difference for the comparison with the models of neutron star cooling.
 
  The spectra emitted from iron atmospheres are much more complex due to
  numerous spectral features produced by iron ions in various stages of ionization 
  (see Fig.~\ref{H_vs_Fe_low_field}). Some of these features  are observable even
  with moderate-resolution (e.g, CCD) spectrometers. On the other hand, when 
  observed with very low energy resolution, the iron atmosphere spectra look 
  very similar to the blackbody spectra. 

  The local specific intensity of radiation
  decreases with the angle between the \NS surface and the wave vector, and the 
  shape of the angular distribution depends on photon energy and chemical 
  composition. This (limb-darkening) effect must be taken into account for
  fitting of both the spectra and the pulse profiles if the radiation is emitted 
  from hot spots on the \NS surface, like in millisecond pulsars 
  (Pavlov \& Zavlin 1997; Zavlin \& Pavlov 1998).

  %******************************************************************************
  %                                                                             %
  % SubsubsubSection:  High-field Neutron Star Atmospheres                            %
  %                                                                             %
  %******************************************************************************

  \subsubsection*{High-field Neutron Star Atmospheres}

   First models of magnetic hydrogen atmospheres with $B\sim 10^{11} - 
   10^{13}$~G have been constructed recently (Shibanov et al.~1992; Pavlov et al.~1994, 
   1995; Zavlin et al.~1995a). These models are based upon simplified opacities of 
   strongly magnetized, partially ionized hydrogen plasma. These opacities do not 
   include the bound-bound transitions, neglect the motional Stark effect, and use a
   simplified model for the ionization equilibrium. Nevertheless, the models provide
   a qualitatively correct description for the magnetic effects on the emergent 
   radiation, and they are accurate enough in the case of high effective 
   temperatures, $\gapr 1$ MK, when the hydrogen is almost completely ionized 
   even in the very strong magnetic fields.

   \begin{figure}[ht]
   \begin{picture}(170,65)(0,-6)
   \put(0,0){\psfig{file=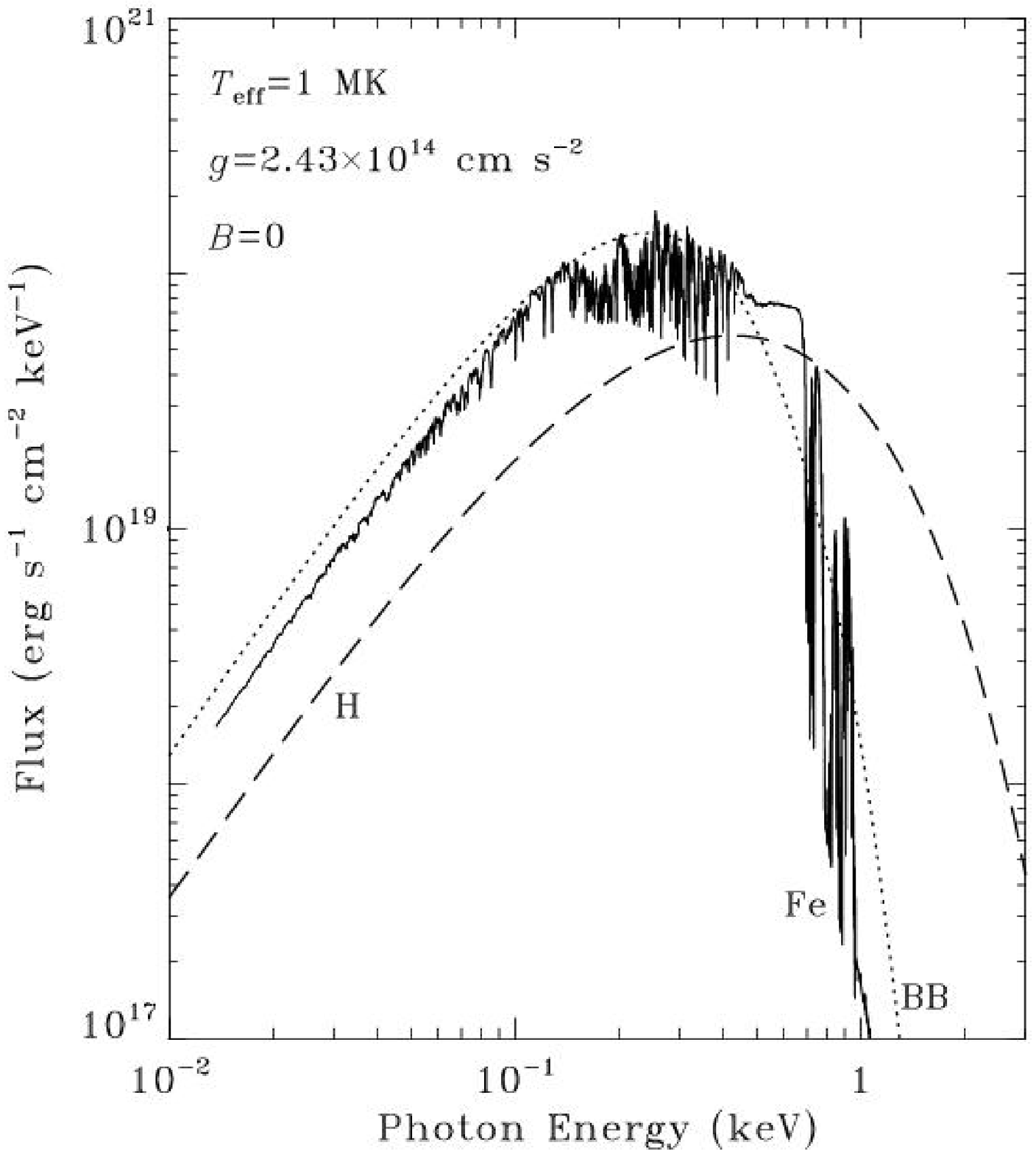,width=8.5cm}}
   \put(70,0){\psfig{file=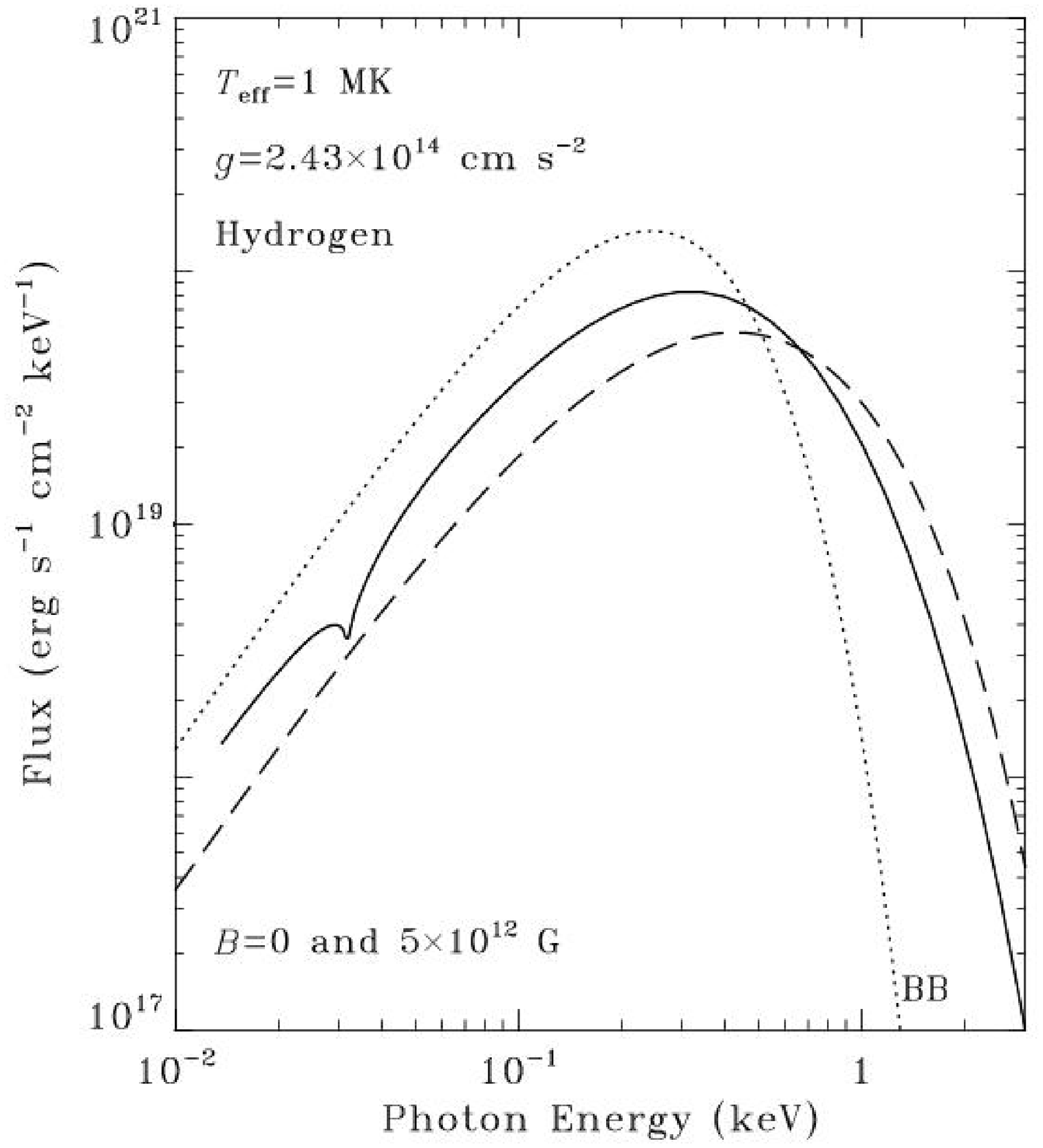,width=8.5cm}}
   \end{picture}
   \caption[]{\small {\sl Left}: Simulated spectra of hydrogen and iron 
   neutron star atmospheres with low magnetic field,
   together with the blackbody spectrum, for $T_{\rm eff}=1$ MK. {\sl Right}:
   The simulated spectra of hydrogen neutron star atmospheres with high
   and low magnetic fields (solid and dashed lines, respectively).
    \label{H_vs_Fe_low_field} \label{H_high_B_spetra}}
   \end{figure}

   Since the magnetic atmospheres are much more transparent in the ``extraordinary'' 
   polarization mode, whose opacity is strongly reduced by the magnetic field, very 
   deep (hot and dense) layers are responsible for the observed radiation. Their
   X-ray spectra are harder than the blackbody spectrum at the same effective 
   temperature, although not as much as the low-field spectra. The only spectral 
   line in the spectra of completely ionized hydrogen atmospheres is the proton 
   cyclotron line at the energy $E_{cp}=(m_e/m_p) E_{ce}=6.3\, (B/ 
   10^{12}~{\rm G})$ eV (see Fig.~\ref{H_high_B_spetra}). The spectra depend not 
   only on strength, but also on direction of the magnetic field, which means  
   that  the radiative flux emitted by a rotating neutron star is pulsed 
   even if the surface temperature is uniform. 
   Angular distribution of the local intensity shows a sharp peak along the 
   magnetic field and a broader peak at intermediate angles (the ``pencil'' and 
   ``fan'' components), the widths of the peaks depend on photon energy. This 
   means that the pulse profiles of radiation emitted from hot polar caps may
   be much sharper than those emitted from low-field atmospheres. The pulse 
   shape strongly depends on the mass-to-radius ratio due to bending of photon 
   trajectories in the strong gravitational fields 
   (Zavlin, Shibanov \& Pavlov 1995b; Shibanov et al.~1995). The 
   radiation emitted from magnetic atmospheres is strongly polarized; the 
   degree of polarization depends on $E$, $B$, and $M/R$ (Pavlov \& Zavlin 2000).

   First results obtained with the improved hydrogen atmosphere models (Pavlov 
   \& Zavlin 2001), which include the bound-bound transitions, show that 
   spectral lines, considerably broadened by the motional Stark effect  
   (Pavlov \& M\'esz\'aros 1993; Pavlov \& Potekhin 1995), become prominent at 
   $T_{\rm eff}\;\lapr 0.5$ MK. The strongest line is observed at $E\simeq
   \{75\,[1+0.13\,\ln(B/10^{13}~{\rm G})] + 63\,(B/10^{13}~{\rm G})\}~{\rm eV}$.

   Magnetic iron atmosphere models have been considered by Rajagopal, Romani, 
   \& Miller (1997). Making use of the
   so-called adiabatic approximation ($\ln\beta\gg 1$),
   these authors calculated the energies and wave functions of the iron ions and 
   the radiative opacities of the polarization modes. Although these models are 
   inevitably rather crude, they provide a baseline for comparison with the magnetic 
   hydrogen atmosphere models and for future work on heavy-element atmosphere 
   modeling. Similar to the low-field case, the magnetic iron atmosphere spectra 
   are fairly close to the blackbody spectra when observed with low-resolution 
   detectors. Developing more accurate iron atmosphere models is important for 
   adequate interpretation of future high-resolution X-ray observations
   of the neutron star thermal radiation.

 %******************************************************************************
 %                                                                             %
 % SECTION:  What we know, or what we believe to know:                         %
 %           The current picture of high energy emission properties of         %
 %           isolated neutron stars                                            %
 %                                                                             %
 %******************************************************************************

 \section{The Current Picture of High-Energy Emission Properties of Isolated 
          Neutron Stars} \label{current_pic}

  As a result of observations with the satellite observatories ROSAT, EUVE,
  ASCA, BeppoSAX, RXTE, CGRO, HST, Chandra and XMM-Newton, the number of
  rotation-powered pulsars seen at X-ray, gamma-ray and optical energies
  has increased substantially in the last decade of the century. For the
  first time it became possible to carry out multi-wavelength studies of 
  the pulsar emission. This is a big advantage as the physical processes 
  which cause the emission in different wavelength bands are obviously
  related to each other. Although the quality of the data obtained with
  different instruments is inevitably rather inhomogeneous, and the 
  conclusions drawn on these data are therefore not fully certain in 
  many cases, there is a general consensus that a first big step towards  
  discrimination between different emission scenarios has been made. In
  this respect, even more is expected from the new observatories,
  Chandra and XMM-Newton, launched at the end of the century. Results
  from the first year of Chandra,  which are briefly mentioned in this
  Section, seem to justify these high expectations.

 %******************************************************************************
 %                                                                             %
 % SubSection:  Young Neutron Stars in Supernova remnants                      %
 %                                                                             %
 %******************************************************************************

 \subsection{Young Neutron Stars in Supernova Remnants} \label{young_ns}

  X-ray observations allow us to find both supernova remnants (SNRs) and 
  the compact objects that may reside within them. In fact, neutron stars 
  and neutron star candidates have been found in a small fraction, $<10\%$, of 
  the 220 known galactic SNRs (Green 1998; Kaspi 2000, and references 
  therein). Less than half of these objects are radio  pulsars, the 
  others are radio-silent (or, at least, radio-quiet) neutron stars 
  which are seen only in X-rays (some of them in gamma-rays). The 
  young radio pulsars can be divided in two groups,  Crab-like and 
  Vela-like pulsars, according to somewhat different observational 
  manifestations apparently associated with the evolution of pulsar 
  properties with age. The radio-silent neutron stars include anomalous 
  X-ray pulsars, soft gamma-ray repeaters, and ``quiescent'' neutron 
  star candidates. We will briefly review all the groups in this subsection.

 %******************************************************************************
 %                                                                             %
 % SubSubSection:  The Crab-like pulsars                                       %
 %                                                                             %
 %******************************************************************************

 \subsubsection{Crab-like Pulsars} \label{crab_like_ns}

\begin{figure}[h!]
   \centerline{\psfig{file=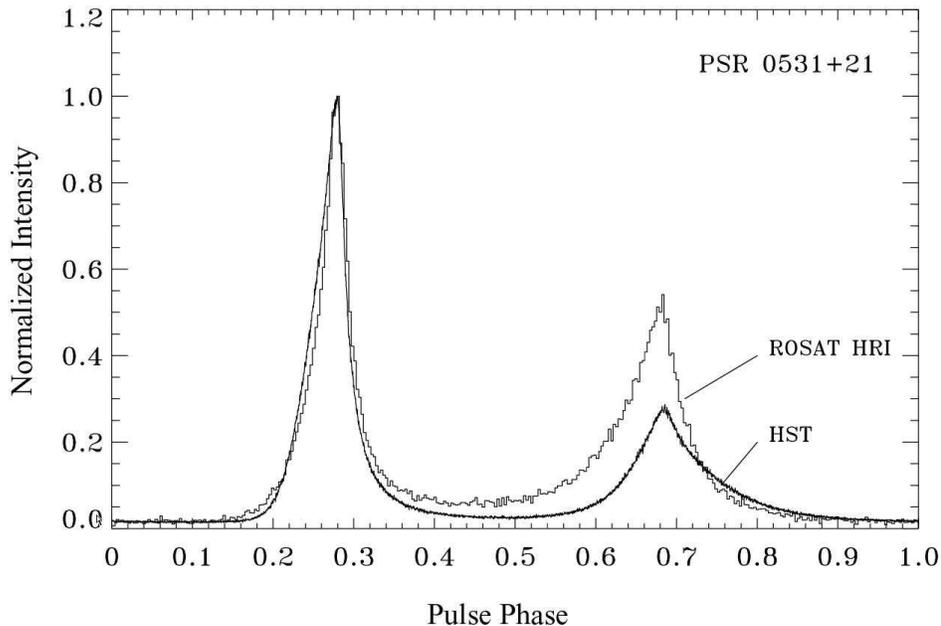,width=12.5cm,clip=}}
   \caption[]{\small The Crab pulse profile as observed with HST and ROSAT
   in the optical and soft X-ray bands. Its characteristic double-peaked
   shape is observed at all wavelengths. The phase difference between the
   first and second peak shows a weak energy dependence.\label{crab_lc}}
  \end{figure}

  It is well established that magnetospheric emission from charged particles,
  accelerated in the neutron star magnetosphere along the curved magnetic field 
  lines, dominates the radiation from young {\em rotation-powered} pulsars 
  with ages $\lapr 5000$ years (cf.~\ref{emission_models}). In the case 
  of the Crab pulsar at least $\sim 75\%$ of the total soft X-ray flux is
  emitted from the co-rotating magnetosphere (Becker \& Tr\"umper 1997).
  Accordingly, its radiation is characterized by a power-law spectrum\footnote{The
  spectrum of the non-thermal radiation is a power-law, ${\rm d}N/{\rm d}E\propto 
  E^{-\alpha}$, as the energy distribution of the particles which emit this 
  radiation follows a power-law in a broad energy range.}, and its spin-modulated 
  lightcurve exhibits two narrow peaks per period (see Fig.~\ref{crab_lc}).
  The Crab pulsar is also a bright gamma-ray and optical-UV source. Its X-ray, 
  gamma-ray, optical and radio pulsations are all phase-aligned, demonstrating 
  that the emission in these bands is clearly non-thermal and originates from 
  the same site in the pulsar magnetosphere. The slope of its flux spectrum 
  slowly increases with photon energy --- the photon index varies from $\alpha=1.1$\, 
   at $E\sim 1$ eV to \,$\alpha=2.1$\,at $E\sim 10^{10}$ eV.

\begin{figure}[h!]
   \centerline{\psfig{file=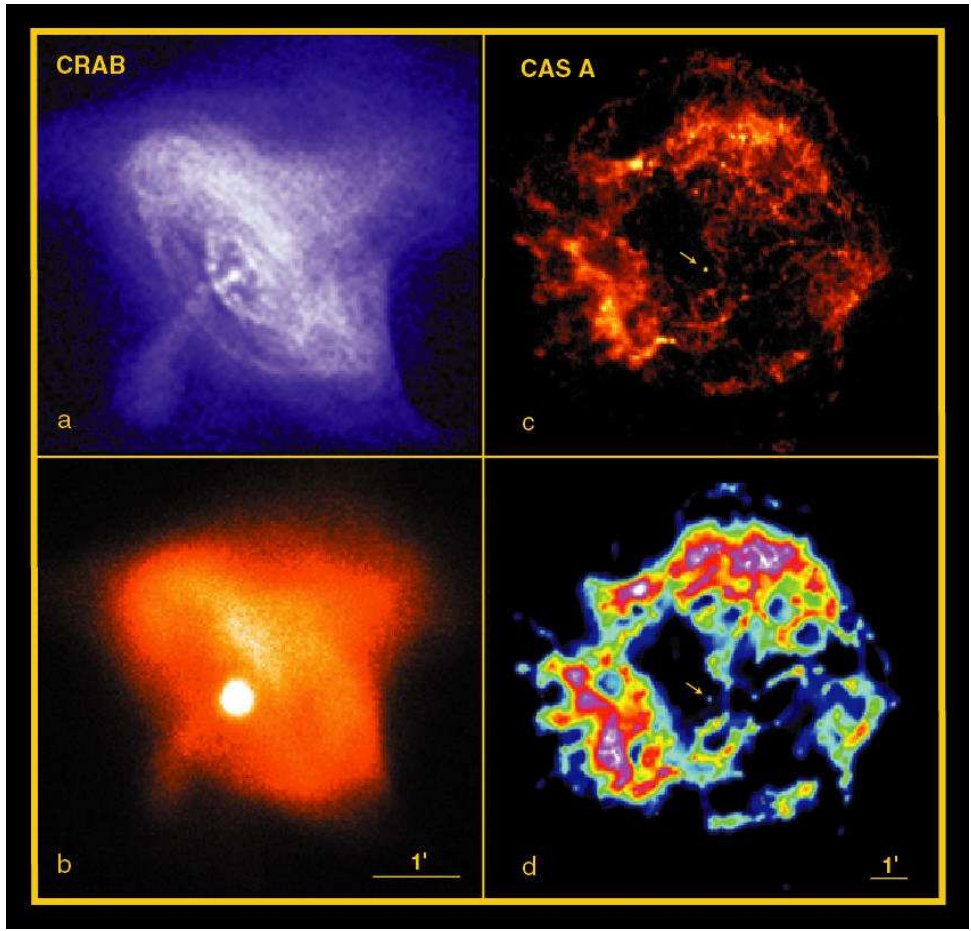,height=12.3cm,clip=}}
   \caption[]{\small The Crab as observed with the Chandra ACIS {\bf (a)} and the
   ROSAT HRI {\bf (b)}. The images demonstrate the improvement of angular 
   resolution between the two detectors by a factor of 10. In the Chandra
   image much more details of the pulsar-driven nebula become visible.
   Image {\bf (c)} shows another recent discovery made by Chandra: the point
   source close to the geometrical center of Cassiopeia A
   (Cas A), a very young (320 yr) supernova remnant. The corresponding ROSAT
   HRI image is shown in {\bf (d)}.  Only the unprecedented high spatial
   resolution provided by Chandra allowed one to detect the point source,
   a young neutron star or a black hole, identified {\em a posteriori} in
   deep Einstein HRI and ROSAT HRI images. \label{hri_vs_chandra}}
  \end{figure}

  As the Crab pulsar is the youngest rotation-powered pulsar and thus 
  should be the hottest neutron star, one could expect to observe its 
  thermal surface emission at the off-pulse phases, when the thermal 
  flux is not buried under the powerful magnetospheric emission.
  However, even the Einstein HRI and ROSAT HRI, despite their high angular
  resolution, were not able to completely get rid of a contribution from 
  the compact synchrotron nebula around the pulsar (see Fig.\ref{hri_vs_chandra}), 
  so that only an upper limit on the thermal flux has been established 
  from the DC level of the soft X-ray pulse profile. Becker \& Aschenbach 
  (1995) found an upper limit of about 2 MK for the surface temperature of
  the Crab pulsar from the ROSAT HRI observations, consistent with the 
  predictions of standard cooling models.

  The ROSAT HRI data taken from the Crab Nebula have been used to improve
  our understanding of this object in many aspects.  Greiveldinger \& 
  Aschenbach (1999), using the HRI observations spanning a period of more 
  than 6 years, have shown that the X-ray intensity of the inner synchrotron 
  nebula varies on time scales of years by about 20\%. The intensity 
  variations are found to be confined to rather large ($\sim 25''\times 25''$,
  or $0.25\times 0.25$ pc) regions in the {\em torus} (its radius is $\approx 0.4$ pc).
  Using the instruments aboard Chandra, it will be easy to further investigate 
  these long-term variations. First images taken with the Advanced CCD Imaging 
  Spectrometer (ACIS) aboard Chandra have already provided spectacular details
  of the inner nebula structure associated with the pulsar-wind outflow --- in 
  addition to the torus ($r\approx 0.38$ pc), the {\em inner ring} ($r\approx 0.14$ pc), 
  {\em jet} and {\em counter-jet} have been identified (Weisskopf et al.~2000). 
  To demonstrate the recent  progress in X-ray astronomy, the images of the Crab 
  pulsar and its plerion, as seen by the Chandra ACIS and the ROSAT HRI, are 
  shown in Figure \ref{hri_vs_chandra}.

  Emission properties similar to those found for the Crab pulsar are observed from 
  the pulsars B0540$-$69, J0537$-$6909 and B1509$-$58 in the supernova
  remnants N158A, N157B and MSH 15$-$52 (the former two are in the Large
  Magellanic Cloud). In particular, PSR B0540$-$69 has a compact X-ray
  nebula strongly resembling that around the Crab pulsar, and $\approx 
  40\%$ of the pulsar's soft X-ray photons are pulsed (Seward \& Harnden 1994; 
  Gotthelf \& Wang 2000). This pulsar has been detected in optical (Boyd
  et al.~1995; Hill et al.~1997) but not in gamma-rays. Like for the 
  Crab-pulsar, its optical pulse profile is very similar to the profile 
  observed at X-ray energies (Gouiffes, Finley \& \"Ogelman 1992; Mineo 
  et al.~1999).

\begin{figure}[ht]
  \centerline{\psfig{file=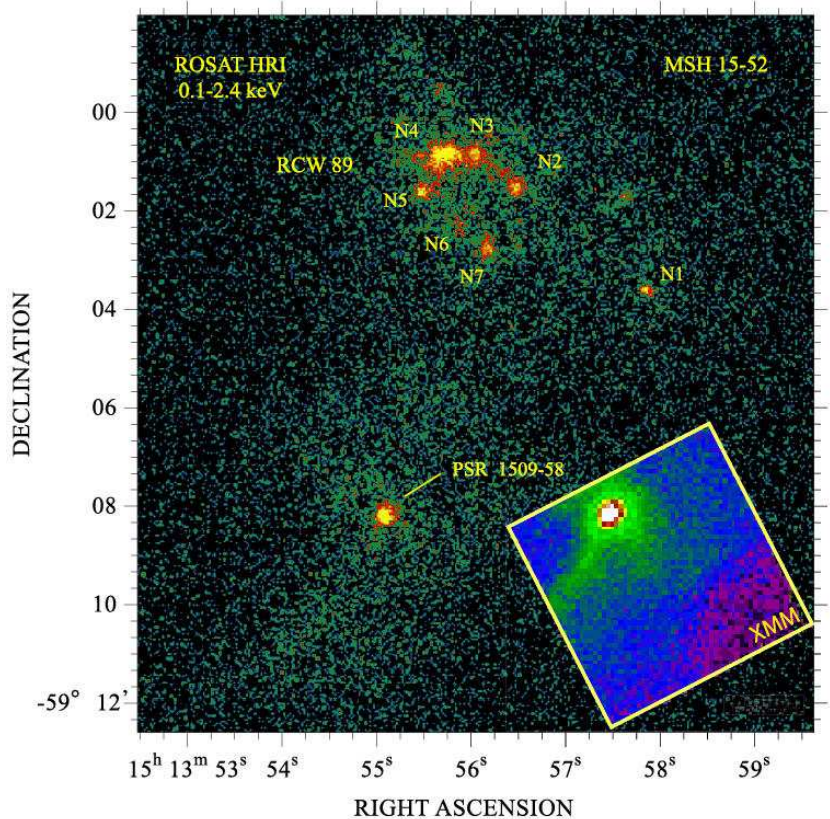,width=11cm,clip=} }
  \caption{\small  Soft X-ray image of MSH 15$-$52 and RCW 89 as seen by the
   ROSAT HRI. The most striking features are the compact knots in RCW 89
   (only the brightest seven are labeled, N1$-$N7), the point source at the
   location of PSR B1509$-$58, the synchrotron nebula around PSR B1509$-$58,
   and the extended diffuse emission in RCW 89. The inset shows a $4' \times 4'$
   area arround the pulsar as seen by the PN-Camera aboard XMM-Newton. \label{1509_image} }
  \end{figure}

  PSR B1509$-$58 has the highest period derivative among the known pulsars.
  Optical radiation from this pulsar has not yet been detected, and only
  upper limits have been obtained for its gamma-radiation above $\sim 10-30$ 
  MeV, suggesting a break in the gamma-ray spectrum somewhere between 10 and
  100 MeV (Kuiper et al.~1999). Its X-ray emission in the ROSAT band is found 
  to have a pulsed fraction of about 65\% (Becker \& Tr\"umper 1997). The soft 
  X-ray pulse is phase-aligned with the hard X-ray and gamma-ray pulses 
  detected by the CGRO detectors BATSE, OSSE and COMPTEL up to at least 10 MeV   
  (Ulmer et al.~1994; Kuiper et al.~1999). These high-energy pulses appear 
  phase-shifted by  $\sim 0.3$ periods relative to the radio pulse.
  Based on the ROSAT HRI observations, Brazier \& Becker (1997) have proposed
  that the X-ray nebula surrounding PSR 1509$-$58 (see Fig.~\ref{1509_image})
  is comprised of a {\it torus} and a {\it jet}, similar to the Crab synchrotron
  nebula. A nearby region of enhanced X-ray emission, RCW 89, may be caused 
  by the collision of the collimated pulsar wind with the outer shell of the
  supernova remnant.

   PSR J0537$-$6909 has been discovered recently (Marshall et al.~1998)
   with RXTE and ASCA and later detected with ROSAT (Wang \& Gotthelf
   1998). It is particularly interesting due to its very short period of 16 ms,
   the shortest one among the ``regular'' pulsars, despite the fact that it is
   older ($\tau\sim 5000$ yr) than the other three members of this subclass
   of pulsars. The ROSAT HRI image shows a bright X-ray nebula whose size
   ($\approx 2$ pc) and cometary shape indicate that the pulsar is moving with
   a  supersonic velocity, $\sim 1000$ km/s, and the X-ray emission of the
   nebula originates mainly from a bow shock.

   Thus, all the very young pulsars show strong non-thermal X-ray emission 
   with an X-ray luminosity $L_x\sim 10^{34}-10^{36}$ erg~s$^{-1}$ in the ROSAT 
   energy range, and they are surrounded by pulsar-powered nebulae (plerions) 
   and supernova ejecta. Presumably, their magnetospheric emission extends 
   from at least infrared to gamma-ray energies, with typical photon indices
   varying between $\approx 1$ and $\approx 2$ (about $1.4-1.7$ in the soft 
   X-ray range).

 %******************************************************************************
 %                                                                             %
 % SubSubSection:  The Vela-like Pulsars                                       %
 %                                                                             %
 %******************************************************************************
 %

 \subsubsection{Vela-like pulsars} \label{vela_like}

  \begin{figure}[ht]
  \centerline{\psfig{file=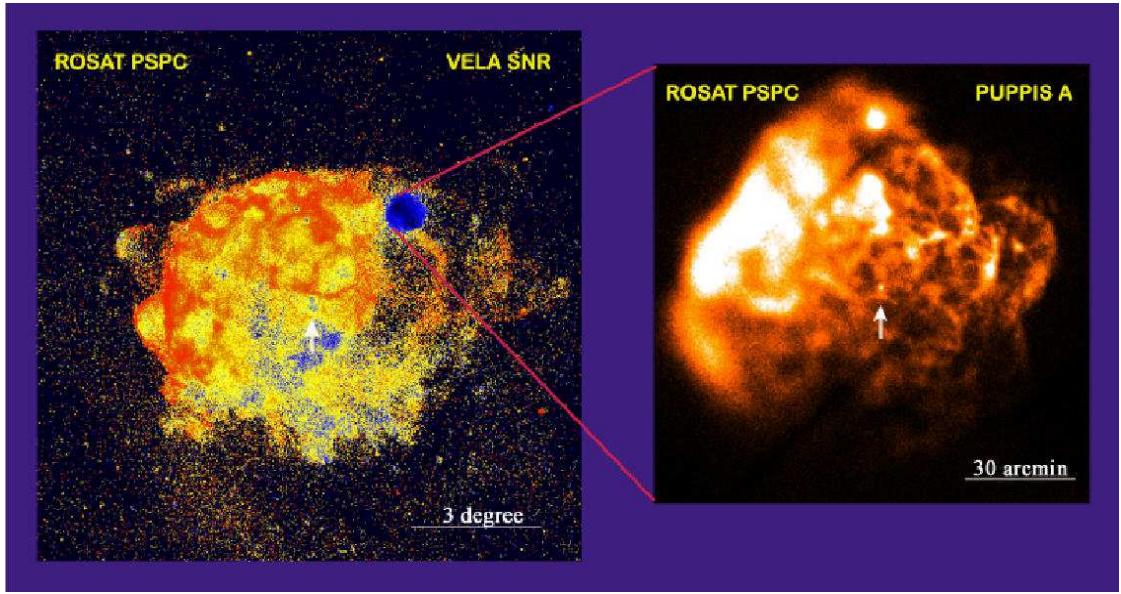,width=14.8cm,cut=}}
  \caption{\small {\sl Left}: 
  ROSAT PSPC image of the Vela SNR. Different colors here correspond to different 
  energies of X-ray photons, from red (lower energies) to blue (higher energies).
  The location of the Vela pulsar is indicated by the arrow. {\sl Right}: Zoomed 
  image of the Puppis A SNR, located at the North-West edge of the Vela remnant. 
  The arrow indicates the point source RX J0820$-$4300 which is a very good 
  candidate for a young cooling neutron star showing photospheric emission (see
  \ref{radio_silent_ns}). \label{vela_pupis_image}}
  \end{figure}
 
\begin{figure}[ht]
   \centerline{\psfig{file=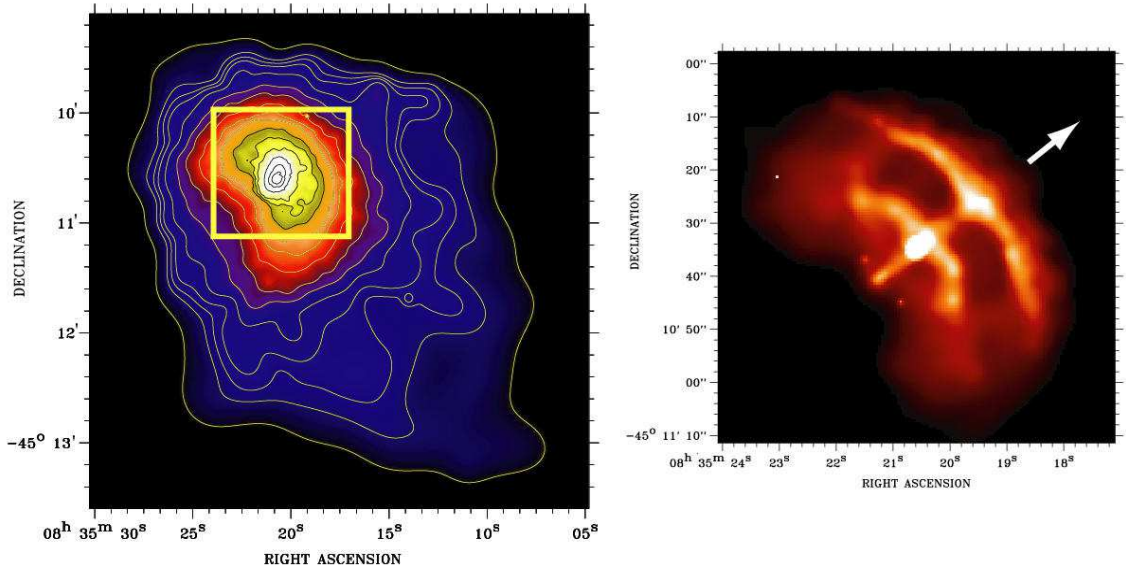,width=15cm,clip=}}
   \caption{\small The Vela pulsar and its X-ray plerion as observed with
   the ROSAT HRI (a) and Chandra ACIS (b). In both images, the pulsar is the 
   brightest source. The Chandra image shows the spatially resolved
   inner part of the plerion, corresponding to the central yellow box in 
   the ROSAT image. The arrow indicates the direction of the pulsar proper 
   motion which is aligned with the rotation axis. \label{vela_plerion}}
  \end{figure}

   Pulsars with a spin-down age of $\sim 10^4-10^5$ years are often referred
   to as Vela-like pulsars, because of their apparent similar emission properties.
   About ten pulsars of this group have been detected in X-rays 
   (cf.~Tab.~\ref{sym_tab}), four of them (the Vela pulsar B0833$-$45, 
   PSR B1706$-$44, B1046$-$58 and B1951$+$32) are gamma-ray pulsars, and only 
   the Vela pulsar has been detected in the optical band. In some respects, 
   these objects appear to be different from the Crab-like pulsars. In particular, 
   their pulses at different energies are not phase-aligned with each other, their 
   optical radiation is very faint compared to that of the very young pulsars,
   and the overall shape of their high-energy spectra looks different. For 
   instance, the closest ($d\approx 300$ pc) and, hence, best-investigated 
   Vela pulsar (see Fig.~\ref{vela_pupis_image} and Fig.~\ref{vela_plerion}) 
   has an optical luminosity four orders of magnitude lower than the Crab 
   pulsar (Manchester et al.~1978; Nasuti et al.~1997), whereas its rotation 
   energy loss is only a factor of 65 lower. Its light curve shows two peaks 
   in the gamma-ray range (Kanbach et al.~1994) and at least three peaks 
   in the X-ray range (Strickman et al.~1999; Pavlov et al.~2000a), versus 
   one peak at radio frequencies, whose phase does not coincide with any of  
   the high-energy pulses. The pulsed fraction in the soft X-ray range, 
   $\approx 12\%$, is much lower than that observed from the Crab-like pulsars.

   In contrast to the young Crab-like pulsars, the soft X-ray spectrum of 
   the Vela pulsar has a substantial thermal contribution with an apparent 
   temperature of $\approx 1$ MK (\"Ogelman, Finley \& Zimmerman 1993; Page, 
   Shibanov \& Zavlin 1996). On the other hand, the spatial structure of the 
   Vela plerion strongly resembles the inner Crab nebula --- it also has 
   a torus-like structure, an inner ring and jets 
   (cf.~Fig.~\ref{hri_vs_chandra}a/b and Fig.~\ref{vela_plerion}). The 
   symmetry axis of the nebula, which can be interpreted as the projection 
   of the pulsar's rotation axis onto the sky plane, is co-aligned with the
   direction of proper motion, exactly as for the Crab pulsar, which indicates
   that the ``natal kick'' of the neutron star occurs along the rotation axis 
   of the neutron star progenitor. The idea of torus configuration formed by a
   shock-confined pulsar wind was first introduced by Aschenbach \& Brinkmann 
   (1975) as a model to explain the shape of the inner Crab nebula. The discovery
   of a similar torus-like structure in the Vela synchrotron nebula hints that 
   this model may be applicable to many young pulsars.
   According to this model, the torus-like structure and its geometrical 
   orientation with respect to the direction of the pulsar's proper motion arise 
   because the interaction of the post-shock plasma with the ambient medium 
   compresses the plasma and amplifies the magnetic field ahead of the moving pulsar.
   This, in turn, leads to enhanced synchrotron emission with the observed 
   torus-like shape.

   Typical sizes of the  X-ray nebula structures scale approximately as 
   $\dot{E}^{1/2}$, as one should expect for relativistic pulsar winds shocked 
   by an ambient medium (Rees \& Gunn 1974). For instance, the inner ring 
   radii for the Crab and Vela nebulae are $0.14~\mbox{pc}$ (for $d=2$ kpc)
   and $0.02~\mbox{pc}$ (for $d=300$ pc), whereas the full extents of the
   X-ray nebulae are $1~\mbox{pc}$ and $0.1~\mbox{pc}$, respectively.
   The X-ray luminosity of the Vela plerion is only 0.04\% (0.1$-$2.4 keV)
   of the pulsar's spin-down energy loss, versus 5\%, 13\% and 1\% for 
   the Crab, B0540$-$69 and B1509$-$58 X-ray nebulae, respectively. 

   Since the other pulsars of this subclass are at much larger distances, 
   it is hard to resolve them from the putative surrounding X-ray nebulae. 
   Therefore, what has been observed is mainly emission from a 
   pulsar-powered synchrotron nebula combined with a small contribution of
   magnetospheric or thermal radiation. The latter is expected to dominate 
   at soft X-ray energies, below 0.5$-$1 keV, hardly observable in distant 
   pulsars because of interstellar absorption.
   The relatively small contribution of the pulsar's radiation, perhaps with
   intrinsically low pulsed fraction, has precluded the detection of pulsed soft 
   X-ray emission from these objects. Compact X-ray nebulae of physical sizes 
   $\sim 0.3 (d/2.4~{\rm kpc})$ pc, $0.4 (d/4~{\rm kpc})$ pc and $0.7 
   (d/2.5~{\rm kpc})$ pc have been observed from PSR B1706$-$44 (Becker, 
   Brazier \& Tr\"umper 1995; Finley et al.~1998), B1823$-$13 (Finley, 
   Srinivasan \& Park 1996), and B1951$+$32 (Safi-Harb, \"Ogelman \& Finley 
   1995), respectively. These sizes exceed that of the Vela X-ray nebula, 
   despite close values of $\dot{E}$, which can be tentatively explained by 
   lower pressure of the ambient medium around these pulsars. It is also 
   possible that future Chandra observations will reveal a fine spatial
   structure of these nebulae, which would lead to the explanation of the 
   apparent difference with the compact nebula around the prototype Vela pulsar.

   Thus, in spite of the apparent differences between the Crab-like and Vela-like 
   pulsars, the sample of well-investigated objects is still too scarce to 
   determine whether these differences are caused by a general evolution of pulsar
   properties during the first millenia of their lives or whether they are due to 
   some incidental properties inherent to the pulsars or their surroundings 
   (for instance, different orientations of the magnetic and rotation axes, or 
   different properties of the ambient medium). The most critical for understanding 
   the nature of these objects will be Chandra observations with high angular 
   resolution, as these observations will allow us to resolve the pulsars 
   from their X-ray plerions.

 %***********************************************************************************
 %                                                                                  %
 % SubSubSection:  ``Quiescent'' Radio-silent Neutron Stars in Supernova Remnants   %
 %                                                                                  %
 %***********************************************************************************
 %
 %
  \subsubsection{Radio-silent Neutron Stars in Supernova Remnants}  \label{radio_silent_ns}

  X-ray images of some young SNRs show bright point sources which have not been 
  detected in radio, optical and gamma-ray bands (see Table \ref{rad_quite_ins}). 
  The youngest among the detected sources of this type is the point source in 
  the very young (320 yr) Cassiopeia A supernova remnant  (cf.~Fig.~\ref{hri_vs_chandra}). 
  This source was discovered in the first light Chandra observation (Tananbaum 
  1999) and subsequently found in archival Einstein HRI and ROSAT HRI images 
  (Aschenbach 1999; Pavlov \& Zavlin 1999). The true nature of this source 
  remains elusive (Pavlov et al.~2000b). It shows no long-term (20 yr) or 
  short-term (days, months) variability, and no X-ray pulsations have been
  detected in the available data.
  The observed spectra do not have enough counts to distinguish 
  between different simple spectral models (e.g., power-law or blackbody,
  corresponding to a non-thermal or thermal origin of the detected emission). 
  However, it turns out that the spectrum is much softer than those of 
  young radio pulsars. If the emission occurs from the neutron star 
  surface, the temperature  distribution over the surface has to be 
  strongly non-uniform. The blackbody fit gives a temperature of 7 MK 
  and a radius of the emitting region of 0.3 km. Assuming that there 
  are magnetically confined hydrogen or helium hot polar caps on a 
  cooler iron surface, Pavlov et al.~(2000b) obtained 2.8 MK and 1 km 
  for the cap temperature and radius, and 1.7 MK for the surface effective 
  temperature. This anisotropic temperature distribution can cause
  a spin-modulation of the X-ray flux, which remains to be detected in
  future observations.

  \begin{figure}[t]
  \centerline{\psfig{file=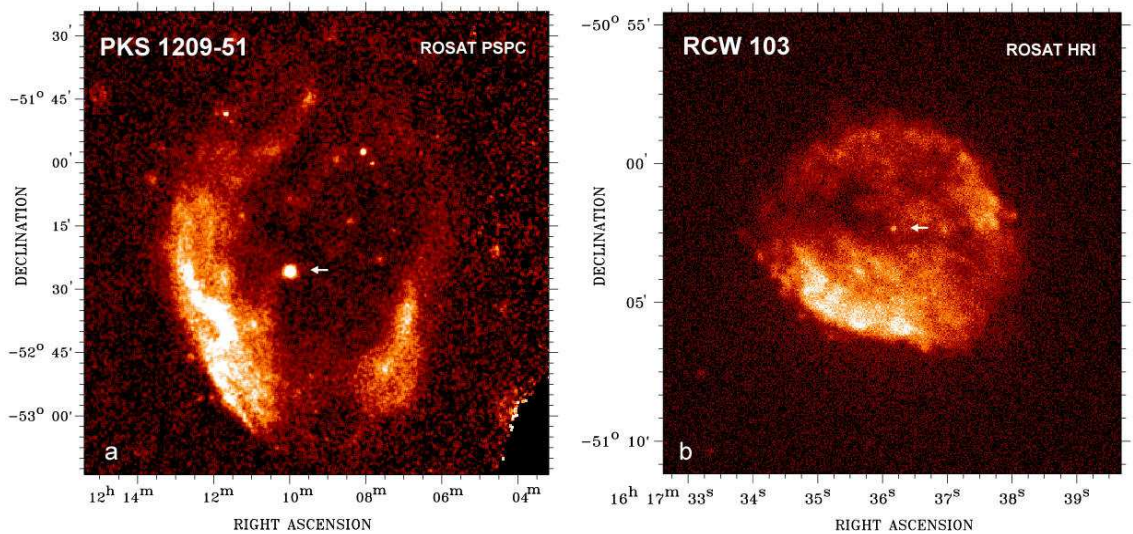,width=15cm}}
  \caption[]{\small ROSAT images of the supernova remnants 
   PKS 1209$-$51/52 and RCW 103. The arrows indicate the positions 
   of the neutron star 1E~1207$-$5209 and the neutron star candidate
   1E~161348$-$5055. Note the different scales of the images. PKS 
   1209-51/52 has an extent of about $1.5^\circ$ whereas the size
   of RCW 103 is about $10'$.\label{pks1209}\label{rcw103}}
  \end{figure}

  A similar point source, 1E~1207$-$5209 at the center of the
  $\approx 7$-kyr-old remnant PKS 1209$-$51/52 (see Fig.~\ref{pks1209}), was
  discovered with HEAO-1 (Tuohy et al.~1979)  and studied with Einstein,
  EXOSAT, ROSAT and ASCA (Helfand \& Becker 1984; Kellett et al.~1987;
  Mereghetti, Bignami \& Caraveo 1996). Its X-ray spectrum suggests that
  the X-rays are emitted from a hydrogen or helium atmosphere of the neutron 
  star, having an effective temperature $1.2-1.3$ MK (Zavlin, Pavlov \& 
  Tr\"umper 1998). The analysis of the Chandra observation of this source 
  has shown that its X-ray flux is modulated with a 424 ms period 
  (Zavlin et al.~2000), which finally proves that it is indeed a 
  neutron star.

  Another example of the radio-silent neutron star candidate is 
  1E~161348$-$5055 at the center of the supernova remnant RCW 103. 
  This source was discovered with the Einstein Observatory (Tuohy \& 
  Garmire 1980) and has an estimated age of $1-3$ kyr. Its X-ray 
  spectrum very strongly resembles that of the Cas A central point source. 
  However, comparing two ASCA observations of RCW 103, Gotthelf, 
  Petre \& Vasisht (1999a) found an order-of-magnitude decrease in its 
  luminosity in four years, which hints that this object may be an 
  accreting source. Even more puzzling is the six-hour periodicity of
  its flux reported by Garmire et al.~(2000) from the Chandra observations
  and archival ASCA data. Further investigations of 1E~161348$-$5055 with 
  Chandra and XMM-Newton are underway and will resolve the true
  nature of this source.

  Similar to the previous examples is the point source in Puppis A
  (cf.~Fig.~\ref{vela_pupis_image}), a supernova remnant located 
  at the edge of the Vela remnant. Puppis A has an age of about 4 kyr
  and harbors a central radio-silent X-ray bright source, RX~J0822$-$4300, 
  which is supposed to be a neutron star candidate (Petre, Becker \& Winkler 
  1996). Contrary to the compact stellar remnants in Cas A and RCW 103, 
  its spectrum and luminosity can be interpreted as  emitted from the 
  entire surface of a neutron star with a 10 km radius and a temperature of  
  $1.6-1.9$ MK, assuming that the surface is covered by a hydrogen 
  or helium atmosphere  (Zavlin, Tr\"umper \& Pavlov 1999). This 
  temperature, like that inferred for 1E~1207$-$5209, is compatible
  with standard neutron star cooling models. It is worth noting that 
  fitting the spectrum with a blackbody model gives an improbably small 
  neutron star radius of $1.0-1.5$ km and a higher temperature $4-5$ MK. 

  From what we know so far about radio-silent neutron stars in supernova remnants,
  one can conclude that such sources are quite different from radio pulsars 
  (in particular, they do not show any activity inherent to radio pulsars). 
  On the other hand, it is very plausible that, in fact, they are more common 
  than radio pulsars, and the relatively small number of the discovered 
  members of this class is due to observational selection --- it is much easier 
  to detect and identify active pulsars than these ``quiet'' sources observable 
  only in the soft X-ray band. 

  For completeness, we should also mention a number of young SNRs whose central 
  parts show bright extended (plerion-like) X-ray sources  with centrally-peaked 
  emission, with properties strongly resembling those observed from the plerions
  around Crab-like pulsars, but without a point source detected. Typical 
  examples of this class are 3C58 and G21.5$-$0.9, with estimated ages of 800 yr
  and $\sim 1$ kyr (Helfand, Becker \& White 1995; Slane et al.~2000).
  It seems very plausible that these SNRs do have active pulsars at the centers of
  their plerions, but an unfavorable direction of the pulsar beam precludes
  detection of pulsations.

 %******************************************************************************
 %                                                                             %
 % SubSubSection:  Soft Gamma-ray repeaters and Anomalous X-ray pulsars        %
 %                                                                             %
 %                                                                             %
 \subsubsection{Anomalous X-ray Pulsars and Soft Gamma-Ray Repeaters} \label{magnetars}

 \begin{table*}
 \begin{picture}(150,200)
 \put(-5,-25){\psfig{figure=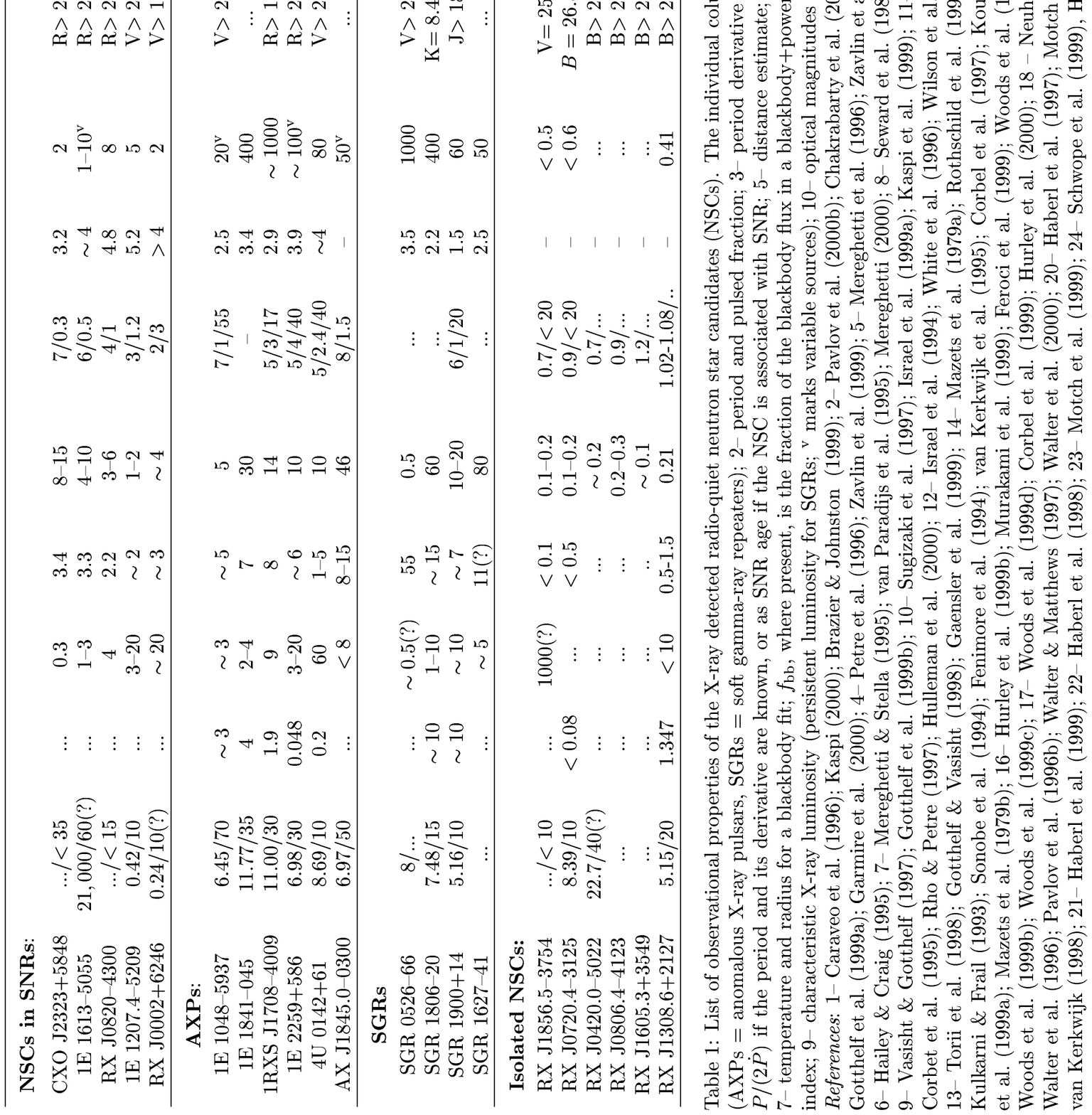,angle=90,width=22cm,height=18.3cm}}
 \end{picture}
 \refstepcounter{table} \label{rad_quite_ins}
 \end{table*}

  \begin{table*}
  \begin{picture}(150,230)
  \put(-20,-25){\psfig{figure=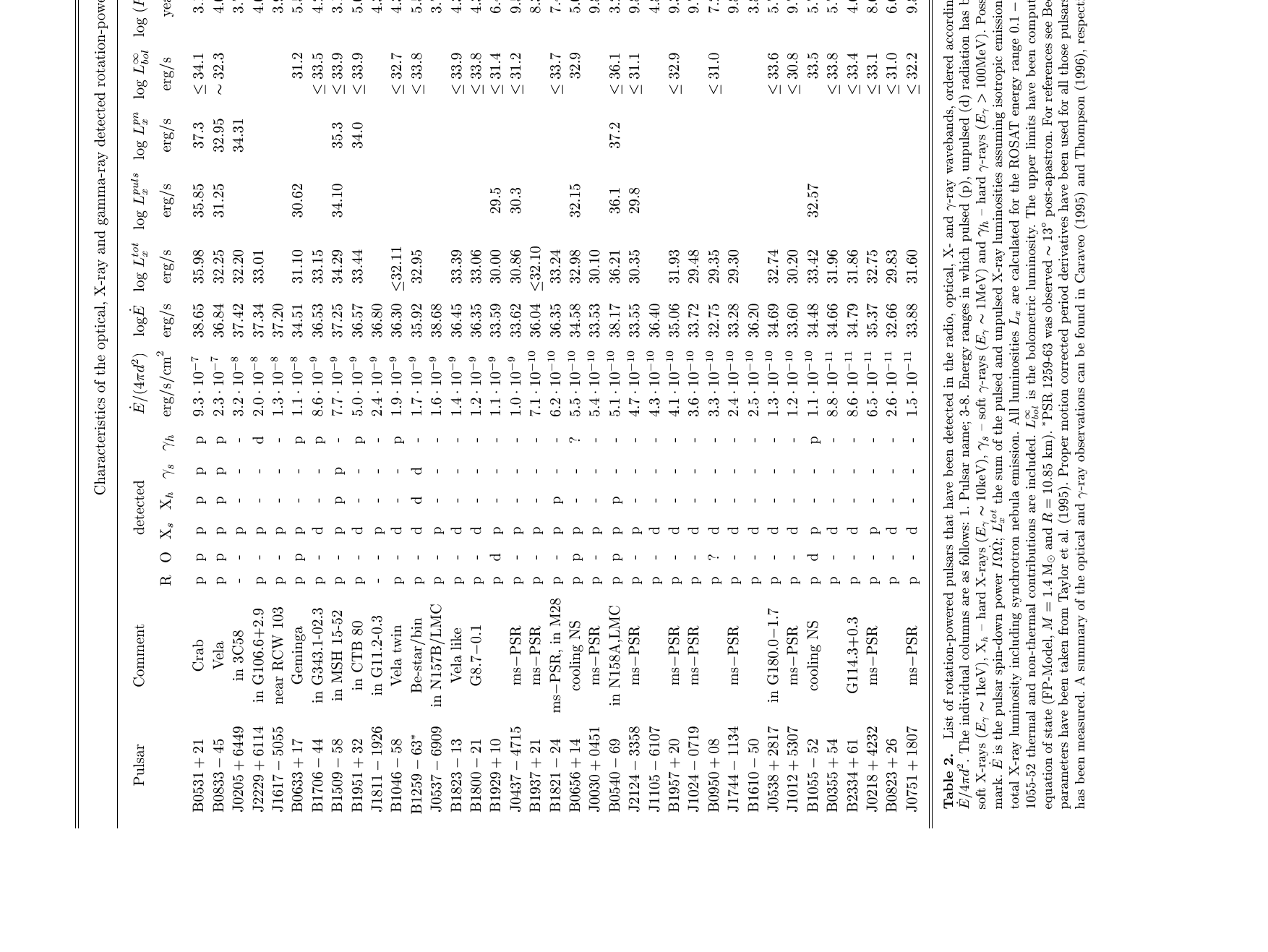,angle=90,width=23cm,height=20cm}}
  \end{picture}
  \refstepcounter{table} \label{sym_tab}
  \end{table*}

  In addition to the above-discussed young, X-ray-bright and radio-quiet neutron star
  candidates which do not show strong pulsations, a number of apparently young neutron
  stars with strong X-ray pulsations, the so-called {\em anomalous X-ray pulsars} (AXPs)
  and {\em soft gamma-ray repeaters} (SGRs), have been discovered recently
  (see Table \ref{rad_quite_ins} for references). At least some of them are believed 
  to be associated with supernova remnants. A common property of these objects is that 
  their periods are in a narrow range of $5-12$ s, substantially exceeding typical
  periods of radio pulsars. The AXPs and SGRs are, however, strongly different in
  their gamma-ray activity. While no gamma-ray emission has been detected from 
  AXPs, SGRs emit occasional gamma-ray bursts of enormous energy, up to 
  $10^{42}$--$10^{44}$ erg.

  Six anomalous X-ray pulsars have been discovered by the end of the century (see 
  Table \ref{rad_quite_ins}). They form a homogeneous class of pulsating neutron stars,
  clearly different from both the accreting pulsars in X-ray binaries and rotation-powered 
  radio pulsars (Mereghetti 2000). AXPs show a relatively stable period evolution with 
  $\dot{P}\approx (0.05$--$4)\times 10^{-11}$ s~s$^{-1}$. Characteristic spin-down 
  ages $\tau \sim 3-100$ kyr and magnetic fields $B\sim 10^{14}$--$10^{16}$~G were 
  estimated under the assumption that the spin-down is due to magneto-dipole braking,
  which is not necessarily correct because these objects are not powered by their
  rotation.
  If the estimated magnetic field strengths are correct, they strongly exceed those
  of radio pulsars, so that it has been suggested that AXPs, as well as SGRs, are 
  {\em magnetars} --- neutron stars with superstrong magnetic fields (Thompson \& 
  Duncan 1995,1996). They have soft X-ray spectra, with characteristic blackbody
  temperatures $T\approx 4-7$ MK and/or power-law indices $\alpha\approx 2.5-4$,
  and typical luminosities $L_x\sim 10^{34}-10^{36}\;\mbox{erg s}^{-1}$. Typical
  blackbody areas are $1-2$ orders of magnitude smaller than the NS surface area.
  At least three AXPs are associated with supernova remnants (see Table \ref{rad_quite_ins}).

  AXPs have been studied with many X-ray observatories, but their nature remains
  elusive. Although it has been widely accepted that these objects are magnetars, no
  direct proof of their superstrong magnetic fields has been obtained. It is not clear
  whether their X-ray emission indeed originates from the neutron star surface or from
  the magnetosphere and/or from a synchrotron nebula, and whether it is due to 
  ``internal'' radiation mechanisms (thermal or magnetospheric emission) or due 
  to accretion from, e.g., a residual disk (van Paradijs, Taam \& van den Heuvel 1995).
  Observations with Chandra and XMM, however, will probably provide the answer. 
  
  Soft gamma-ray repeaters (SGRs) are among the most fascinating galactic objects. 
  After the discovery of periods in the range of $P=5-8$ s and period derivatives
  $\dot{P}\sim 10^{-10}\;\mbox{s s}^{-1}$ (in two of the four known SGRs), it has 
  been suggested that these sources are associated with young, $\sim 1-10$ kyr, 
  neutron stars in supernova remnants. The energy released during the most
  powerful bursts of these sources is enormous --- e.g., an energy of $\;\gapr 
  10^{44}\mbox{erg}$ was estimated for the August 27, 1998 outburst of SGR 1900+14 
  (Inan et al.~1999). 

  SGRs are not only extremely powerful sources of gamma-ray bursts, but also 
  bright quiescent X-ray sources, with typical luminosities $L_x\sim 10^{34}-
  10^{36}\;\mbox{erg s}^{-1}$. The origin of the quiescent radiation remains 
  unclear. Statistically acceptable fits of the quiescent spectra can be obtained 
  with a combined blackbody plus power-law model, with typical parameters  
  $T\sim 5$ MK, $R\sim 1$ km and a photon-index of $\alpha=1-4$. The blackbody
  component might be interpreted as thermal radiation from the neutron star
  surface, but the area of the emitting region is two orders of magnitude 
  smaller than the neutron star surface area. The power-law component
  might hint that ultra-relativistic particles are involved, but no models 
  have been suggested to explain their origin 
  and acceleration mechanisms. However, with the poor angular resolution of 
  the ASCA telescopes, it is difficult to separate the point source radiation 
  from the diffuse SNR radiation, whereas the ROSAT count rates are too low 
  for a precise spectral analysis. Chandra and XMM-Newton observations will 
  yield much more definitive results and will allow one to reveal the nature 
  of the quiescent emission from SGRs and to elucidate the properties of
  the ultra-magnetized neutron stars apparently responsible for their radiation.

 %******************************************************************************
 %                                                                             %
 % SubSection:  The  Cooling Neutron Stars                                     %
 %                                                                             %
 %******************************************************************************
 %
 \subsection{Thermal Emission from Middle-Aged Pulsars} \label{cooling_ns}

\begin{figure}[th]
  \centerline{\psfig{figure=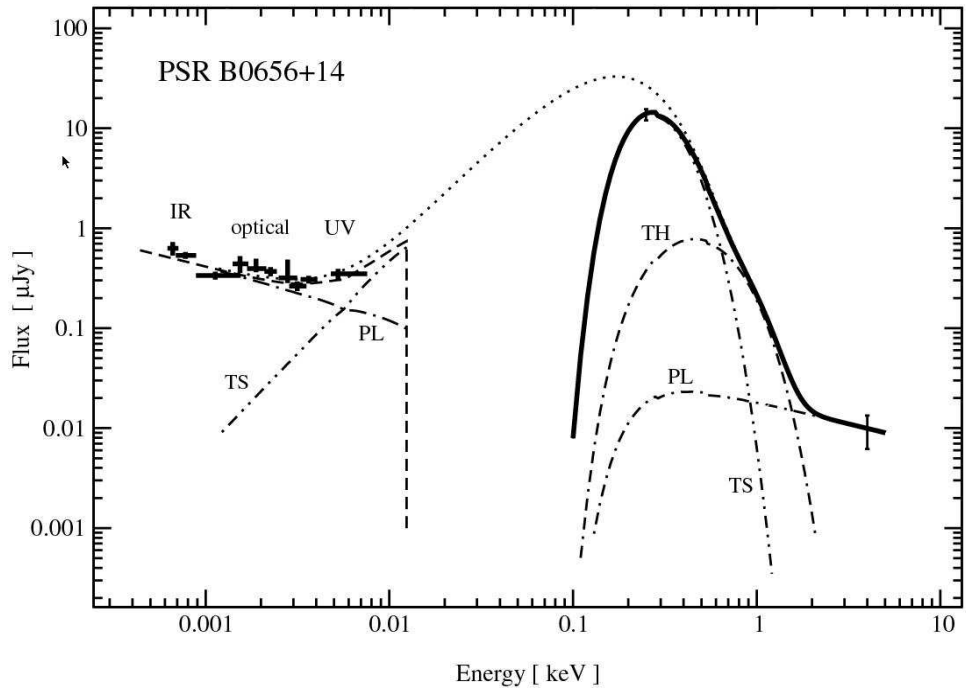,width=13cm}}
  \caption[]{\small Energy spectrum of PSR B0656+14, a prototype
   middle-aged pulsar with thermal radiation dominating in soft X-rays.
   Shown are the X-ray (ROSAT and ASCA) spectrum fitted with a model
   consisting of thermal soft (TS), thermal hard (TH) and 
   power-law (PL) components, and IR-optical-UV fluxes measured with the HST and
   ground-based telescopes. The error bars in the X-ray range show
   typical uncertainties in the ROSAT and ASCA bands.  The dashed and
   dotted lines show the continuation of the X-ray spectrum to the optical
   band with and without allowance for interstellar absorption.\label{0656_spec}}
  \end{figure}

 As we have discussed above, soft X-ray radiation of rotation-powered pulsars
 in an age interval of $\sim 10^5-10^6$ yrs should be dominated by thermal
 emission from the neutron star surface. These pulsars are old enough for their
 magnetospheric emission to become fainter than the thermal surface emission, 
 but they are still young and hot enough to be detectable in the soft X-ray range. 
 There are three middle-aged pulsars, Geminga, PSR B0656+14 and B1055$-$52, 
 from which thermal X-ray radiation from the surface of the cooling neutron 
 star has certainly been observed. Because of the similarity of their emission
 properties, they were dubbed {\em the three Musketeers} (Becker \& Tr\"umper 1997).
 The high-energy (IR through gamma-ray) spectra of these pulsars consist of
 two components. The thermal component dominates in the UV through soft X-ray
 range (up to $1-2$ keV), whereas the non-thermal component with approximately
 power-law (PL) spectrum prevails in IR, optical, hard X-ray and gamma-ray
 ranges. It follows from the ROSAT and ASCA observations of the brightest
 middle-aged pulsar B0656+14 that the thermal component cannot be described
 by a single temperature, i.e. the neutron star surface temperature is not
 uniform (Greiveldinger et al.~1996; Zavlin, Pavlov \& Halpern 2001). In the
 simplest model, the thermal component is comprised of a soft thermal 
 component (TS) from most of the neutron star surface (at $E\lapr 
 0.5-1$ keV) and a hard thermal component (TH) from polar caps 
 heated by relativistic particles. An example of a TS+TH+PL fit to the 
 IR-optical-X-ray spectrum of PSR B0656+14 is shown in Fig.~\ref{0656_spec}. 
 Alternatively, the temperature non-uniformity can be due to anisotropic 
 heat conductivity of the neutron star crust caused by anisotropic 
 magnetic field --- the heat flux across the field is suppressed so that 
 the magnetic poles are hotter than the equator (Greenstein \& Hartke 1983).

 \begin{figure}[th]
  \centerline{\psfig{figure=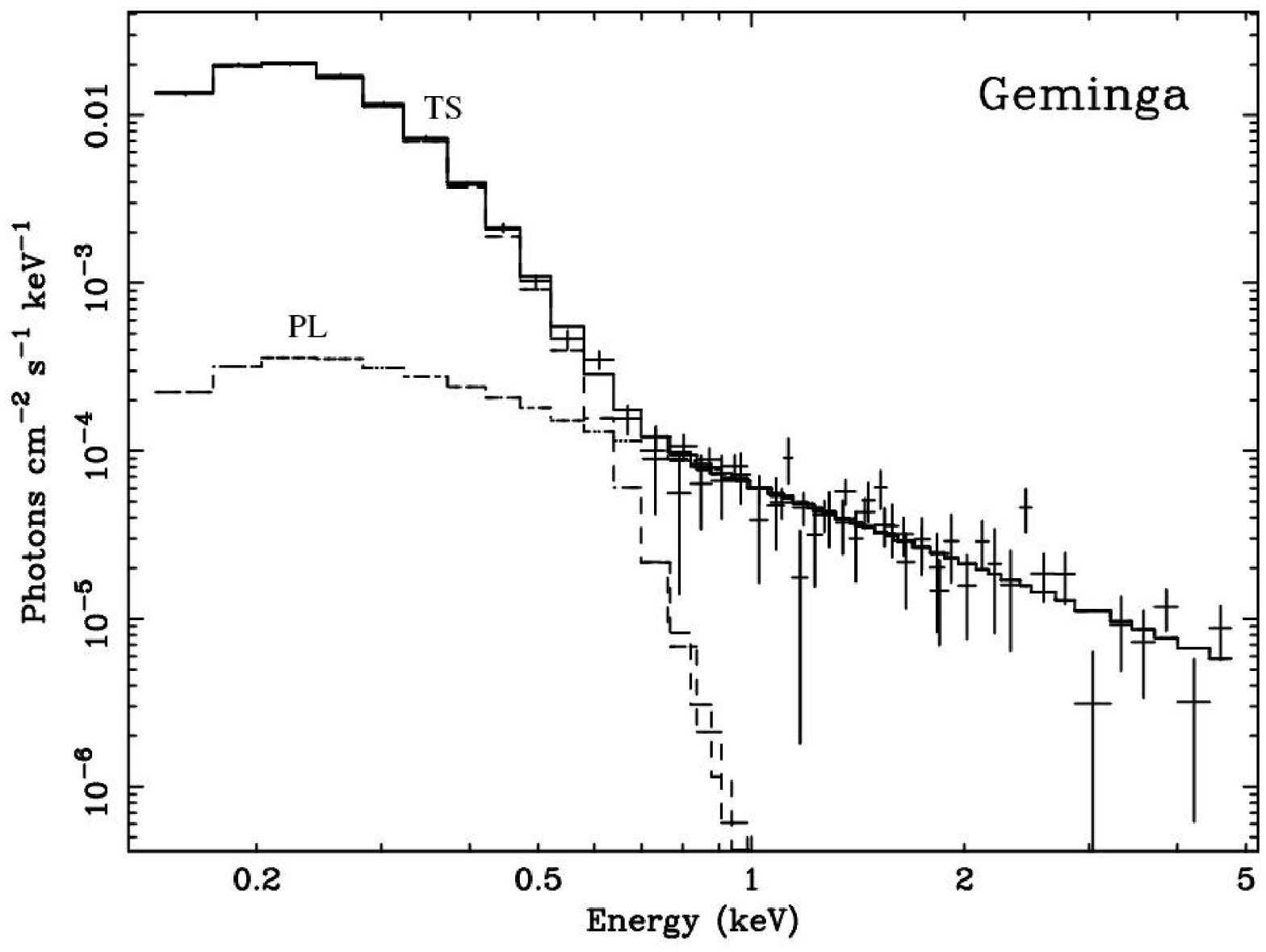,width=13cm,clip=}}
  \caption[]{\small X-ray spectrum of Geminga --- a typical spectrum of a
   middle-aged pulsar, consisting of a thermal component interpreted as
   emission from the neutron star surface, and a harder power-law (non-thermal)
   component, dominating  beyond $\approx 0.7$ keV. The soft part of the spectrum
   was obtained with ROSAT whereas the harder emission was observed by
   ASCA. (From Halpern \& Wang 1997.)\label{cooling_ns_spectrum} }
  \end{figure}

  The other two pulsars are not so bright as B0656+14, and their thermal components
  can be fitted with a single-temperature model (see Fig.~\ref{cooling_ns_spectrum}
  for the X-ray spectrum of Geminga).

  The existence of at least two spectral components is also confirmed by a   
  phase-resolved analysis of the X-ray emission (\"Ogelman 1995). All the 
  three pulsars show a phase shift of $\sim 100^\circ$ at an energy $0.4-0.6$ keV, 
  accompanied by an increase in the pulsed fraction from $\sim 10-30\%$ to 
  $\sim 20-65\%$. The X-ray pulse profiles for both the thermal and non-thermal 
  components are found to be approximately sinusoidal. The weak modulation of 
  the thermal soft component can be explained by the above-mentioned 
  non-uniformity of the surface temperature due to the presence of a 
  strong magnetic field.
  The surface temperatures of the three pulsars, obtained from blackbody fits,
  are in the range $T\sim 0.3-1.2$ MK. The radii of the emitting areas cannot
  be found without knowing the distances to these objects. Adopting the distances
  estimated from the radio-pulsar dispersion measure (which may be off by a
  factor of $\sim 2$), the radii are in the range $R\sim 7-30$ km, in rough
  agreement with the canonical neutron star radius of 10 km.
  The hard X-ray spectral components, dominating at energies above $\sim 1-2$ keV, 
  can be interpreted as magnetospheric emission (Halpern \& Wang 1997; Wang et 
  al.~1998; Greiveldinger et al.~1996; Zavlin et al.~2001).

\begin{figure}[ht]
  \centerline{\psfig{figure=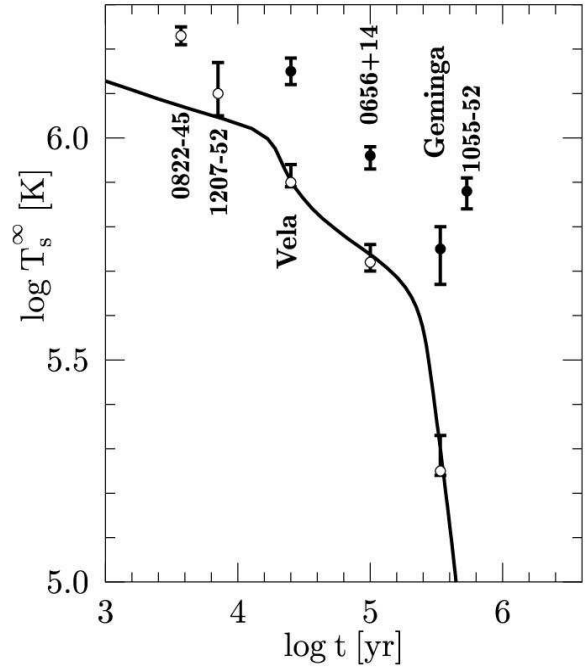,height=9cm,clip=}}
  \caption[]{\small Surface temperatures for the three musketeers 
   (PSR B0656+14, Geminga and B1055$-$52), the Vela pulsar and the 
   radio-quiet neutrons stars RX~J0822$-$4300 and 1E~1207$-$52. 
   The hatched regions indicate the possible ranges of $T_s^\infty$ 
   as predicted by the standard (double hatched) and accelerated 
   (single hatched) cooling models for different critical temperatures 
   of the superfluid neutrons and protons, $T_{cn}$ and $T_{cp}$  
   ($\sim 10^6 - 10^{10}$ K). The solid line shows the standard cooling 
   curve for a $1.30 M_\odot$ neutron star with $T_{cn}=200$ MK, 
   $T_{cp}=130$ MK. Filled and open circles indicate temperatures obtained from
   the blackbody and hydrogen-atmosphere fits, respectively.
   (From Yakovlev 
   et al.~1999). \label{cooling}}
  \end{figure}

  It should be stressed that the inferred effective temperatures, and hence the
  radius-to-distance ratios, depend on the model of thermal component. For
  instance, if one assumes that the neutron star surface is covered by a
  hydrogen or helium atmosphere, the effective temperatures are lower than
  those derived from the simple blackbody fits by a factor of $1.5-3$
  (Pavlov et al.~1995; see also Section \ref{photospheric}). An example
  demonstrating the difference of the temperatures inferred for the blackbody 
  and hydrogen atmosphere model fits is shown in Figure \ref{cooling}. We see 
  that the different spectral models correspond to quite different cooling 
  scenarios and, hence, to different properties of the neutron star interiors.
  Heavy-element atmospheres give temperatures close to the blackbody temperatures.
  However, the heavy-element atmosphere spectra should show numerous absorption
  lines and photoionization edges (Rajagopal \& Romani 1996; Zavlin et al.~1996).
  Because of low energy resolution of the ROSAT PSPC and low sensitivity
  of the ASCA SIS in soft X-rays, it has been impossible to detect such lines
  and thus to determine the chemical composition of neutron star atmospheres.
  We hope that this problem will be solved with Chandra and XMM-Newton.
  Without knowing the surface chemical composition, any conclusions about
  the effective temperatures and radii should be considered with caution.

  Important information on the emission mechanisms of middle-aged pulsars
  can be obtained from observations in the optical and gamma-ray ranges.
  PSR B0656+14 and Geminga have been observed at near-IR, optical and near-UV
  frequencies (Pavlov et al.~1996a;  Bignami et al.~1996; Shearer et al.~1996; 
  Pavlov, Welty \& C\'ordova 1997; Koptsevich et al.~2000), and PSR B1055$-$52 
  has been detected in a near-UV band (Mignani, Caraveo \& Bignami 1997). For all
  the three pulsars, the IR-optical flux is clearly non-thermal, while the
  thermal component starts to dominate at UV frequencies (see an example
  in Fig.~\ref{0656_spec}). For Geminga, a broad optical emission feature
  at $\sim 6000$ \AA$\;$ was reported by Bignami et al.~(1996), who
  attributed it to proton cyclotron emission from an atmospheric plasma.
  This interpretation does not look plausible because it requires an
  artificial velocity distribution for atmospheric electrons to explain
  the lack of electron cyclotron line in the hard X-ray spectrum. The
  nature of this feature, and the overall optical spectrum of Geminga,
  can hardly be understood without spectroscopic observations (Martin,
  Halpern \& Schiminovich 1998).

\begin{figure}[ht!]
  \hspace{2cm}{\psfig{figure=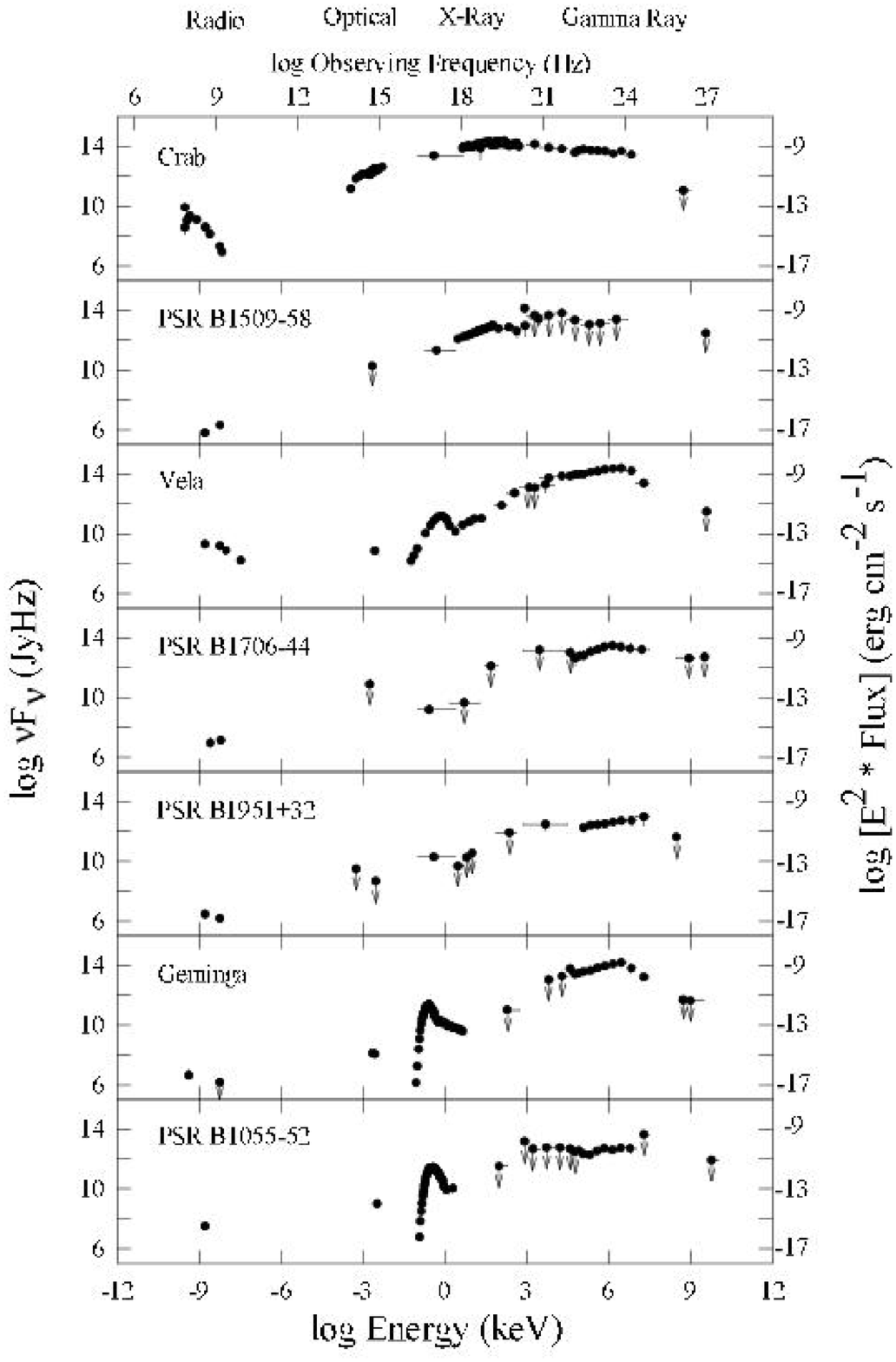,height=19cm,clip=}}
  \caption{\small Multiwavelength spectra for the known gamma-ray pulsars,
   showing the observed power per logarithmic energy interval. All the spectra have 
   in common that the high-energy radiation power rises from the optical to the
   X-ray band and that the maximum observed energy output is in the gamma-ray 
   band, which demonstrates that emitting particles are accelerated to very 
   high energies. (From Thompson et al.~1999). \label{nu_f_nu}}
   \end{figure}

   The two older middle-aged pulsars, Geminga and B1055$-$52, are bright
   gamma-ray pulsars in the CGRO EGRET energy range, 30 MeV -- 20 GeV,
   which gives the main contribution to their photon luminosity (see
   Thompson et al.~1999, and references therein).  Gamma-ray emission
   from B0656+14 has been marginally detected, at a 3$\sigma$ level
   (Ramanamurthy et al.~1996). The  gamma-ray spectra are close to
   power-laws, with photon indices of about $1.4-1.8$ (see Fig.~\ref{nu_f_nu}).
   A spectral turnover at about 3 GeV has been observed in the Geminga
   spectrum. The data can be interpreted with both the polar cap and
   outer gap models (see Section \ref{magnetospheric_emission_models}).
   Observations in a broader energy range with more sensitive gamma-ray
   detectors are required to construct a detailed model of the gamma-radiation.

\begin{figure}[th]
   \centerline{\psfig{file=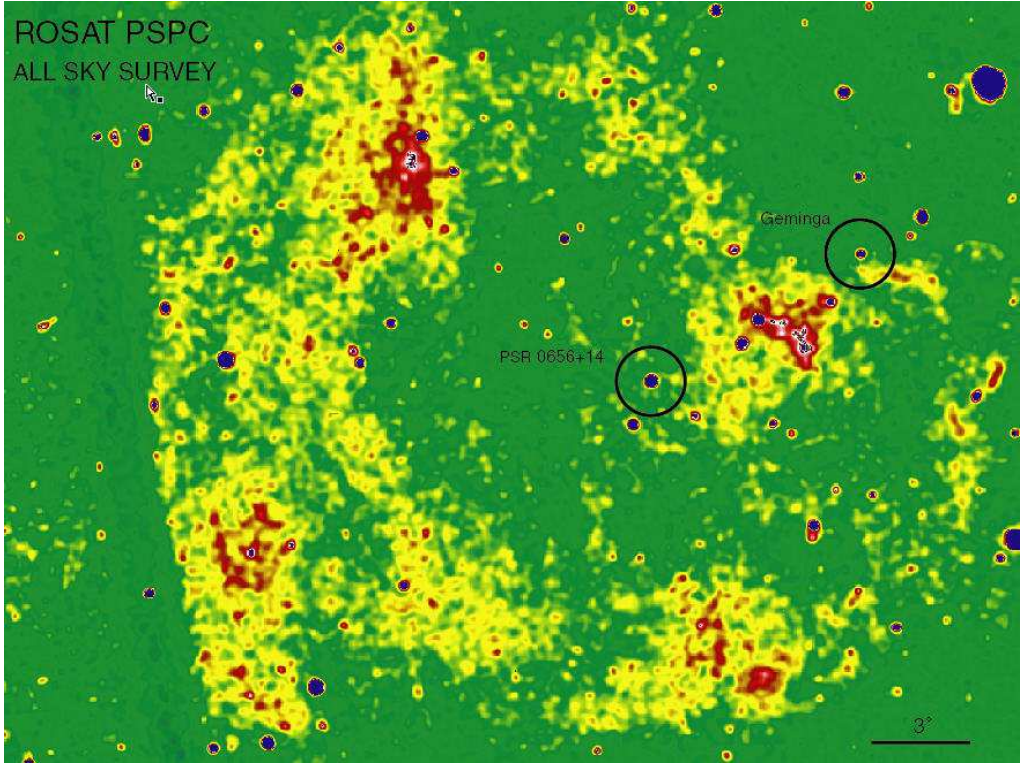,width=13.5cm,clip=}}
   \caption{\small The $20^\circ$-wide Monogem Ring as observed
   in the ROSAT all-sky survey. The ROSAT PSPC full fields of view
   during the pointed observations of Geminga and PSR 0656+14 are
   indicated by circles. The image demonstrates that both pulsars
   are located in crowded regions with patchy background emission,
   strongly blurred with ASCA spatial resolution of $\sim
   3'$. The image demonstrates the power of the first
   all-sky survey with an imaging X-ray telescope, providing X-ray
   images of extended celestial objects of very large sizes.
   \label{monogem}}
   \end{figure}

   Since all active pulsars are powerful sources of relativistic winds,
   one should expect that they generate pulsar-wind nebulae (PWNe), similar
   to those observed around the Crab-like and Vela-like pulsars. The PWN
   sizes should scale as $(\dot{E}/p_0)^{1/2}$, where $p_0$ is the pressure
   of the ambient medium. The existence of X-ray bright PWNe (albeit of
   much larger sizes) around several pulsars, including the three musketeers,
   was reported by Kawai \& Tamura (1996) based on ASCA observations.
   However, the analysis of the ROSAT and  BeppoSAX observations of these
   sources  by Becker et al.~(1999)  did not confirm the ASCA results ---
   the extended emission observed with ASCA was resolved in a number of
   unrelated background objects. In particular, Geminga and PSR B0656+14
   are located  in  the Monogem  ring (see Fig.\ref{monogem}), a
   $\sim 20^\circ$ wide object which is believed to  be an old and nearby
   supernova remnant (see Plucinsky et al.~1996). A large  fraction of the
   sources detected by ASCA are found to  be diffuse and fuzzy emission of a 
   small part of the Monogem ring rather than pulsar-powered nebulae.

   In addition to PSR B0656+14, B1055$-$52 and Geminga, one could expect
   thermal radiation from the cooling neutron star surface to dominate
   in soft X-ray emission from two more middle-aged pulsars detected with
   ROSAT, PSR B0538+28 and B0355+54. Both have spin parameters similar to 
   those observed for Geminga and B1055$-$52, and both appear to be good 
   candidates for gamma-ray pulsars. However, these sources are approximately 
   a factor of 10 more distant than Geminga, so that the limited photon statistics
   has hampered a spectral or temporal analysis.

 %******************************************************************************
 %                                                                             %
 % SubSection:  Pulsars old and close in Space                                 %
 %                                                                             %
 %******************************************************************************
 %
 \subsection{Old Nearby Radio Pulsars} \label{old_ns}

\begin{figure}[ht!]
  \centerline{\psfig{figure=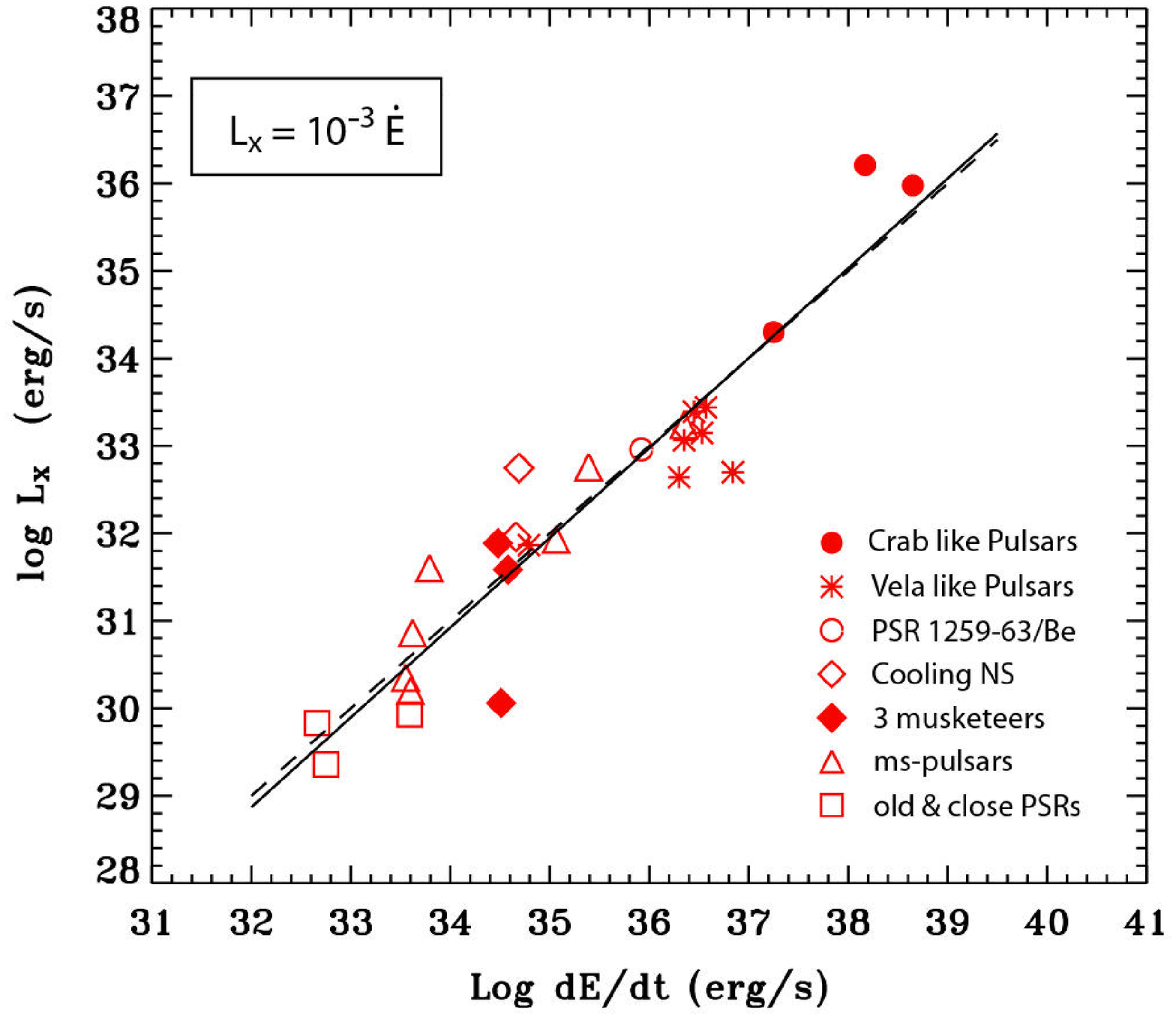,width=12cm,clip=}}
  \caption{\small X-ray luminosity vs.~spin-down energy loss for all
  rotation-powered pulsars detected by ROSAT. For the ``three musketeers'',
  Geminga, PSR 0656+14 and 1055-52, the low energy thermal component
  has been subtracted from the data. The solid line represents
  $L_x(\dot{E}) \propto \dot{E}^{1.03}$, the dashed line $L_x(\dot{E})
   = 10^{-3} \dot{E}$. Remarkably, all the detected pulsars, from the young 
  Crab-like to the $10^9$ year old millisecond pulsars follow the
  linear trend. (From Becker \& Tr\"umper 1997). \label{lx_vs_edot}}
   \end{figure}

  When the age of a neutron star reaches $\sim 10^6$ yr, its temperature becomes
  too low to be observed in X-rays. At the same time, the energy loss
  rate $\dot{E}$, and hence the luminosity of non-thermal radiation and
  thermal radiation from polar caps of radio pulsars, also decrease with
  age. Therefore, old pulsars are faint in the X-ray range and can be
  observed only at small distances.  ROSAT and ASCA have detected X-ray 
  emission from three old and close pulsars: PSR B1929+10, B0950+08 and 
  B0823+26. All the three are characterized by a spin-down age of $2-30$ 
  Myr and are at distances of $\sim 0.2-0.4$ kpc. Temporal and spectral 
  information, however, is only available for PSR B1929+10 (Yancopoulos, 
  Hamilton \& Helfand 1994), whereas for the other two pulsars the 
  sensitivity of ROSAT and ASCA was not sufficient to collect enough
  photons for a detailed analysis. The pulse profile of PSR B1929+10 
  is very broad, with a single pulse stretching across almost the entire 
  phase cycle. Becker \& Tr\"umper (1997) and Wang \& Halpern (1997) 
  found that both the power-law and  black-body models fit the observed spectrum 
  equally well, leaving the origin of the detected X-rays unconstrained. 
  If the observed radiation is interpreted in terms of thermal emission 
  from hot polar caps ($T\sim 3$ MK), the caps appear to be very small 
  ($A\sim 100$ m$^2$), and their X-ray luminosity is much lower than
  predicted by many polar cap heating models.  If this radiation is
  non-thermal, its luminosity is consistent with the general trend, 
  $L_x\sim 10^{-3} \dot{E}$, found by Becker \& Tr\"umper (1997) for the
  non-thermal emission from those radio pulsars which are detected 
  in the X-ray range (see Fig.\ref{lx_vs_edot}). The sensitivity of 
  XMM-Newton and Chandra is required to finally identify the emission 
  mechanism.

  Although the thermal radiation from the surface of an old neutron star
  does not peak anymore in the X-ray band, the power of the Hubble Space 
  Telescope allows one to observe it in the optical-UV range. Pavlov 
  et al.~(1996a) detected PSR B1929+20 at near-UV frequencies and showed 
  that the observed flux corresponds to a temperature of about 0.2 MK, if 
  the radiation is thermal. A candidate for the optical counterpart of 
  PSR B0950+08 was also detected by the same authors, but its identification 
  is less certain. Studying optical radiation from old nearby pulsars is 
  very useful for understanding their thermal and non-thermal evolution.

  No gamma-radiation has been detected even from the nearest old 
  radio pulsars. Since the gamma-ray efficiency, $\epsilon_\gamma = 
  L_\gamma/\dot{E}$,  grows with age for young and middle-aged pulsars 
  (for PSR 1055$-$52 it is almost 20\%), it must have a break at about
  $\sim 1$ Myr not to exceed 100\%. A reason for this break could
  be that the thermal surface photons are involved in the production
  of the observed gamma-rays, via Compton up-scattering of thermal
  photons by ultrarelativistic particles. As a neutron star cools down 
  with growing age, the productions of gamma-rays, and hence the 
  gamma-ray efficiency, decreases.

\subsection{Isolated Radio-quiet Neutron Stars}

  Analyzing the ROSAT PSPC observations of a field containing
  the molecular cloud R~CrA ($d\approx 130$ pc), Walter, Volk \& Neuh\"auser 
  (1996) noticed a bright point source, RX~J1856$-$3754, projected onto the 
  cloud. Its spectrum is very soft -- the best fit with a blackbody model 
  gives a temperature of 0.66 MK and a luminosity of $\sim 5\times 10^{31}\;
  \mbox{erg s}^{-1}$ for a distance of 100 pc. Based on the lack of an optical 
  counterpart brighter than V$\sim 23$, Walter et al.~(1996) suggested that 
  the source is a nearby isolated neutron star. As the objects described 
  in 8.3.1.3, it is radio-quiet, but, contrary to those objects, it is
  not associated with any SNR (i.e., it is ``truly isolated''). Since 
  its temperature is much lower than the temperatures of the isolated
  neutron stars in supernova remnants, it is natural to assume
  that this object is much older, i.e. the neutron star has cooled down.
   Surprisingly, no pulsations of the X-ray radiation,
  expected from a neutron star with a ``typical'' magnetic field
  and favorable orientation of the magnetic and spin axes, were 
  found. Pavlov et al.~(1996b) fitted the X-ray spectrum with
  neutron star atmosphere models (see Section \ref{photospheric}) 
  and showed that different chemical compositions and different 
  magnetic fields of the surface layers correspond to quite 
  different optical magnitudes, V=$22-28$, and distances, $5-200$ 
  pc (for a neutron star radius of 10 km). Therefore, optical 
  detection of this source (and other similar sources) would be 
  a powerful tool to investigate the properties of neutron star 
  atmospheres, while measuring the distance would allow one to 
  evaluate the neutron star radius and constrain equation of 
  state of the neutron star interiors.

  A very faint, blue optical counterpart of RX J1856$-$3754
  was discovered with the HST by Walter \& Matthews (1997). The 
  extremely large X-ray-to-optical flux ratio of $\sim 75,000$
  proves unequivocally that this is indeed a neutron star.
  However, the optical magnitude, V=25.7, is considerably different 
  from the predictions of the four atmosphere models considered 
  by Pavlov et al.~(1996b), which means that either the atmosphere
  has a different chemical composition and magnetic field or the 
  temperature distribution is essentially non-uniform, e.g., because 
  of anisotropy of heat conduction in a very strong magnetic field. 
  The nature of RX~J1856$-$3754 became even more puzzling after its 
  proper motion, 0.33 arcsec/yr, was measured (Walter et al.~2000). 
  This proper motion corresponds to a transverse velocity of 140 km/s 
  at $d=100$ pc, too fast for accretion from ISM to be a major heating 
  source. This means that RX~J1856$-$3754 is a cooling neutron star, 
  and with the apparent surface temperature of $\approx 0.7$ MK it 
  should be younger than 1 Myr, for the standard (slow) 
  cooling models.  On the other hand, projecting the proper motion 
  backward, Walter et al.~(2000) suggest that RX~J1856$-$3754 and 
  the well-known runaway O star $\zeta$ Oph originated from the same 
  binary system disrupted by a supernova explosion about 2 Myr ago. 
  A neutron star of such an age should have an apparent effective 
  temperature $<0.4$ MK, in contradiction with the current data.

  The lack of pulsations, which could be explained by co-alignment
  of the magnetic and rotation axes, or the rotation axis and
  the line of sight, does not allow one to measure the spin period
  of RX~J1856$-$3754. Fortunately, two other objects with similar
  properties, for which pulsations have been measured, were 
  discovered with ROSAT -- RX J0720$-$3125 with $P=8.37$ s and
  RX J0420$-$5022 with $P=22.7$ s (Haberl et al.~1996,1997,1999).
  Future measurements of their period derivatives will allow one 
  to estimate their ages and elucidate the nature of these neutron
  stars. Particularly important will be deep optical/UV observations of
  these objects (a viable candidate for the optical counterpart
  of RX J0720$-$3125, with B$\simeq 26.6$, has been found by
  Motch \& Haberl 1998 and Kulkarni \& van Kerkwijk 1998).

  Three more objects of apparently the same class are known at the
  time of writing this article (see Table \ref{rad_quite_ins} 
  and references therein). We expect that the number of detected 
  radio-silent neutron stars  will grow considerably in the near 
  future, and we will be able to compare their properties with 
  predictions of different models of  neutron star evolution. One 
  of the most important problems related  to these objects is the 
  source of energy which heats the radiating layers of the neutron
  stars up to $0.7-1.4$ MK -- it may be either the internal heat 
  of relatively young cooling neutron stars, presumably with large
  magnetic fields, or accretion from the ISM onto old neutron stars 
  (e.g., Treves et al.~2000). Presently, we cannot exclude the  
  possibility that the observed six sources belong to two quite 
  different classes, young coolers and old accretors.

  %******************************************************************************
  %                                                                             %
  % SubSection:  The recycled millisecond Pulsars                               %
  %                                                                             %
  %******************************************************************************
  %

  \subsection{Recycled Millisecond Pulsars} \label{ms_psr}

  In the $P$-$\dot{P}$ parameter space, millisecond pulsars (ms-pulsars) are
  distinguished from the majority of ordinary-field pulsars by their short
  spin periods and small period derivatives, corresponding to very old
  spin-down ages of typically $10^9-10^{10}$ years and low magnetic field
  strengths of $\sim 10^8 - 10^{10}$ G (cf.~Fig.~\ref{p_pdot}). More
  than $\sim 75\%$ of the known disk ms-pulsars are in binaries with a compact
  companion star, compared to $\cong 1\%$ binaries among the ordinary pulsars.
  This gives support to the idea that these neutron stars have been spun-up
  by angular momentum transfer during a past mass accretion phase
  (Bisnovatyi-Kogan \& Komberg 1974; Alpar et al.~1982;  Bhattacharya \& van
  den Heuvel 1991). Indeed, the first accreting ms-pulsar, SAX J1808.4$-$3658,
  has been discovered with BeppoSAX (see van der Klis, this book). Presumably,
  these pulsars were originally among ordinary pulsars which would have
  turned off because of the loss of their rotational energy if they were not
  in close binaries; they are therefore often called the ``recycled'' pulsars.

  By the end of 2000, about 100 recycled radio pulsars are known, of which 57
  are located in the galactic plane (Camilo 1999; Edwards et al.~2000;
  Lommen et al.~2000; Lyne et al.~2000; Manchester et al.~2000). The others
  are in globular clusters (Kulkarni \& Anderson 1996; Camilo et al.~2000)
  which provide a favorable environment for the recycling scenario (Rasio,
  Pfahl \& Rappaport 2000). Only 10 of the 57 ms-pulsars in the galactic
  plane are solitary (including PSR B1257+12 which has a planetary system); the
  rest are in binaries, usually with a low-mass white dwarf companion. The
  formation of solitary recycled pulsars is not well-understood, but it is
  widely believed that either the pulsar's companion was evaporated
  or the system was tidally disrupted after the formation of the ms-pulsar.

  Recycled pulsars had been studied exclusively in the radio domain until
  the early 1990's, when ROSAT, ASCA, EUVE, RXTE and BeppoSAX were launched.
  The first millisecond pulsar discovered as pulsating X-ray source was
  PSR J0437$-$4715 (Becker \& Tr\"umper, 1993), a nearby 5.75 ms pulsar
  which is in a binary with a low-mass white dwarf companion. Further
  detections followed, which, by the end of the century, sum up to almost
  1/3 of all X-ray detected rotation-powered pulsars (cf.~Table \ref{sym_tab}).

\begin{figure}[h]
  \centerline{\psfig{figure=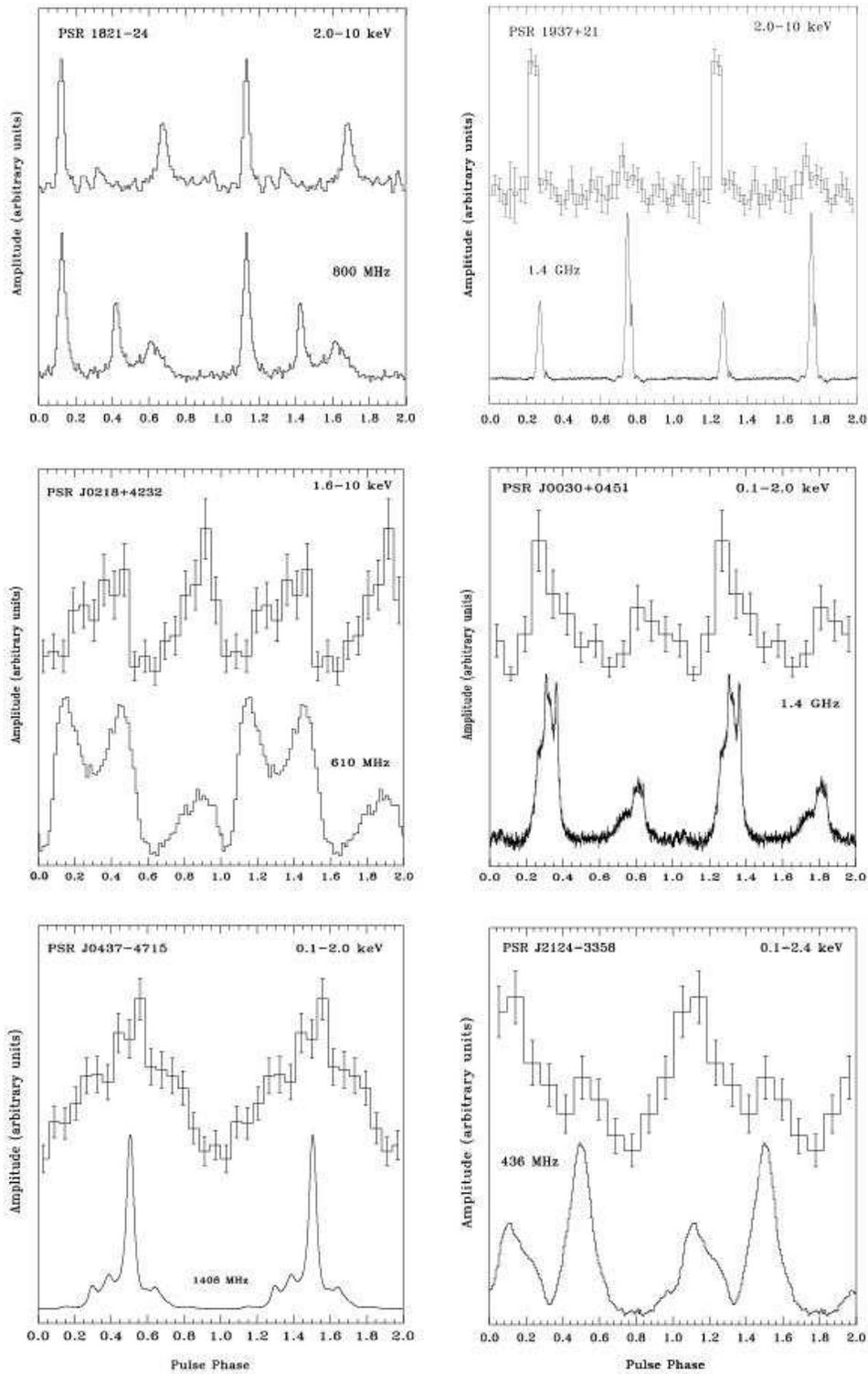,height=20cm,clip=}}
  \caption{\small Integrated lightcurves for all ms-pulsars for which spin-modulated
   X-ray emission is detected. The upper phase histograms show the X-ray profiles
   in the given energy bands. The radio lightcurves are shown for comparison. Two
   phase cycles are shown for clarity. The relative phase between the radio and X-ray
   pulses is only known for PSR 1821-24 and PSR B1937+21 (Takahashi et al.~2001).
   In all other cases the phase alignment is arbitrary because of the lack of
   accurate satellite clock calibration.}
   \label{1821_1937_lc}\label{0218_BeppoSax}\label{0437_lc}\label{ms_profiles}
   \label{2124_lc}\label{0030_lc}
  \end{figure}

  The available data suggest that the observed X-ray emission is likely to be
  generated by non-thermal processes in most of ms-pulsars. This is supported 
  by observations of the 3.05 ms pulsar B1821$-$24 which is located in the 
  globular cluster M28 (Kawai \& Saito 1999), PSR B1937+21 -- the fastes 
  ms-pulsar known (Takahashi et al.~1999), and PSR J0218+4232 (Mineo et 
  al.~2000). For these objects, power-law spectra and/or pulse profiles with 
  narrow peaks have been measured (see Fig.~\ref{ms_profiles}). For PSR 
  J0437$-$4715, the results of a recent Chandra observation suggest that the
  emission contains both the thermal component from the hot polar caps (Zavlin 
  \& Pavlov 1998) and a non-thermal component from the magnetosphere. The data 
  on J2124$-$3358 do not allow to determine unambiguously which of the two 
  components, thermal or nonthermal, is present in the observed emission.
  The 4.86 ms pulsar J0030+0451, which has spin parameters similar to those of
  J2124$-$3358, shows a high pulsed fraction of $69\pm 18\%$. This, together 
  with its Crab-like pulse profile and the gross similarity between its radio 
  and X-ray profiles (cf.~Fig.~\ref{0030_lc}), suggests that the X-ray emission 
  of this pulsar is dominated by the non-thermal component (Becker et al.~2000).

  All other X-ray detected ms-pulsars (B1957+20, J1012+5307, B0751+18,
  J1744$-$1134 and J1024$-$0719) are identified only by their positional
  coincidence with the radio pulsar (Becker \& Tr\"umper 1999) and, in 
  view of the low number of detected counts, do not provide much more than 
  flux estimates. The power of XMM-Newton and Chandra is needed to explore 
  their emission properties in more detail. However, the fact that all 
  millisecond pulsars have roughly the same X-ray efficiency ($L_X/\dot{E} 
  \sim 10^{-3}$) as ordinary pulsars (cf.~Fig.\ref{lx_vs_edot}) supports 
  the conclusion that, as a rule, the non-thermal X-ray radiation from their 
  magnetospheres prevails over the thermal radiation from their polar caps 
  (cf.~Becker and Tr\"umper 1997).

  As far as the emission of gamma-rays from ms-pulsars is concerned, PSR
  J0218+4232 has been proposed to be the counterpart of the EGRET source
  2EG J0220$+$4228 (Verbunt et al.~1996; Kuiper et al.~2000). The  final
  verification, however, has to await the next gamma-ray missions, GLAST,
  and INTEGRAL, which are scheduled for the first decade of the new millennium.
  If J0218+4232 is indeed a gamma-ray pulsar, then, depending on the assumed emission
  model,  $7-33\%$ of its spin-down energy would go into the production of
  gamma-rays. No other ms-pulsars have been identified with gamma-ray sources
  so far, although according to the polar-cap and outer-gap emission models
  their predicted efficiencies should be even higher than that estimated for
  J0218+4232.

\begin{figure}[t]
  \centerline{\psfig{figure=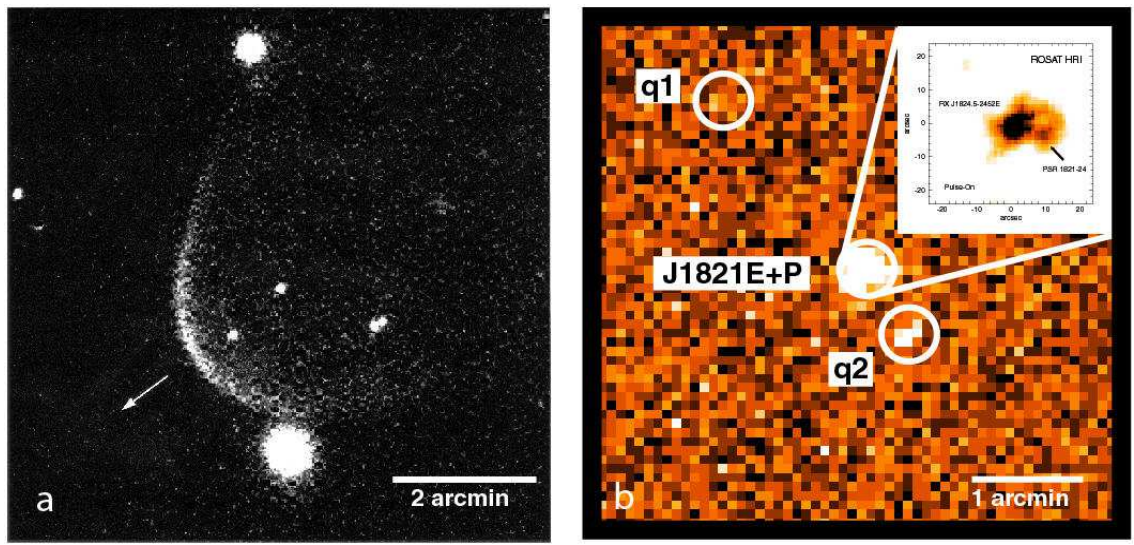,width=15cm}}
  \caption[]{\small ({\bf a}) Bow-shock nebula around PSR J0437$-$4715 as observed
   in the H$_\alpha$ emission (courtesy of A.~Fruchter). The arrow indicates the
   direction of the pulsar's proper motion. ({\bf b}) $5'\! \times\! 5'$ ROSAT HRI
   image of the globular cluster M28. \rxe, \rxp, q1, and q2 indicate the positions
   of four X-ray sources, of which q1 and q2 are globular cluster background sources.
   The upper-right inset magnifies the core encompassing J1824E+P. Here, the ROSAT
   HRI data are oversampled at $1''$ bins and temporally phased to emphasize
   ``pulse-on'' events from the millisecond pulsar B1821-14 which is the faint
   source indicated by the arrow.\label{0437_bow_shock} \label{m28_hri}}
  \end{figure}

  An important aspect of pulsar studies is searching for pulsar-wind nebulae
  (PWNe). So far, bow-shock PWNe have been firmly detected in H$_\alpha$
  emission around PSR 1957+20 (Fruchter et al.~1992) and PSR J0437$-$4715 (Bell,
  Bailes \& Bessel 1993). The bow-shock stand-off distance found in J0437$-$4715
  is about $7''$ (Bell et al.~1995).  Observations with the ROSAT HRI yielded
  a 3$\sigma$ upper limit of $0.2\times 10^{-3} \dot{E}$ for the X-ray emission
  from the nebula (Becker \& Tr\"umper 1999).
  Another PWN candidate is an object RX J1824.5$-$2452E near PSR B1821$-$24 (see
  Fig.~\ref{m28_hri}). However, as the pulsar is located in a globular cluster,
  it is quite likely that this extended X-ray source is a superposition of
  spatially unresolved globular cluster sources (cataclysmic variables or
  low-mass X-ray binaries) rather than a plerion powered by the pulsar. We
  expect the true nature of RX J1824.5$-$2452E will be established in a deep
  Chandra observation of the globular cluster M28.

\section{Impressive Achievements and Great Expectations}

 Astronomers of our generation have been truly lucky --- their 
 collective efforts, generously supported by tax-payers of different 
 countries, revolutionized our understanding of the Universe and its 
 constituents, from clusters of galaxies to neutron stars. Fifty years 
 ago, it was hard to imagine that neutron stars, very hypothetical 
 objects at that time, not only would be discovered, but will also be studied 
 in such a detail\footnote{Sachiko Tsuruta, who devoted her scientific 
 life to studying the thermal evolution of neutron stars, recalls an 
 episode of the mid-sixties, when she had finished her PhD thesis on the
 thermal evolution of neutron stars, which by that time were not expected
 to be observable. Says Tsuruta: ``.... By 1965, Bahcall and  Wolf 
 published papers that a neutron star cannot be seen if there are pions 
 in the neutron star core. Soon after that I met John Bahcall at some 
 conference, and he urged me to bet for discovering neutron stars, while 
 he would bet against it. To my regret, I replied that a good Japanese 
 woman should not bet. Then in 1967 a pulsar was discovered!''}. 
 The new vision of the Universe in general, and neutron stars in particular,
 has become possible only due to opening the new windows for observing
 the fascinating Cosmos --- now we can study the Universe not only through 
 the traditional, very narrow optical window, but also in radio, X-rays 
 and gamma-rays. Without the X-ray and gamma-ray space observatories, our 
 understanding of neutron stars, and many other objects virtually unknown 
 half the century ago, would be much less complete. In particular, 
 high-energy observations of the last decade of the 20-th century allowed 
 us to understand that the world of neutron stars is not as simple as 
 many astronomers had believed in the seventies and eighties (and some of
 them still believe). Neutron stars are not ``just dim, heavy balls 
 of ten kilometer radius'',  as an expert in extragalactic astronomy claimed, 
 explaining why proposals to observe neutron stars with the Hubble Space 
 Telescope have been rejected so often. The neutron stars are not all 
 alike --- on the contrary, their properties and observational manifestations 
 are no less diverse than those of usual stars and galaxies.

 Amongst the more than 1,000 neutron stars discovered, about 100, including 
 60 isolated neutron stars (see Tables \ref{rad_quite_ins} and \ref{sym_tab}) 
 have been observed at high energies with space observatories. These 
 observations have firmly established that the properties of neutron stars are 
 indeed highly unusual, particularly, their gravitational and magnetic fields 
 are truly immense. We dare to predict that such exotic properties will never 
 be achieved in terrestrial laboratories. Thus, neutron stars provide a unique 
 opportunity to study the matter under extreme conditions. In particular, 
 neutron stars can be viewed as cosmic laboratories for studying nuclear 
 interactions, general relativity and superstrong magnetic and electric 
 fields. This is the point where astrophysics and physics merge and cannot 
 be separated from each other. 

 In spite of the impressive achievements of the neutron star physics/astrophysics,
 a lot of work still remains to be done in this recently emerged field. First, 
 the evolution of neutron stars, starting from their violent birth in supernova
 explosions, is far from being well understood. Until very recently, a common
 prejudice had been that all neutron stars are born as active, rotation-powered
 pulsars, which slow down their rotation, eventually stop their activity and, 
 after crossing a ``death line'', get into the ``pulsar graveyard''. A former 
 pulsar remains in the graveyard forever, cool and quiet,  unless it is captured
 by a flying-by star (e.g., in a globular cluster) and forms a close binary, 
 where accretion onto the neutron star can spin it up (recycle) to so short 
 periods that it again becomes an active pulsar. 
  
 The recent high-energy observations, however, show that the picture may not 
 be so simple. In particular, it appears that many very young neutron stars are
 not active pulsars at all. The most recent example may be the central source
 of the 320-year-old Cassiopeia A supernova remnant 
 (see Fig.~\ref{hri_vs_chandra}; 
 although at the time of writing of this article it is still not completely 
 clear whether it is a neutron star or a black hole). Since such objects are not
 seen in radio, and are extremely faint in optical, they could not be observed
 until the onset of the X-ray astronomy era, which means that our perception
 of neutron star early evolution was very strongly biased in favor of much
 easier observable rotation-powered pulsars. Why are many (perhaps, the majority 
 of) nascent neutron stars not active pulsars? Is it because they are 
 indeed magnetars, whose superstrong magnetic field inhibits the pulsar 
 activity? Or, on the contrary, their magnetic fields are so weak and/or 
 rotation is so slow that the pulsar does not turn on? Or the pulsar activity
 is quenched by accretion of debris of the supernova explosion? Are the 
 (apparently young) anomalous X-ray pulsars and soft gamma-ray repeaters 
 indeed the magnetars or their unusual observational properties are due to 
 quite different reasons, like a residual disk? To answer these questions, 
 further observations, with more sensitive instruments of higher angular 
 and energy resolution are needed. 

 One more set of evolutionary problems is associated with the generation 
 and evolution of neutron star magnetic fields. Although there are no doubts
 that the very strong fields exist in many (if not all) neutron stars, there 
 is no clear understanding of how they are generated. Why they are so different 
 in different kinds of neutron stars (e.g., regular and recycled pulsars), 
 what is their geometry, and do they decay during the neutron star life time? 
 It should be mentioned that the direct measurements of the magnetic field 
 have been possible only for neutron stars in binaries. What is called the 
 ``magnetic field'' in, e.g., radio pulsars, is only an order-of-magnitude 
 model-dependent estimate. Direct measurements of magnetic fields in isolated 
 neutron stars, e.g. with the aid of spectral lines formed in their photospheres 
 or from X-ray polarimetry, is one of very important goals for future 
 observations.

 One of the most important goals in studying isolated neutron stars is 
 elucidating their internal composition (neutrons, quark-gluon plasma, 
 strange matter, meson condensate ?) and the properties of the superdense
 matter (equation of state, nucleon superfluidity). Different equations of 
 state correspond to different  mass-radius dependences. Hence, the most 
 direct way to determine the equation  of state (which, in turn, depends 
 on the internal composition) is measuring the masses and radii of neutron 
 stars.  One can constrain the radius from the  star's bolometric flux and 
 effective temperature, provided the thermal radiation is not strongly 
 ``contaminated'' by magnetospheric radiation of relativistic particles. 
 The effective temperature can be determined from fitting the spectrum of
 the thermal radiation to neutron star atmosphere models. This method 
 requires a good knowledge of the distance to the neutron star,  which
 can be estimated from the radio dispersion measure if the neutron star 
 is an active radio pulsar, or, much more precisely, from measuring its 
 parallax.

 The $M/R$ ratio can be directly measured from the gravitational redshifts of 
 spectral lines in the X-ray range. Measuring the redshifts would require 
  X-ray detectors with high 
 energy resolution and reliable computations for the 
 energies of various atoms and ions in strong magnetic fields. Since the  
 atomic states are greatly distorted by typical magnetic fields of neutron 
 stars (and, consequently, spectral lines are strongly shifted from their 
 zero-field positions), an independent measurement of the magnetic field 
 (e.g., from fitting the continuum radiation to neutron star atmosphere 
 models) would be necessary.
 The $M/R$ ratio can also be evaluated from the analysis of the X-ray pulse 
 profiles. Since the temperature is not uniform over the neutron star surface
 (due to anisotropy of the heat transfer in the crust or, for active pulsars, 
 due to the accretion of relativistic particles onto the pole regions), 
 the observed X-ray flux should vary with the rotation period. Because of the 
 gravitational bending of photon trajectories, the shape of the pulse 
 profile substantially depends on $M/R$. The observations will require high 
 detector sensitivity ($\sim 10^4-10^5$ counts are needed to obtain accurate 
 pulse profiles in a few energy ranges) and good time resolution ($\sim 10^{-5}$ 
 s for millisecond pulsars). Finally, the properties of the internal matter 
 can be constrained from measuring the effective temperatures of neutron stars 
 of different ages. The thermal evolution of neutron stars depends 
 substantially on the internal composition, equation of state, and nucleon 
 superfluidity.

  Of course, no firm conclusions about the internal properties can be drawn
  without studying the physical conditions in the surface layers of neutron
  stars:  magnetic fields, temperatures, densities and, in particular, the 
  chemical composition.  Elucidating the chemical composition is also important 
  in order to understand how neutron stars interact with their environment, 
  both in their very young age, when a fraction of the supernova ejecta can 
  fall back on the star's surface, and during their whole life which may 
  include some accretion episodes. The investigation of the surface layers 
  (atmospheres) will require the analysis of soft X-ray spectra in terms of 
  atmosphere models (hence, high sensitivity and spectral resolution of the 
  detectors are needed). Important information about neutron stars can be 
  obtained  from their transverse velocities (proper motion) and parallaxes. 
  These quantities have been measured for a handful of radio pulsars. X-ray 
  telescopes with sub-arcsecond angular resolution will allow us to measure 
  astrometric characteristics of nearby radio-quiet neutron stars. 

  After the 32 years of radio pulsar investigations, we still lack a consistent
  theory of the pulsar activity. New X-ray and gamma-ray data are expected to 
  close this gap. In particular, X-ray radiation of many pulsars is due to 
  relativistic particles in their magnetospheres, and studying the spectra 
  and the pulse profiles of this radiation will allow us to determine the 
  energy spectrum and directional pattern of the relativistic particles and, 
  consequently, conditions in the pulsar acceleration zones and their temporal 
  evolution. Furthermore, the X-ray range is most convenient for investigating 
  the hot polar caps of radio pulsars, inevitable companions of the pulsar 
  activity.

  We expect that many of the above-formulated goals will be achieved with the
  aid of the satellite X-ray observatories Chandra and XMM-Newton, launched
  in 1999. First few months of Chandra observations have brought several 
  important discoveries: the central compact object in Cas A, the unusual 
  six-hour period of the central source of RCW 103, the discovery of the 
  small-scale structure in the compact nebulae around the Crab and 
  Vela pulsars. Much more discoveries, from both Chandra and XMM-Newton,
  will have been done by the time when this article is published.
  Furthermore, a number of new high-energy missions, GLAST, INTEGRAL,
  Constellation-X, and XEUS are being planned and, hopefully, will be
  launched within the next $10-20$ years. Particularly useful
  for studying isolated neutron stars will be the Constellation-X and XEUS
  missions. For instance, the Constellation-X mission is planned to consist of
  six X-ray telescopes to be
  launched to the libration point in 2007--2008. Each of these payloads will
  combine the excellent angular resolution of Chandra with the large
  collecting area of XMM-Newton, so that we may expect a new revolution
  in X-ray astronomy in the second decade of the third millennium.
  Thus, we are looking forward to new discoveries which will raise additional questions
  --- as such is the nature of the scientific cognition.

 %******************************************************************************
 %                                                                             %
 % Section:  REFERENCES                                                        %
 %                                                                             %
 %******************************************************************************
 %
 \section{References}

 \begin{verse}
 \begin{small}

 Alpar M.A., Cheng A.F., Ruderman M.A., Shaham J., 1982, Nature, 300, 728
 % ms-pulsars are recycled

 Arons, J., Scharlemann, E.T. 1979, ApJ, 231, 854
 % polar cap models

 Arons, J., Tavani, M., 1993, ApJ, 403, 249
 % High-Energy emission from the Eclipsing millisecond pulsar PSR 1957+20

 Aschenbach, B. 1999, IAU Circ. \#7249
 % Cas A in ROSAT images
 
 Aschenbach B., Brinkmann W. 1975, A\&A, 41, 147
 % A model of the X-ray structure of the Crab Nebula

 Baade, W., Zwicky, F., 1934, Proc.~Nat.~Acad.~Sci.,20, 254
 % On super-novae, see also: Bade, W., Zwicky, F., 1934,
 % Phys.~Rev., 45, 138, Supernova and Cosmic rays
 
 Bhattacharya D., van den Heuvel E.P.J., 1991, Phys.~Rep., 203, 1
 %ms-pulsars are recycled

 Becker, W., \& Tr\"umper, J. 1993, Nat, 365, 528
 % 0437 paper

 Becker, W., Tr\"umper, J., \"Ogelman, H.B., 1993, in {\it Isolated
 Pulsars}, (eds K.A.~Van Riper, R.~Epstein \& C.~Ho), 104-109,
 (Cambridge University Press)
 % survey paper on neutron star observation

 Becker, W., 1995, Thesis, MPE-Report 260

 Becker, W., Aschenbach, B., in: {\it The Lives of the Neutron Stars}, eds. 
 A.Alpar, U.Kilizoglu \& J.van Paradijs, Kluwer Academic Publishers, p47

 Becker, W., Brazier, K., Tr\"umper, J., 1995, A\&A, 298, 528 

 Becker, W., Tr\"umper, J., 1997, A\&A, 326, 682
 % The X-ray luminosity of rotation-powered pulsars

 Becker, W., Tr\"umper, J.  1999, A\&A, 341, 803
 % X-ray emission properties of ms-pulsars                  

 Becker, W., Kawai, N., Brinkmann, W., Mignani, R., 1999, A\&A, 352, 532
 % The putative pulsar-wind nebula of the three musketeers

 Becker, W., Tr\"umper, J., Lommen, A.N., Backer, D.C., 2000, ApJ, 545, 1015
 % 0030 paper.

 Bell, J. 1977, Ann. NY Acad. Sci., 302, 685
 % After dinner speech about the discovery of radio pulsars by J.Bell on the
 % 8th Texas Symposium on Relativistic Astrophysics

 Bell, J.F., Bailes, M., Bessel, M.S. 1993, Nature, 364, 603
 % Optical detection of the companion of the millisecond pulsar J0437-4715
 % [discovery of Halpha nebula]
 
 Bell, J.F., Bailes, M., Manchester, R.N., Weisberg, J.M., Lyne, A. 1995,
 ApJ, 440, L81
 % The proper motion and wind nebula of the nearby millisecond pulsar
 % J0437-4715

 Bertsch, D.L., et al. 1992, Nature, 357, 306
 % Pulsed gamma-radiation from Geminga

 Beskin, V.S., Gurevich, A.V., Istomin, Ya.N., 1993, Physics of the 
 Pulsar Magnetosphere, Cambridge University Press, ISBN 0-521-41746-5

 Bignami, G.F., Caraveo, P.A., 1996, Ann.~Rev.~Astron.~Astrophys.~, 34, 331-381
 % explanation on the meaning of the name Geminga, complete, overall summary

 Bignami, G.F., Caraveo, P.A., Mignani, R., Edelstein, J., Bowyer, S., 1996, ApJ, 456, L111  
 % Multiwavelength data suggest a  cyclotron feature on  the hot thermal continuum  of Geminga.

 Bisnovatyi-Kogan G.S., Komberg B.V., 1974, Sov.~Astron., 18, 217
 % ms-pulsars are recycled

 Boyd P.T., van Citters G.W., Dolan J.F., et al., 1995, \ApJ 448, 365
 % braking index of 0540-69

 Bowyer, S., 1990, in {\em Observatories in Earth Orbit and Beyond},
 eds. Y.~Kondo, Kluwer Academic Publishers, p.153
 % EUVE Mission paper

 Bowyer, C.S., Byram, E.T., Chubb, T.A., Friedman, H., 1964, Nature, 201, 1307
 % detection of X-rays from the Crab in a Moon occultation experiment in the
 % first   rocket flight.

 Bradt, H.V., Rappaport, S., Mayer, W., Nather, R.E., Warner, B., 
 Macfarlane, M.,
 Kristian, J., 1969, Nature, 222, 728
 % X-ray and optical observations of the pulsar NP 0532 in the Crab Nebula

 Bradt, H.V., Swank, J.H., Rothschild, R.E. 1990, Adv. Space Res., 10, 297
 % The X-ray timing explorer

 Brazier, K.T.S., Becker, W. 1997, MNRAS, 284, 335
 % High-resolution X-ray imaging of the SNR MSH 15-52

 Brazier, K.T.S., Johnston, S. 1999, MNRAS, 305, 671
 % The implications of radio-quiet neutron stars

 Brinkmann, W., \"Ogelman, H. 1987, A\&A, 182, 71
 % X-ray Observations of the Radio Pulsar 1055--52 [EXOSAT]

 Bulik, T., Pavlov, G.G. 1996, ApJ, 469, 373
 %Polarization modes in a strongly magnetized hydrogen gas

 Butler, R.C., Scarsi, L., 1990, in {\em Observatories in Earth Orbit and Beyond},
 eds Y.~Kondo, Kluwer Academic Publishers, p.141 
 % BeppoSAX mission paper

 Camilo, F. 1999, In {\em Pulsar Timing, General Relativity and the Internal
 Structure of Neutron Stars}, ed. Z. Arzoumanian, F.~Van der Hooft, \&
 E.P.J.~van den Heuvel, [Amsterdam: Koninklijke Nederlandse Akademie van
 Wetenschappen], p. 115

 Camilo, F., Lorimer, D.~R., Freire, P., Lyne, A.~G., \& Manchester,
 R.~N., 2000, ApJ, 535, 975
 % Observation of 20 millisecond pulsars in 47 Tuc at 20 cm
 
 Caraveo, P.A., Bignami, G.F., Tr\"umper, J. 1996, A\&A Rev., 7, 209
 % Radio-silent isolateted NSs as a new astyronomical reality

 Chakrabarty, D., Pivovaroff, M.J., Hernquist, L.E., Heyl, J.S., Narayan, R.
 2001, ApJ, in press
 % Cas A

 Chandrasekhar, S., 1931, ApJ, 74, 81 
 % The maximum mass of ideal white dwarfs

 Cheng, A.F., Helfand, D.J. 1983, ApJ, 271, 271
 % Detection of 1055--52 with Einstein

 Cheng, K.S., Ho, C. \& Ruderman, M.A., 1986a, ApJ, 300, 500
 % Energetic radiation from rapidly spinning pulsars. I. Outer magnetosphere gaps

 Cheng, K.S., Ho, C. \& Ruderman, M.A., 1986b, ApJ,  300, 522
 % Energetic radiation from rapidly spinning pulsars. II. Vela and Crab

 Chiu, H.Y., Salpeter, E.E., 1964, Phys.~Rev.~Letters, 12,413
 % proposed cooling neutron star as X-ray counter part for X-ray source in Crab
 % detected by Bowyer et al

 Clear, J., et al. 1987, A\&A, 174, 85-99
 % Spectral studies of the Crab using COS-B

 Cocke, W.J., Disney, M.J. Taylor, D.J. 1969 Nature, 221, 525
 % optical pulses of the Crab pulsar

 Corbel, S., Wallyn, P., Dame, T.M., Durouchoux, P., Mahoney, W.A.,
 Vilhu, O., Grindlay, J.E. 1997, ApJ, 478, 624
 % Distance to SGR 1806-20

 Corbel, S., Chapus, C., Dame, T.M., Durouchoux, P. 1999, ApJ, 526, L29
 % Distance to SGR 1627-41

 Corbet, R.H.D., Smale, A.P., Ozaki, M., Koyama, K., Iwasawa, K. 1995,
 ApJ, 443, 786
 % Spectrum and pulses of 1E 2259+586 from ASCA and BBXRT observations

 C\'ordova F.A., Hjellming R.M., Mason K.O., Middleditch J., 1989, ApJ, 345, 451\\
 % 0656+14 Einstein paper

 Danner, R., Kulkarni, S.R., Tr\"umper, J. 1998, AAS Meeting 192, \#43.09
 % The x-ray counterpart to SGR 0526-66: Monitoring N49 with ROSAT

 Daugherty, J.K., Harding, A.K. 1996, ApJ, 458, 278
 % polar cap model -- cascade

 Edwards, R.~T., 2000, in {\it Pulsar Astronomy - 2000 and Beyond},
 ed. M.Kramer, N.Wex, and R.Wielebinski, [San Francisco : ASP], p.33
 % ms pulsars

 Fenimore, E.E., Laros, J.G., Ulmer, A. 1994, ApJ, 432, 742
 % X-ray spectrum of SGR 1806-20 [ICE; during bursts]

 Feroci, M., Frontera, F., Costa, E., Amati, L., Tavani, M., Rapisada, M.,
 Orlandini, M. 1999, ApJ, 515, L9
 % A ginat outburst from SGR 1900+14 observed with BeppoSAX GRB monitor

 Fichtel, C.E., Hartman, R.C., Kniffen, D.A., Thompson, D.J., Bignami, G.F.,
 Ogelman, H., Ozel, M.E., Tumer, T., 1975, ApJ, 198, 163
 % Mission Overview SAS 2

 Forman, W. et al., 1978, ApJS, 38, 357
 % The 4th UHURU catalog of X-ray sources

 Fritz, G., Henry, R.C., Meekins, J.F., Chubb, T.A., Friedman, H., 1969, Science, 164, 709
 % X-ray pulsar in the Crab Nebula

 Fruchter, A.S., Bookbinder, J., Garcia, M.R., Bailyn, C.D., 1992, Nature, 359, 303
 % PWN arroud 1957+20 in H_alpha 

 Gaensler, B.M., Gotthelf, E.V., Vasisht, G. 1999, ApJ, 526, L37
 % A new SNR coincident with slow pulsar AX J1845-0258

 Garmire, G.P., Pavlov, G.G., Garmire, A.B., Zavlin V.E. 2000, IAU Circ. \#7350
 % 6-hr periodicity in CCO of RCW 103

 Giacconi, R., Gursky, H., Paolini, F.R., Rossi B.B., 1962, Phys.~Rev.~Lett., 9, 439
 % Evidence for X-rays from sources outside the solar system

 Giacconi, R., Kellogg, E., Gorenstein, P., Gursky, H., Tananbaum, H., 1971, ApJ, 165, L27
 % UHURU Mission Overview

 Giacconi, R., 1974, in {\em X-ray Astronomy}, eds R.~Giacconi and H.~Gursky,
 R.~Reidel Publishing Company 1974, Holland, ISBN 90 277 02950

 Giacconi, R., et al, 1979, ApJ, 230, 540
 % Einstein mission paper

 Glen, G., Sutherland, P. 1980, ApJ, 239, 671
 %Cooling of NSs; "isothermal approximation"

 Glendenning, N.K., 1996, Compact Stars, Springer, ISBN-0-387-94783-3
 % book

 Gnedin, Yu.N., Pavlov, G.G. 1974, Sov.~Phys.--JETP, 38, 903
 %Transfer equations for normal waves and polarization of radiation in an
 %anisotropic medium (first formulation of radiative transfer in NS atmospheres)

 Gold, T., 1968, Nature, 218, 731
 % Rotating neutron stars and the origin of the pulsating radio sources

 Gold, T., 1969, Nature, 221, 25
 % Rotating neutron stars and the nature of pulsars

 Gotthelf, E.V., Petre, R., Hwang, U. 1997, ApJ, 487, L175
 % The nature of central source of RCW 103

 Gotthelf, E.V., Petre, R., Vasisht, G. 1999a, ApJ. 514, L107
 % Varibility of CCO of RCW 103 

 Gotthelf, E.V., Vasisht, G. 1998, New Astronomy, 3, 293
 % Discovery of AX J1845-0258 (AXP)

 Gotthelf, E.V., Vasisht, G., Dotani, T. 1999b, ApJ, 522, L49
 % Spin history of the AXP in Kes 73

 Gotthelf, E.V., Wang, Q.D. 2000, ApJ, 532, L117
 % Plerionic nebula around 0540-69 (HRC/Chandra)

 Gouiffes, C., Finley, J.P., \"Ogelman, H., 1992, ApJ, 394, 581
 % optical pulse profile of PSR 0540

 Green D.A., 1998, `A Catalogue of Galactic Supernova Remnants
 (1998 September version)', Mullard Radio Astronomy Observatory,
 Cambridge, United Kingdom
 %SNR catalog

 Greenstein G., Hartke, G.J., 1983, ApJ, 271, 283
 % Pulse-like character of blackbody radiation from neutron stars

 Greiveldinger, C., Camerini, U., Fry, W., et al. 1996, ApJ, 465, L35
 % Heated polar caps in PSR 0656+14 and 1055-52

 Gregory, P.C., Fahlman, G.G. 1980, Nature, 287, 805
 % Discovery of 1E2259+586 in CTB109 with Einstein

 Gudmundsson, E.H., Pethick, C.J., Epstein, R.I. 1983, ApJ, 272, 286
 % relation between T_s and T_i

 Haberl, F., Motch, C., Buckley, D.A.H., Zickgraf, F.-J., Pietsch, W. 1997,
 A\&A, 326, 662
 % RX J0720.4-3125: Strong eveidence for an INS

 Haberl, F., Motch, C., Pietsch, W. 1998, Astron. Nachr., 319, 97
 % INSs in the ROSAT Survey [RX J0806-4123]

 Haberl, F., Pietsch, W., Motch, C. 1999, A\&A, 351, L53
 % RX J0420.0-5022: an INSC with evidence for 22.7 s X-ray pulsations

 Hailey, C.J., Craig, W.W. 1995, ApJ, 455, L151
 % Discovery of NSC in a new SNR near CTB1

 Halpern, J.P.,  Holt, S.S., 1992, Nature, 357, 222
 % Identification of Geminga

 Halpern, J.P., Wang, F.Y.-H., 1997, ApJ, 477, 905
 % A Broad-band X-ray study of the Geminga Pulsar

 Hambarian, V., Hasinger, G., Schope, A.D., Schulz, N.S., 2001,  A\&A, in press
 % Discovery of 5.16s pulsations from the isolated neutron star RBS 1223 / RX 1308.6+2127

 Harding, A.K., Muslimov, A.G. 1998, ApJ, 500, 862
 % Pulsar x-ray and gamma-ray pulse profiles: Constraint on
 % obliquity and observer angles

 Harnden, F.R., Seward, F.D. 1984, ApJ, 283, 274
 %Einstein observations of the Crab pulsar and its compact nebula

 Helfand, D.J., Becker, R.H. 1984, Nature, 307, 215
 % Einstein observations of PKS 1209--52

 Hill, R.J., et al. 1997, ApJ, 486, L99
 % HST/FOS observation of 0540-69

 Hillier , R.R., Jackson, W.R., Murray, A., Redfern, R.M., Sale, R.G., 1970, ApJ, 162, L177
 % discovery of gamma-ray pulses up to 0.6 MeV from Crab in balloon experiment

 Hoyle, R.A., Narlikar, J., Wheeler, J.A., 1964, Nature, 203, 914
 % Electromagnetic waves from very dense stars

 Hulleman, M.H., van Kerkwijk, M.H., Verbunt, F.V.M., Kulkarni, S. 2000,
 A\&A, 358, 605
 % A deep search for the optical counterpart to the AXP 1E 2259+586

 Hurley, K., Kouveliotou, C., Cline, T., Mazets, E., Golenetskii, S.,
 Frederiks, D.D., van Paradijs, J. 1999a, ApJ, 523, L37
 % Where is SGR 1806-20

 Hurley, K., Kouveliotou, C., Murakami, T., et al. 1999b, ApJ, 510, L111
 % ASCA discovery of an x-ray pulsar in the error box of SGR 1900+14

 Hurley, K., Strohmayer, T., Kouveliotou, C., et al. 2000, ApJ, 528, L21
 % ASCA observations of the quiescent x-ray counterpart to SGR 1627-41

 Inan, U.S., Lehtinen, N.G., Lev-Tov, S.J., Johnson, M.P.,
 Bell, T.F., Hurley, K. 1999, Geophys. Res. Lett., 26, 3357
 % Ionization of lower ionosphere by gamma-rays from a magnetar:
 % Detection of a low-energy (3-10 keV) component (SGR 1900+14)

 Iglesias, C.A., Rogers, F.J. 1996, ApJ, 464, 943
 % Opacities

 Israel, G.L., Mereghetti, S., Stella, L. 1994, ApJ, 433, L25
 % Discovery of AXP 0142+61

 Israel, G.L., Covino, S., Stella, L., Campana, S., Haberl, F.,
 Merghetti, S. 1999a, ApJ, 518, L107
 % Further evidence that 1RXS J170849.0-400910 is an AXP

 Israel, G.L., Ooosterbroek, T., Angelini, L., Campana, S.,
 Mereghetti, S., Parmar, A.N., Segreto, A., Stella, L., van Paradijs, J.,
 White, N. 1999b, A\&A, 346, 929
 % BeppoSAX monitoring of AXP 4U 0142+61

 Kanbach, G., et al. 1980, A\&A, 90, 163
 %Detailed characteristics of the high-energy
 %gamma radiation from PSR 0833-45 measured by COS-B

 Kanbach, G., et al. 1994, A\&A, 289, 855
 % Vela pulsar light curves with EGRET
 Kaspi, V.M. 2000, in Pulsar Astronomy -- 2000 and Beyond,
   eds. M.~Kramer, N.~Wex and R.~Wielebinski,
   ASP Conference Series, v.202, p.485
 % NS/SNR Associations

 Kaspi, V.M., Chakrabarty, D., Steinberger, J. 1999, ApJ, 525, L33
 % Precision timing of two AXPs [J1708-4009 and 2259+586]

 Kaspi V.M. Johnston S., Bell J., et al., 1994 \ApJ 423, L43
 % braking index of 1509-58

 Kawai, N.,  Tamura, K., 1996, in {\em IAU Colloquium 160}, eds S.Johnston,
 M.A.Walker and M.Bailes, p367 
 % X-ray bright PWN arround several pulsars

 Kawai, N., Saito, Y., 1999, Astro. Lett. and Communications, 38, 1
 % 1821

 Kellett, B.J., Branduardi-Raymont, G., Culhane, J.L., Mason, I.M.,
 Mason, K.O., Whitehouse, D.R. 1987, MNRAS, 225, 199
 % EXOSAT observations of PKS 1209--52

 Kendziorra, E., Staubert, R., Pietsch, W., Reppin, C., Sacco, B., 
 Tr\"umper, J., 1977, ApJ, 217, L93
 %Timing analysis of Her X-1 data

 Kniffen, D.A., 1990, in {\em Observatories in Earth Orbit and Beyond},
 eds Y.~Kondo, Kluwer Academic Publishers, p.63
 % GRO Mission paper

 Kniffen, D.A., Hartman, R.C., Thompson, D.J., Bignami, G.F., Fichtel, C.E.,
 T\"umer T., \"Ogelman, H., 1974, Nature, 251, 397
 % Confrimation of the Crab gamma-ray pulses using SAS-2

 Koptsevich, A.B., Pavlov, G.G., Shibanov, Yu.A., Sokolov, V.V., Zharikov, S.V.,
 Kurt, V.G. 2000, A\&A, accepted
 % Optical spectrum of 0656+14

 Kouveliotou, C., Dieters, S., Strohmayer, T., van Paradijs, J.,
 Fishman G.L., Meegan, C.A., Hurley, K., Kommers, J., Smith, I.,
 Frail, D., Murakami, T. 1998, Nature, 393, 235
 % An X-ray pulsar with superstrong magnetic field in the SGR 1806-20

 Kriss, G.A., Becker, R.H., Helfand, D.J., Canizares, C.R. 1985, ApJ, 288, 703
 % Einstein observation of the Kes 73 central source

 Kuiper, L., Hermsen, W., Krijger, J.M., Bennett, K.,  et al., 1999, A\&A, 351,
 119
 %  Gamma-ray paper of 1509-58

 Kuiper L., Hermsen W., Verbunt F., Thompson, D.J., Stairs, I.S.,
 Lyne, A.G., Strickman, M.S., Cusumano, G. 2000, A\&A, 359, 615
 % The likely detection of pulsed high-energy gamma-ray emission from ms-psr 0218

 Kulkarni, S.~R.,  Anderson, S.~B. 1996, in {\em Dynamical Evolution of
 Star Clusters -- Confrontation of Theory and Observations}: IAU
 Symposium 174, Kluwer Academic Publisher, p.181  

 Kulkarni, S.R., Frail, D.A. 1993, Nature, 356, 33
 % association of SGR 1906-20 with SNR G10.0-0.3

 Kulkarni, S.R., van Kerkwijk, M.H. 1998, ApJ, 507, L49
 % Optical observations of INS RX J0720.4-3125

 Kundt, W., Schaaf, R., 1993, Ap \& SpSc, 200, 251
 %Hot polar caps

 Landau, L., 1932, Phys.Z. Sowjetunion, 1, 285
 % On the theory of stars

 Large, M.I., Voughan, A.E., Mills, B.Y., 1968, Nature, 220, 340
 % A pulsar supernova association (Vela pulsar discovery)

 Lattimer, J.M., Pethick, C.J., Prakash, M., Haensel, P. 1991,
 Phys. Lett., 66, 2701

 Lommen, A.N., Zepka, A., Backer, D.C., Cordes, J.M., Arzoumanian, Z.,
 McLaughlin, M., \& Xilouris, K. 2000, ApJ, submitted

 Long K.S., Helfand D.J., 1979, ApJ, 234, L77
 % Supernova remnants in the large magelanic cloud

 Lyne A. G, Pritchard R. S, Smith F. G., 1988, \MNRAS, 233, 667
 % braking index of Crab

 Lyne A. G, Pritchard R. S, Graham-Smith F., Camilo F., 1996, Nature, 381, 497
 % braking index of Vela

 Lyne, A.~G., Camilo, F., Manchester, R.~N., Bell, J.~F., Kaspi, V.~M., 
 D'Amico, N., McKay, N. P.~F., Crawford, F., Morris, D.~J., Sheppard, D.~C., 
 \& Stairs, I.~H., 2000, MNRAS, 312, 698
 
 Manchester, R.N., et al. 1978, MNRAS, 184, 159
 % Optical observations of the Vela pulsar

 Manchester, R.~N., Lyne, A.~G., Camilo, F., Kaspi, V.~M., Stairs,
  I.~H., Crawford, F., Morris, D.~J., Bell, J.~F., \& D'Amico, N.,
  2000, in {\it Pulsar Astronomy - 2000 and Beyond}, ed. M.Kramer, 
  N.Wex, and R.Wielebinski, [San Francisco : ASP], p.49

 Marsden, D., Rotschild, R.E., Lingenfelter, R.E. 1999, ApJ, 520, L107
 % Is SGR 1900+14 a magnetar?

 Marshall, F.E., Gotthelf, E.V, Zhang, W., Middleditch, J., Wang, Q.D.,
 1998, ApJ, 499, L179
 %Discovery of an Ultrafast X-Ray Pulsar in the Supernova Remnant N157B

 Martin, C., Halpern, J.P., Schiminovich, D. 1998, ApJ, 494, L211
 % Optical spectrum of Geminga

 Mazets, E.P., et al. 1979a, Nature, 282, 587
 % Discovery of 8 sec pulsations in SGR 0526-66

 Mazets, E.P., Golenetskii, S.V., Guryan, Yu. 1979b, Sov. Astron. Lett., 5(6), 343
 % Discovery of SGR 1900+14

 Mereghetti, S. 2000, In The Neutron Star -- Black Hole Connection,
 NATO ASI Series, to be published (astro-ph/9911252)
 % The anomalous X-ray pulsars (review)

 Mereghetti, S., Stella, L. 1995, ApJ, 442, L17
 % AXPs as a new class of objects

 Mereghetti, S., Bignami, G.F., Caraveo, P.A. 1996, ApJ, 464, 842
 % Central source of PKS 1209-52

 Michel, F.C., 1991, Theory of Neutron Star Magnetospheres, University of 
 Chicago Press, ISBN 0-226-52331-4

 Mignani, R., Caraveo, P.A., Bignami, G.F. 1997, ApJ, 474, L51
 % HST discovers optical radiation from PSR 1055-52

 Mineo, T., Cusumano, G., Kuiper, L., Hermsen, W., Massaro E.,
 Becker, W., Nicastro, L., Sacco, B., Verbunt, F., Lyne, A.G.,
 Stairs, I.H., Shibata, S. 2000, A\&A, 355,  1053
 % BSAX observation of PSR J0218

 Mineo, T., Cusumano, G., Massaro, E., Nicastro, L., Parmar, A.N., Sacco, B., 1999, A\&A, 348,  519
 % X-ray paper of PSR 0540-69

 Morrison, P., 1958, Nuovo Cimento, 7, 858
 % On gamma-ray astronomy

 Morrison, P., Olbert, S., Rossi, B., 1954, Phys.Rev., 94,440
 % The origin of cosmic rays

 Motch, C., Haberl, F. 1998, A\&A, 333, L59
 % Constraints on optical emission from the INSC RX J0720.4-3125

 Motch, C., Haberl, F., Zickgraf, F.-J., Hasinger, G., Schwope, A.D. 1999,
 A\&A, 351, 177
 % The INS candidate RX J1605.3+3249

 Murakami, T., Kubo, S., Shibazaki, N., Takeshima, T., Yoshida, A.,
 Kawai, N. 1999, ApJ, 510, L119
 % Accurate position of SGR 1900+14 and changes in pulse period and folded
 % pulse profile with ASCA

 Murray, S.S., Slane, P.O., Seward, D., Ransom, S.C., Gaensler, B.M., 2001, to appear in ApJ
 % Discovery of X-ray Pulsations from the Compact Central Source in the SNR 3C38

 Neuh\"auser, R., Tr\"umper, J. 1999, A\&A, 343, 151
 % On the number of accreting and cooling isolated NSs detectable with the RASS

 \"Ogelman, H., Finley, J.P., Zimmerman, H.U. 1993, Nature, 361, 136
 % ROSAT PSPC observations of the Vela pulsar

 \"Ogelman, H., 1995, in {\em The Lives of Neutron Stars}
 eds A.~Alpar, U.~Kiliz\'oglu \&  J.~van Paradijs, Kluwer Academic Publishers, p.101
 % X-ray observations of Cooling Neutron stars

 Oosterbroek, T., Parmar, A.N., Mereghetti, S., Israel, G.L. 1998,
 A\&A, 334, 925
 % The two-component x-ray spectrum of the 6.4 s pulsar 1E 1048.1-5937

 Oppernheimer, J.R., Volkoff, G.M., 1939, Phys.~Rev., 55, 374
 % On Massive Neutron Cores

 Ostriker, J.P., \& Gunn, J.E. 1969, ApJ, 157, 1395

 Pacini, F., 1967, Nature, 216, 567
 % Energy emission from a neutron star

 Pacini, F., 1968, Nature, 219, 145
 % Rotating neutron stars, pulsars and supernova remnants

 Page, D., Applegate, J.L. 1992, ApJ, 394, L17
 %Standard and accelerated cooling (modified and direct Urca)

 Page, D., Shibanov, Yu.A., Zavlin, V.E. 1996, in {\it R\"ontgenstrahlung
 from the Universe}, MPE Report 263, p.173.

 Pavlov, G.G., M\'esz\'aros, P. 1993, ApJ, 416, 752
 %Finite-velocity effects on atoms in strong magn fields

 Pavlov, G.G., Potekhin, Y.A. 1995, ApJ, 450, 883
 % Bound-bound transitions in strong magn field

 Pavlov, G.G., Shibanov, Yu.A. 1978, Sov.~Astron., 22, 214
 %Thermal emission of an optically thick plasma with a magnetic field
 %(first work on emission from a neutron star atmosphere)

 Pavlov, G.G., Zavlin, V.E. 1997, ApJ, 490, L91
 % Mass-to-radius ratio for 0437-4715

 Pavlov, G.G., Zavlin, V.E. 1999, IAU Circ. \#7270
 % Cas A in Einstein images

 Pavlov, G.G., Zavlin, V.E. 2000, ApJ, 529, 1011
 % Polarization of thermal radiation from NSs

 Pavlov, G.G., Zavlin, V.E. 2001, ApJ, to be published
 %Hydrogen spectral lines from high-B NS atmospheres

 Pavlov, G.G., Shibanov, Yu.A., Ventura, J., Zavlin, V.E. 1994, A\&A,
 289, 847
 % Anisotropy of polar cap emission in strongly magnetized NSs
 % "Model atmospheres and radiation of magnetic NSs: anisotropic thermal
 % emission" (radiation from polar caps)

 Pavlov, G.G., Shibanov, Y.A., Zavlin, V.E., Meyer, R.D., 1995, in
 {\em The Lives of Neutron Stars} eds A.~Alpar, U.~Kiliz\'oglu \&
 J.~van Paradijs, Kluwer Academic Publishers, p.71
 % A review on NS atmospheres

 Pavlov, G.G., Stringfellow, G.S., C\'ordova, F.A. 1996a, ApJ, 467, 370
 % HST observations of 0656+14, 1929+10, 0950+08

 Pavlov, G.G., Zavlin, V.E., Tr\"umper, J., Neuh\"auser, R. 1996b, ApJ, 472, L33
 % Multiwavelength observations of NSs as a tool ... [RX J1856-3754]

 Pavlov, G.G., Welty, A.D., C\'ordova, F.A. 1997, ApJ, 489, L75
 % HST observations of 0656+14

 Pavlov, G.G., Sanwal, D., Garmire, G.P., Zavlin, V.E., Burwitz, V,,
 Dodson, R. 2000a, AAS Meeting 196, \#37.04
 % Chandra observations of the Vela pulsar and its nebula

 Pavlov, G.G., Zavlin, V.E., Aschenbach, B., Tr\"umper, J., Sanwal, D.
 2000b, ApJ, 531, L53
 % Cas A

 Petre, R., Kriss, G.A., Winkler, P.F., Canizares, C.R. 1982, ApJ, 258, 22
 % Einstein observation of the central source of Pup A

 Petre, R. Becker, C.M., Winkler, P.F. 1996, ApJ, 465, L43
 %A Central Stellar Remnant in Puppis A

 Plucinsky P.P., Snowden S.L., Aschenbach B., et al. 1996, ApJ,  463, 224
 % Monogem Ring

 Rajagopal, M., Romani, R.W. 1996, ApJ, 461, 327
 % Nonmagnetic NS atmospheres

 Rajagopal, M., Romani, R.W., Miller, M.C. 1997, ApJ, 479, 347
 % Magnetic Fe atmospheres

 Ramanamurthy, P.V., Fichtel, C.E., Harding, A.K., et al. 1996, A\&AS, 120, 115
 % gamma-rays from 0656+14

 Rasio, F.A., Pfahl, E.D., \& Rappaport, S. 2000, ApJ, 532, L47

 Romani, R.W. 1987, ApJ, 313, 718
 % Thermal emission, ns atmosphere 

 Romani, R.W. 1996, ApJ, 470, 469
 % Gamma-ray pulsars: Radiation processes in the outer magnetosphere

 Romani, R.W., Yadigaroglu, I.-A. 1995, ApJ, 4838, 314
 % Gamma-ray pulsars: Emission zones and viewing geometries

 Rosenfeld, L., 1974, in the Proceedings 16th Solvay Conference on Physics,
 eds F.~Pacini, Editions de l'Universite de Bruxelles, p174
 % a comment made after a talk held at the conference

 Rothschild, R.E., Kulkarni, S.R., Lingenfelter, R.E. 1994, Nature, 368, 432 
 % Discovery of x-ray source coincident with SGR 0525-66

 Ruderman, M., Sutherland, P.G., 1975, ApJ, 196, 51
 % Theory of pulsars: polar gaps, sparks and coherent microwave radiation

 Schwope, A.D., Hasinger, G., Schwarz, R., Haberl, F., Schmidt, M. 1999,
 A\&A, 341, L51
 % The INS candidate RBS1223 (1RXS J130848.6+212708)

 Seward, F.D. 1990, ApJSS, 73, 781   
 % Review on Einstein observations of SNRs in our Galaxy

 Seward, F.D., Harnden, F.R. 1982, ApJ, 256, L45
 % Discovery of 1509--58 with Einstein

 Seward, F.D., Harnden, F.R. 1994, ApJ, 421, 581
 % ROSAT/HRI observation of PRS 0540-69

 Seward, F.D., Wang, Z.-R. 1988, ApJ, 332, 199
 % Pulsars, X-ray Synchrotron Nebulae, and Guest Stars. 
 % Einstein detections of 1929+10 and 0950+08 reported

 Seward, F.D., Charles, P.A., Smale, A.P. 1986, ApJ, 305, 814
 % Discovery of 1E 1048-5937 (AXP)

 Seward, F.D., Harnden, F.R., Helfand, D.J. 1984, ApJ, 287, L19
 %Discovery 0f 0540--69 with Einstein

 Shearer, A., Redfern, R.M., Gorman, G,, et al. 1997, ApJ, 487, L181
 % Pulsed optical emission from 0656+14

 Shibanov, Yu.A., Yakovlev, D.G. 1996, A\&A, 309, 171
 %Temperature distribution over the surface of NS with a dipole magn field

 Shibanov, Yu.A., Zavlin, V.E., Pavlov, G.G., Ventura, J. 1992,
 A\&A, 266, 313
 % NS atmosphere models

 Shibanov, Yu.A., Pavlov, G.G., Zavlin. V.E., Qin, L., Tsuruta, S. 1995,
 in Proc. 17-th Texas Symposium on
        Relativistic Astrophysics, ed.~H.~Bohringer, G.~Morfill, \&
        J.~Tr\"umper, Ann.~NY Acad.~Sci., 759, 291
 % Emission from the NS surface with anisotropic T distribution

 Sonobe, T., Murakami, T., Kulkarni, S.R., Aoki, T., Yoshida, A. 1994,
 ApJ, 436, L23
 % persistent emission of SGR 1806-20 [ASCA]

 Staelin, D.H., Reifenstein, III, E.C., 1968, Science, 162, 1481
 % Pulsating radio sources near the Crab nebula

 Strickman, M.S., Harding, A.K., de Jager, O.C. 1999, ApJ, 524, 373
 % RXTE observations of the Vela pulsar

 Sturner, S.J., Dermer, C.D. 1994, ApJ, 420, L79
 %polar cap models -- inverse Compton included

 Sturner, S.J., Dermer, C.D., Michel, F.C. 1995, ApJ, 445, 736
 %Magnetic Comptob-induced pair cascade model for gamma-ray pulsars

 Sugizaki, M., Nagase, F., Torii, K., Kunigasa, K., Asanuma, T.,
 Matsuzaki, K., Koyama, K., Yamauchi, S. 1997, PASJ, 49, L25
 % Discovery of an 11 s X-ray pulsar [J170849.0-400910]

 Takahashi M., Shibata S., Torii K., Saito Y., Kawai N. 1998, IAU Circ. 7030
 % PSR 1937+21 detection with ASCA

 Tanaka, Y., Inoue, H., Holt, S., 1994, PASJ, 46, L37
 % ASCA Mission paper

 Tananbaum, H. 1999, IAU Circ. \#7246
 % Discovery of the Cas A central source

 Tananbaum, H., Gursky, H., Kellogg, E.M., Levinson, R., Schreier, E.,
 Giacconi, R. 1972, ApJ, 174, L143
 % Discovery of Her X-a with Uhuru

 Taylor, B.G., Andersen, R.D., Peacock, A., Zobl, R. 1981, Space Sci.~Rev., 30, 479
 % The EXOSAT Mission

 Taylor, J.H., Manchester, R.N., Lyne, A.G., 1993, ApJSuppl., 89, 189
 % Catalog of 558 pulsars

 Thompson, C., Duncan, R.C. 1995, MNRAS, 275, 255
 % SGRs as Magnetars. I.
 
 Thompson, C., Duncan, R.C. 1996, ApJ, 473, 322
 % SGRs as Magnetars. II

 Thompson, D.J., Fichtel, C.E., Kniffen, D.A., \"Ogelman, H.B., 1975, ApJ, 200, L79
 % SAS-2 high-energy gamma-ray observations of the Vela pulsar^M

 Thompson, D.J., et al. 1999, ApJ, 516, 297
 %"Gamma Radiation from PSR B1055--52", with a review on other gamma pulsars

 Torii, K., Kunigasa, K., Katayama, K., Tsunemi, H., Yamauchi, S. 1998,
 ApJ, 503, 843
 % Discovery of a 7 s AXP AX J1845.0--0300

 Treves, A., Turolla, R., Zane, S., Colpi, M. 2000, PASP, 112, 297
 % INSs: Accretors and Coolers

 Tr\"umper, J., Pietsch, W., Reppin, C., Voges, W., Staubert, R., 
 Kendziorra, E.
 1978, ApJ, 219, L105
 %Evidence for strong cyclotron line emission in the hard X-ray spectrum of Hercules X-1

 Tr\"umper, J., 1983, Adv.~Space Res., 2, 241
 % A ROSAT Mission paper

 Tsuruta S., 1998, Physics Reports, 292, 1
 % Review on cooling neutron stars

 Tucker, W. 1984, The Star Splitters, NASA SP-466
 % A description of the HEAO project

 Tuohy, I.R., Garmire, G.P. 1980, ApJ, 239, L107
 % Einstein observations of RCW 103 (central source)

 van Kerkwijk, M.H., Kulkarni, S.R., Matthews, K., Neugebauer, G. 1995, ApJ, 444, L33
 % A luminous companion to SGR 1806-20

 van Paradijs, J., Taam, R.W., van den Heuvel, E.P.J. 1995, A\&A, 299, L41
 % AXPs as accreting objects

 Vasisht, G., Gotthelf, E.V. 1997, ApJ, 486, L129
 % Discovery of AXP in Kes 73

 Verbunt F., Kuiper L., Belloni T., et al. 1996, A\&A, 311, L9
 % PSR J0218 detection with HRI

 Walter, F.M., Matthews, L.D. 1997, Nature, 389, 358
 % HST observations of 1856-3754

 Walter, F.M., Volk, S.J., Neuh\"auser, R. 1996, Nature, 379, 233
 % Discovery of INS RX J1856-3754

 Walter, F.M., An, P., Lattimer, J., Prakash, M. 2000, in Highly Energetic
 Physical Processes and Mechanics for Emission from Astrophysical Plasmas,
 eds. P.C.H. Martens, S. Tsuruta and M.A. Weber, IAU Symp. 195, p.437
 % The isolated NS RX J185635-3754

 Wang, Z.-R., Seward, F.D. 1984, ApJ, 285, 607
 % Detection of 1951+32 with Einstein

 Wang, Q.D., Gotthelf, E.V. 1998, ApJ, 509, L109
 % ROSAT HRI detection of the 16 ms pulsar J0537-6910

 Wang, F.Y.-H., Ruderman, M., Halpern, J.P., Zhu, T. 1998, ApJ, 498, 373
 % Models for x-ray emission from isolated pulsars

 Weber, F. 1999, Pulsars as Astrophysical Laboratories
 for Nuclear and Particle Physics, Institute of Physics,
 ISBN 0-7503-0332-8

 Weisskopf, M.C., Hester, J.J., Tennant, A.F., et al. 2000, ApJ, 536, L81
 % Spatial and spectral structure of the x-ray Crab nebula

 White, N.E., Angelini, L., Ebisawa, K., Tanaka, Y., Ghosh, P. 1996,
 ApJ, 463, L83
 % The spectrum of the 8.7 s pulsar 4U 0142+61

 Wills, R.D., Bennett, K., Bignami, G.F., Buccheri, R., Caraveo, P.A., Hermsen, W.,
 Kanbach, G., Masnou, J.L., Mayer-Hasselwander, H.A., Paul, J.A., Sacco, B., 1982,
 Nature, 296, 723
 %COS-B observations of the Crab-Pulsar

 Woods, P.M., Kouveliotou, C., van Paradijs, J., Finger, M.H., Thompson, C.
 1999a, ApJ, 518, L103
 % BeppoSAX observations of SGR 1900+14 in quiescence and during an active period

 Woods, P.M., Kouveliotou, C., van Paradijs, J., et al. 1999b, ApJ, 524, L55
 % Variable spin-down in SGR 1900+14 and correlation with the burst activity

 Woods, P.M., Kouveliotou, C., van Paradijs, J., et al. 1999c, ApJ, 527, L47
 % Hard burst emission from SGR 1900+14

 Woods, P.M., Kouveliotou, C., van Paradijs, J., et al. 1999d, ApJ, 519, L139
 % Discovery of a new SGR 1627-41

 Yakovlev, D.G., Levenfish, K.P., Shibanov, Yu.A. 1999, Physics-Uspekhi, 169, 825
 %Cooling NSs and superfluidity in their interiors

 Yancopoulos, S., Hamilton, T.T., Helfand, D.J., 1994, \ApJ, 429, 832
 %1929+10 ROSAT paper

 Zavlin V.E., Pavlov G.G., 1998, A\&A, 329, 583
 % Thermal x-rays from PSR 0437-4715

 Zavlin, V.E., Pavlov, G.G.,  Shibanov, Yu.A., Ventura, J., 1995a, A\&A, 297, 441 
 % Thermal radiation  from rotating neutron star:  effect of the magnetic 
 % field and surface temperature distribution.
 % Thermal emission from NS atmosphere with anisotropic temperature

 Zavlin, V.E., Shibanov, Yu.A., Pavlov, G.G. 1995b, Astron. Lett., 21, 168
 % Effect of the Neutron Star Gravitational Field on
 %  Radiation from the Hot Polar Spots of Radio Pulsar

 Zavlin, V.E., Pavlov, G.G., Shibanov, Yu.A. 1996, A\&A, 315, 141
 % Nonmagnetic NS atmospheres

 Zavlin, V.E., Pavlov, G.G., Tr\"umper, J. 1998, A\&A, 331, 821
 % Central source of PKS 1209-52

 Zavlin, V.E., Tr\"umper, J., Pavlov, G.G. 1999, ApJ, 525, 959
 % X-ray emission from radio-quiet NS in Puppis A

 Zavlin, V.E., Pavlov, G.G., Sanwal, D., Tr\"umper, J. 2000, ApJ, 540, L25
 % Discovery of 424 ms pulsations in CCO of PKS 1209-51/52
 
 Zavlin, V.E., Pavlov, G.G., Halpern, J.P. 2001, ApJ, submitted
 % A broad-band spectrum of PSR 0656+14

 Zhang, B., \& Harding, A.K. 2000, ApJ, 532, 1150

\end{small}
\end{verse}

\end{document}